\documentclass[aps,pre,twocolumn,10pt]{revtex4-2}
\usepackage[colorlinks = True, linkcolor = blue, citecolor = blue, breaklinks=True]{hyperref}
\usepackage{times}
 \usepackage{graphicx}
\usepackage{placeins}  
\usepackage{setspace}
\usepackage{calc}
\usepackage{calrsfs}
\usepackage{float}
\usepackage{mathtools}
\usepackage{ulem}
\usepackage{amssymb,amsmath,amsfonts}
\usepackage{graphics,epsfig}
\usepackage{color,bm,hyperref}
\usepackage{xcolor}
\usepackage{tikz}
\usepackage{yfonts}
\usepackage{comment}
\usepackage{multirow}
\usepackage{array}

\usetikzlibrary{shapes.geometric,arrows}
\usetikzlibrary{decorations.markings}
\DeclareMathAlphabet{\pazocal}{OMS}{zplm}{m}{n}

\definecolor{lime}{HTML}{A6CE39}
\DeclareRobustCommand{\orcidicon}{\hspace{-4pt}
\begin{tikzpicture}
\draw[lime, fill=lime] (0,0)
circle [radius=0.16]
node[white] {\hspace{0.1mm}{\fontfamily{qag}\selectfont \tiny ID}};
\draw[white, fill=white] (-0.07,0.1)
circle [radius=0.01];
\end{tikzpicture}
\hspace{-3.2mm}
}
\foreach \x in {A, ..., Z}{\expandafter\xdef\csname
orcid\x\endcsname{\noexpand\href{https://orcid.org/\csname orcidauthor\x\endcsname}
{\noexpand\orcidicon}}
}

\begin{document}

\title{Analytical approach to subsystem resetting in generalized Kuramoto models}
\author{Rupak Majumder\orcidA{} \thanks{These authors contributed equally to this work.}}
\email{Equal Contribution: rupak.majumder@tifr.res.in}
\affiliation{Department of Theoretical Physics, Tata Institute of Fundamental Research, Homi Bhabha Road, Mumbai 400005, India}
\author{Anish Acharya\orcidB{} \thanks{These authors contributed equally to this work.}}
\email{Equal Contribution: anish.acharya@tifr.res.in}
\affiliation{Department of Theoretical Physics, Tata Institute of Fundamental Research, Homi Bhabha Road, Mumbai 400005, India}
\author{Shamik Gupta\orcidC{}}
\email{shamik.gupta@theory.tifr.res.in}
\affiliation{Department of Theoretical Physics, Tata Institute of Fundamental Research, Homi Bhabha Road, Mumbai 400005, India}

\begin{abstract}
Stochastic resetting has emerged as a powerful mechanism for driving systems into nonequilibrium stationary states with tunable properties. While most existing studies focus on global resetting, where all degrees of freedom are simultaneously reset, recent work has shown that resetting only a subset of degrees of freedom (subsystem resetting) can qualitatively alter collective behavior in interacting many-body systems. In this work, we develop a general theoretical framework for analysing subsystem resetting in Kuramoto-type coupled-oscillator systems. Building on a continued-fraction approach, we derive self-consistent equations for the stationary-state order parameter of the non-reset subsystem, applicable to both noisy and noiseless dynamics and to models with arbitrary interaction harmonics. Using this framework, we systematically investigate how the stationary state and phase transitions depend on the resetting rate, the size of the reset subsystem, and the reset configuration. We show that subsystem resetting can shift or even suppress synchronization transitions, and can give rise to nontrivial features such as re-entrant behavior and restructuring of phase boundaries. In specific cases, including the noiseless Kuramoto model with a Lorentzian frequency distribution, our results recover known analytical predictions and extend them to more general settings. These results establish subsystem resetting as a versatile control protocol for engineering collective dynamics in nonequilibrium interacting systems.
\end{abstract}
\maketitle


\section{\label{sec:level1}Introduction}
Stochastic resetting has emerged as a versatile paradigm for driving generic systems, be they classical or quantum, single or many-particle, into nonequilibrium stationary states with nontrivial properties. Originally introduced in the context of single-particle diffusion, where resetting to a fixed configuration at random times yields a nonequilibrium stationary state with significantly-modified first-passage behavior~\cite{evans2011diffusion}, the framework has since been broadened to cover a wide spectrum of systems and applications~\cite{evans2020stochastic,gupta2022stochastic,nagar2023stochastic,pal2022inspection, evans2024stochasticresettinglargedeviations}. It is worth noting the significant growth the field of resetting has experienced in recent years, drawing increasing attention across multiple and diverse domains: classical~\cite{evans2013optimal,PhysRevLett.113.220602,gupta2014fluctuating, PhysRevE.92.062148,christou2015diffusion, pal2016diffusion,PhysRevE.96.022130,majumdar2018spectral,masoliver2019telegraphic,boyer2019anderson,basu2019symmetric,mendez,chechkin,karthika2020totally,magoni2020ising,ciliberto,PhysRevE.104.024105,PhysRevResearch.4.013161,PhysRevResearch.2.043390,PhysRevE.101.062147,rosemary,gupta2021resetting,de2021optimization,evans2022exactly,rksingh,sarkar2022,ahmad2022first,PhysRevE.106.044127,bressloff2024global,magoni2020ising,montero2024effect,PhysRevResearch.6.033189, aron2024controlspatiotemporalchaosstochastic,acharya2025_kuramoto}, quantum~\cite{mukherjee2018quantum,perfetto2021designing,fazio,das2022quantum,acharya2023tight,yin2023restart,PhysRevA.108.062210}, chemical~\cite{PhysRevE.92.060101,PhysRevResearch.3.013273},  biological~\cite{ramoso2020stochastic}, financial~\cite{stojkoski2022income,jolakoski2023first}. These studies have established resetting as a powerful mechanism for controlling fluctuations and relaxation in dynamical systems.

An important direction concerns interacting many-body systems, where resetting competes with intrinsic interactions and collective effects; see Ref.~\cite{nagar2023stochastic} on resetting effects in interacting systems. We now highlight a representative subset of works most relevant to our focus, while noting that many other contributions exist in this rapidly-developing area. Early work in this context focused on fluctuating interfaces, demonstrating that resetting leads to stationary states with modified scaling properties and non-Gaussian fluctuations~\cite{gupta2014fluctuating}, including cases with non-Poissonian resetting protocols~\cite{Gupta_2016}. Resetting effects have also been investigated in driven interacting particle systems such as the totally asymmetric simple exclusion process, where the interplay of resetting with nonequilibrium transport leads to nontrivial stationary states and modified current–density relations~\cite{karthika2020totally}. More recently,  resetting has been explored in spin systems and in coupled nonlinear oscillator systems in the framework of the celebrated Kuramoto model. In particular, the Ising model with resetting exhibits a nonequilibrium stationary state with a nontrivial phase diagram~\cite{magoni2020ising,k2026isingmodelpowerlaw}, while in coupled oscillator systems, resetting promotes phase synchronization among the oscillators and alters collective dynamics~\cite{sarkar2022,acharya2025_kuramoto}. A common feature of these studies is that resetting is implemented globally, i.e., all the constituent degrees of freedom of the system are reset simultaneously. This leads to complete erasure of memory between reset events, which renders the dynamics amenable to analytical approaches involving the renewal theory~\cite{evans2020stochastic}, but at the same time leads to the rounding of phase transitions into smooth crossovers.

A qualitatively different scenario arises when resetting is applied only to a subset of degrees of freedom~\cite{PhysRevE.109.064137}. In such subsystem resetting protocols, memory is only partially erased, and its effects persist through the non-reset components. As a result, the renewal structure breaks down, and the interplay between resetting and interactions gives rise to fundamentally new behavior~\cite{PhysRevE.109.064137}. Recent work has shown that subsystem resetting can be used in a diverse range of interacting systems to manipulate phase behavior, including shifting, splitting, eliminating, or inducing phase transitions~\cite{acharya2025manipulating}. In contrast to global resetting, subsystem resetting can thus qualitatively restructure phase diagrams and generate phenomena absent in the underlying dynamics. From a theoretical standpoint, subsystem resetting poses significant challenges due to the absence of renewal structure and the coupled evolution of reset and non-reset sectors. 

The particular class of interacting many-body  systems we are interested in this work concerns coupled nonlinear oscillators, which exhibit the collective behavior of synchronization, a ubiquitous phenomenon in nature  manifesting in diverse settings, ranging from the collective flashing of fireflies~\cite{buck1966biology} and the coordinated firing of neurons~\cite{Gray1989} to the rhythmic applause of audience~\cite{neda2000sound}. This phenomenon has been observed across domains, from classical~\cite{PhysRevLett.64.821, PhysRevLett.76.1816, PhysRevLett.94.163901, PhysRevLett.96.024101, boda2013rhythm}, quantum~\cite{PhysRevLett.111.234101, PhysRevResearch.1.033012, PhysRevLett.125.013601, PhysRevResearch.5.033209}, electrical~\cite{PhysRevLett.102.074101, motter2013spontaneous, sajadi2022synchronization}, chemical~\cite{zaikin1970concentration, doi:10.1126/science.1070757, doi:10.1126/science.1166253}, biological~\cite{ winfree1980geometry, Cobb1995, Rubchinsky2012, Sarfati2020FireflySync, Weber2020YeastSync, 10.7554/eLife.78908}, to financial~\cite {WALTI201196, GUO2021101934, Baltakiene2022}. Modern theoretical study of synchronization began with the pioneering work of Winfree, who introduced a mathematical framework to describe coupled oscillators in biological contexts~\cite{winfree1980geometry}. Building on this foundation, Kuramoto proposed a remarkably simple yet powerful model to capture the onset of collective synchronization in large populations of nonlinear oscillators with distributed natural frequencies~\cite{kuramoto1984chemical}. The Kuramoto model has since become the canonical paradigm for studying synchronization. It is simple enough to permit analytical predictions, yet rich enough to capture the essential features of synchronization~\cite{Strogatz1990, STROGATZ20001, 
 OA, Gupta_2014, pikovsky2015dynamics, gupta2018statistical}. Despite its simplicity, the Kuramoto model appears in widely different contexts across length and timescales, for example, in collective flavor oscillations of supernova neutrinos~\cite{PhysRevD.58.073002} and in superradiance within optical cavities~\cite{PhysRevA.94.023807, PhysRevX.11.021052}.

The original Kuramoto model describes globally-coupled phase oscillators with distributed natural frequencies, interacting through the sine of their phase differences (also known as first-harmonic interaction). Subsequent generalizations incorporated higher-harmonic interactions as well as stochastic noise~\cite{PhysRevE.88.052111}. In these models, the stationary state typically exhibits a transition from an incoherent (disordered) phase to a synchronized (ordered) one as system parameters are varied. In the former phase, the system does not exhibit any macroscopic synchronization of phases of the oscillators, in contrast to the latter phase. The transition points, as well as the nature of the transition, depend on the system parameters, including the parameters of the frequency distribution, the strength of the inter-oscillator interactions, and the noise intensity. Among the two aforementioned phases, one is often more desirable than the other from a practical standpoint. For example, in the synchronization of neuronal populations associated with Parkinson’s disease, excessive synchrony is pathological and undesirable, whereas for cardiac cells, synchronization is crucial for proper heart function and efficient contractions. Hence, a natural question arises: If the system resides in a parameter regime where the desired phase is unstable dynamically, causing the system to settle into the undesirable phase, what is an efficient way to drive the system into the desired phase when system parameters cannot be tuned at one's will? In general, the question is how to stabilize a dynamically and thermodynamically unstable phase without directly tuning the microscopic interactions of the system. The subsystem resetting protocol achieves precisely this goal, as has been first demonstrated in Refs.~\cite{PhysRevE.109.064137,acharya2025manipulating}.


Following Refs.~\cite{PhysRevE.109.064137, acharya2025manipulating}, in the subsystem resetting protocol, the dynamics of the system is interrupted repeatedly at random times at which a subpart of the system is reset to the desired state, while the rest evolves undisturbed~\cite{ZHAO2025130416, 10.1063/5.0246886, wz1y-hgnk, LI2025130903}. Between successive resets, the system evolves according to its intrinsic (bare) dynamics. The part of the system undergoing bare dynamics interspersed with random-time resetting is called the reset subsystem, while the remainder of the system, which evolves according to the intrinsic dynamics, is called the non-reset subsystem. Our central question is then the following: How do the properties of the non-reset subsystem, such as the amount of order (synchrony) in the stationary state, the nature of transitions and transition points, change depending on the characteristics of the subsystem resetting protocol? In particular, we investigate how these features can be controlled by varying (i) the size of the reset subsystem, (ii) how often the reset happens (rate of resetting), and (iii) the nature of
the reset configuration (the specific configuration the reset subsystem is reset to during each reset event)?


Reference~\cite{PhysRevE.109.064137} studied subsystem resetting in the noiseless Kuramoto model for specific frequency distributions (e.g., Lorentzian and Gaussian) and considered only resetting to the fully-synchronized state. The work reported that the continuous transition of the bare model gets converted into a crossover for any resetting rate and any size of the reset subsystem. The analysis was performed using the so-called Ott-Antonsen ansatz~\cite{OA, Ott_2009}, which yields a tractable low-dimensional description of the dynamics of the order parameter in the thermodynamic limit (the limit of the total number of oscillators $N\to \infty$). However, this ansatz is inherently restrictive as it applies only to a specific invariant manifold of initial conditions and ceases to hold in the presence of noise. To address these limitations, Ref.~\cite{acharya2025manipulating} introduced a continued-fraction method~\cite{Risken1980,acharya2025_kuramoto} to study subsystem resetting in the noisy Kuramoto model with only first-harmonic interaction. It considered a range of reset configurations, from asynchronous to fully synchronized, and studied how the phase transition in the non-reset subsystem depends on the degree of synchrony in the reset configuration.


In this paper, we first elaborate on the continued-fraction method introduced in Ref.~\cite{acharya2025manipulating} and apply it to various Kuramoto models with first-harmonic interaction, deriving a self-consistent relation for the average order parameter of the non-reset subsystem. We demonstrate that this method is applicable to both noisy and noiseless Kuramoto models. Using this approach, we analytically compute how the stationary-state phase distribution of the non-reset subsystem changes as we vary the resetting rate, the size of the reset subsystem, and the reset configuration across different system parameters. In particular, for resetting to the incoherent state, we explicitly determine how the transition points shift as functions of the resetting rate, the size of the reset subsystem, and the system parameters. For the noiseless Kuramoto model with a Lorentzian frequency distribution, we further show that our approach reproduces the results of Ref.~\cite{PhysRevE.109.064137} obtained using the Ott–Antonsen ansatz. We then extend this method to Kuramoto models with higher-harmonic interactions and apply it to the noisy Kuramoto model with first- and second-harmonic interactions to derive a self-consistent relation for the average order parameter of the non-reset subsystem. For this case also, we compute the transition points for resetting to the incoherent state. Finally, we validate all our analytical predictions through explicit numerical simulations. Our results, shown in Figs.~\ref{fig: model 1},~\ref{fig: model 2},~\ref{fig: model 4}, and~\ref{fig: reentrant_window}, and summarized in Table~\ref{tab:comparison_resetting}, illustrate novel features unique to subsystem resetting, including nontrivial restructuring of phase boundaries (and even its suppression!), and a remarkable re-entrant transition. Our work, therefore, establishes subsystem resetting as a powerful control protocol for engineering collective behavior in nonequilibrium many-body systems.

\begin{table*}[htbp]
	\caption{Summary of main results of different models in presence of subsystem resetting: Here, $\lambda$ is the resetting rate, $f$ is the fraction of the total number of oscillators that are being reset, $r_0$ is the order parameter for the reset configuration, $2\sigma$ gives the width of the frequency distribution, while $\mu_+(\mathbf{J})$ denotes the largest eigenvalue of the $2\times 2$ matrix $\mathbf{J}$, with $\mathbf{J}$ given by Eq.~\eqref{eq:jacobian}. The re-entrant transition is depicted in Fig.~\ref{fig: reentrant_window} and discussed in detail in Sec.~\ref{subsec: noisy-harmonic-biharmonic Bare}.}
	\label{tab:comparison_resetting}
	\begin{ruledtabular}
		\begin{tabular}{lccc}
			\multirow{2}{*}{Model} 
			& \multicolumn{2}{c}{\parbox{9cm}{\centering Reset to $r_0=0$:\\ Phase transition between incoherent and synchronized phases}} 
			& \multirow{2}{*}{\parbox{3cm}{\centering Reset to\\ $r_0\neq0$}} \\
			\cline{2-3}
			\noalign{\vspace{1.2ex}}
			& Transition point $K_1^\mathrm{c}(\lambda)$ 
			& Behavior of $K_1^\mathrm{c}$ against $\lambda$ 
			& \\
			\noalign{\vspace{0.6ex}}
			\hline
			Model~\ref{subsec: Iso-noisy-harmonic Bare}
			& $\displaystyle K_1^\mathrm{c}(\lambda) = \frac{2D(D+\lambda)(4D+\lambda)}{(1-f)\lambda^2+D(5-3f)\lambda+4D^2}$
			& Monotonic
			& \multirow{4}{*}{\parbox{3.5cm}{\centering No phase transition,\\ but only a crossover}} \\[4.5ex]
			
			Model~\ref{subsec: noiseless-harmonic-lorentzian Bare}
			& $\displaystyle K_1^\mathrm{c}(\lambda) = \frac{2\sigma(\sigma+\lambda)(2\sigma+\lambda)}{(1-f)\lambda^2+(3-f)\sigma\lambda+2\sigma^2}$
			& Monotonic
			& \\[4.5ex]
			
			Model~\ref{subsec: noiseless-harmonic-uniform Bare}
			& $\displaystyle K_1^\mathrm{c}(\lambda) = \frac{4\sigma}{(1-f)\pi+2f\tan^{-1}(2\sigma/\lambda)}$
			& Monotonic
			& \\[4.5ex]
			
			Model~\ref{subsec: noisy-harmonic-biharmonic Bare}
			& $\displaystyle K_1^\mathrm{c}(\lambda) = \frac{2}{\mu_+(\mathbf{J})}$
			& Non-monotonic (re-entrant)
			& \\
		\end{tabular}
	\end{ruledtabular}
\end{table*}

The paper is organized as follows. In Sec.~\ref{sec:level2}, we define the generalized Kuramoto model with general harmonic interactions and its particular variants for which we will elucidate the effects of subsystem resetting. In Sec.~\ref{con lim}, we discuss the continuum limit of the bare Kuramoto model. In Sec.~\ref{sec:level3}, we discuss in detail the specifics of the subsystem resetting protocol. The case of first-harmonic interaction is taken up in Sec.~\ref{sec:first-harmonic}, which we study using our developed analytical formalism discussed in detail in this same section. The effects of additional second-harmonic interaction is discussed in detail in Sec.~\ref{sec: app to gen} in which we also spell out the extension of our analytical formalism for general interactions. The paper ends with conclusions. Some of the technical details of the main text are presented in the Appendixes.

\section{\label{sec:level2}Generalized Kuramoto model}
As mentioned in the Introduction, the Kuramoto model involves a system of $N$ globally-coupled phase-only oscillators. We denote the phase of the $j$th oscillator at time $t$ by the angle variable $\theta_j(t) \in [0,2\pi)$, with $j=1,2,\ldots,N$. In the
following, the word ‘phase’ will be used to also refer to a thermodynamic phase of a macroscopic system. To avoid any possible confusion between the two different usages
of the word `phase', we will from now on use the term `angle' to mean oscillator phase, while
the term `phase' will be exclusively used to mean a thermodynamic phase. Considering the interaction between the oscillators to be all-to-all, a general first-order time evolution that one can define within the framework of the Kuramoto model is given by
\begin{equation}
    \frac{d \theta_j}{dt} = \omega_j + \sum_{k = 1}^N F(\theta_k-\theta_j) + \sqrt{2D}\, \eta_j(t). \label{eq:GeneralizedKuramoto-0}
\end{equation}
Here, the term $F(\theta_k-\theta_j)~\forall~k,j$ denotes the interaction between the $k$th and the $j$th oscillator. 

In Eq.~\eqref{eq:GeneralizedKuramoto-0}, the natural frequencies $\omega_j \in (-\infty,\infty)$ of the oscillators are quenched-disordered random variables distributed according to a given probability distribution $g(\omega)$, which has a finite mean $\omega_0 \geq 0$ and a finite width $\sigma \geq 0$. Examples of such distributions are uniform, Lorentzian, and Gaussian. Note that for the Lorentzian, for which the mean is infinite, the quantity $\omega_0$ would stand for the location of the peak of the distribution.  Note that the dynamics~\eqref{eq:GeneralizedKuramoto-0} has $O(2)$ symmetry, whereby it remains invariant when all the oscillator angles are rotated by the same angle. In case of $g(\omega)$ being one-humped and
symmetric about its mean, the $O(2)$ symmetry allows us to go into a rotating frame with the transformation $\theta_j(t) \mapsto \theta_j(t) - \omega_0 t$ and $\omega_j \mapsto \omega_j - \omega_0~\forall~j$, to convert the problem into a simplified version in which the distribution $g(\omega)$ is centered around $\omega = 0$, i.e., $\omega_0=0$. In the remainder of the paper, we will always consider such a  $g(\omega)$ (note that we will also consider a uniform distribution that is symmetric about zero). In the third term on the right-hand side (rhs) of Eq.~\eqref{eq:GeneralizedKuramoto-0}, the quantity $\eta_j(t)$ is a Gaussian, white noise acting on the $j$th oscillator, with the properties
\begin{eqnarray}
    \langle \eta_j(t) \rangle &=& 0~\forall~j~\mathrm{and}~\forall~t,\\
    \langle \eta_j(t)\eta_k(t') \rangle &=& \delta_{jk}\delta(t-t')~\forall~j,k~\mathrm{and~}\forall~t,t'.
\end{eqnarray}
Here, the angular brackets denote averaging over noise realizations. The parameter $D$ denotes the strength of the noise in the time evolution of the system. 

Since $(\theta_k-\theta_j)$ is an angle-like variable, the inter-oscillator interaction function $F(q)$ is $2\pi-$periodic in $q$. Hence, $F(q)$ may be written as a Fourier expansion as follows:
\begin{eqnarray}
    F(q) = \sum_{l = -\infty}^{+\infty} \frac{F_l}{2} e^{ilq}.
\end{eqnarray}
Now, $F(q)$ being a real-valued function, we must have $(F_l)^{*} = F_{-l}$, where star denotes complex conjugation. If we further assume that these $F_l$'s are purely imaginary, so that $\widetilde{K}_l = -i F_l~\forall~l$ are purely real quantities, we get the inter-oscillator interaction function to be of the following form:
\begin{eqnarray}
    F(q) = \sum_{l = 1}^{\infty} \widetilde{K}_l \sin{(lq)}. \label{eq: Inter-Oscillator Interaction Sine Expanson}
\end{eqnarray}
The above form implies that the interaction is reciprocal: the effect on the $k$th oscillator due to the $j$th one is equal in magnitude but opposite in sign to the one on the $j$th oscillator due to the $k$th one. Using the above expansion, Eq.~\eqref{eq:GeneralizedKuramoto-0} rewrites as
\begin{equation}
    \frac{d \theta_j}{dt} = \omega_j + \frac{1}{N}\sum_{k = 1}^N \sum_{l=1}^\infty K_l \sin[l(\theta_k-\theta_j)] + \sqrt{2D}\, \eta_j(t), \label{eq: Generalized Kuramoto}
\end{equation}
where we have defined $K_l \equiv N\widetilde{K}_l$ to ensure effective competition between the first two terms on the rhs of the above equation in the thermodynamic limit $N \to \infty$. Here, the real parameters $K_l$ denote coupling constants, which we take to be non-negative quantities. 
As representative examples, we have for the case of the Kuramoto model that $F(q) = K_1 \sin{q}$, i.e., here, one has only first-harmonic interaction. On the other hand, in the case of the Kuramoto model with both first and second-harmonic interactions, we have $F(q) = K_1 \sin{q}+K_2 \sin{2q}$. 

The Kuramoto system is capable of exhibiting rich dynamics due to the interplay between randomness and coupling. The source of randomness can be due to the variation in the natural frequency among the oscillators and/or due to the random noise acing independently on each of the oscillators. In the absence of the inter-oscillator interaction, each oscillator angle tends to rotate independently in time. This results in the individual angles being scattered uniformly and independently in $[0, 2\pi)$ at large times, leading to an unsynchronized/incoherent state. A counter effect is provided by the inter-oscillator interaction, which tends to make the oscillators acquire the same angle, thereby leading to a synchronized state. Depending upon the relative magnitude of these competing effects, one observes within the Kuramoto dynamics in the limit $N \to \infty$ and in the stationary state a synchronization phase transition, or, a bifurcation. The nature of this bifurcation will depend on the specifics of the model.

The aforementioned phase transition can be characterized by a number of synchronization order parameters $z_l(t) = r_l(t) e^{i\psi_l(t)}$, defined as
\begin{eqnarray}
    z_l(t) \equiv r_l(t) e^{i\psi_l(t)} \equiv \frac{1}{N} \sum_{j = 1}^N e^{il\theta_j(t)}, \label{eq: order parameter any order}
\end{eqnarray}
where $l = 1, 2, \ldots$. Clearly, we have $0\leq r_l(t) \leq 1$ and $\psi_l(t) \in [0,2\pi)~\forall~l$ and $\forall~t$. How many of these order parameters we will need to correctly identify all the phases of the system will depend on how many $K_l$'s are non-zero in the expansion of the inter-oscillator interaction given by Eq.~\eqref{eq: Inter-Oscillator Interaction Sine Expanson}. 

\subsection{\label{subsec: Iso-noisy-harmonic Bare}Noisy Kuramoto model with first-harmonic interaction and identical frequencies}

In this case, $F(q) = K_1 \sin{q}$ and $g(\omega) = \delta(\omega)$. The evolution equation of the $j$th oscillator becomes~\cite{RevModPhys.77.137}
\begin{equation}
    \frac{d \theta_j}{dt} =  \frac{K_1}{N} \sum_{k = 1}^N \sin{(\theta_k-\theta_j)} + \sqrt{2D}\,\eta_j(t). \label{eq: Kuramoto case I}
\end{equation}
The different phases of this model can be characterized by only a single order parameter, i.e., $z_1(t) = r_1(t) e^{i\psi_1(t)}$. In the stationary state (st), attained as $t\to \infty$, the stationary value of $r_1(t)$, i.e., $r_1^\mathrm{st} \equiv r_1(t \to \infty)$, characterizes the different phases of the system. It is known that for a fixed noise strength $D$, the quantity $r_1^\mathrm{st}$ shows a supercritical bifurcation (a continuous phase transition) from $r_1^\mathrm{st} = 0$ (unsynchronized/incoherent phase) to $r_1^\mathrm{st} \neq 0$ (synchronized/coherent phase) as $K_1$ is increased beyond a critical value $K_1^\mathrm{c}$ given by~\cite{RevModPhys.77.137}
\begin{eqnarray}
    K_1^\mathrm{c} = 2D. \label{eq: Kc model 1}
\end{eqnarray}

\subsection{Noiseless Kuramoto model with first-harmonic interaction and distributed frequencies}

\subsubsection{Unimodal Lorentzian $g(\omega)$\label{subsec: noiseless-harmonic-lorentzian Bare}}

In this case, we have $F(q) = K_1 \sin{q}$ and $D = 0$. The distribution $g(\omega)$ of the natural frequencies has the following form:
\begin{eqnarray}
    g(\omega) = \frac{\sigma}{\pi} \frac{1}{\omega^2+\sigma^2}, \label{eq: Lorentzian Dist}
\end{eqnarray}
with $2\sigma \geq 0$ being the full-width-at-half-maximum of the distribution. The evolution equation of the $j$th oscillator reads as~\cite{RevModPhys.77.137}
\begin{equation}
    \frac{d \theta_j}{dt} =  \omega_j + \frac{K_1}{N} \sum_{k = 1}^N \sin{(\theta_k-\theta_j)}. \label{eq: Kuramoto case II}
\end{equation}
Here also the stationary value $r_1^\mathrm{st}$ characterizes the different phases of the system. Namely, for a fixed width $\sigma$, the quantity $r_1^\mathrm{st}$ shows a supercritical bifurcation from $r_1^\mathrm{st} = 0$  to $r_1^\mathrm{st}  \neq 0$  as $K_1$ is increased beyond a critical value $K_1^\mathrm{c}$ given by~\cite{RevModPhys.77.137} 
\begin{eqnarray}
    K_1^\mathrm{c} = 2\sigma. \label{eq: Kc model 2}
\end{eqnarray}
For any general unimodal $g(\omega)$, this bifurcation continues to be supercritical, and the corresponding critical coupling is given by~\cite{RevModPhys.77.137}
\begin{eqnarray}
    K_1^\mathrm{c} = \frac{2}{\pi g(0)}. \label{eq: Kc general g omega harmonic }
\end{eqnarray}
\subsubsection{Uniform $g(\omega)$ \label{subsec: noiseless-harmonic-uniform Bare}}
Another case of interest in the setting of Eq.~\eqref{eq: Kuramoto case II} is when the distribution is not unimodal, but is instead a uniform distribution of the form~\cite{PhysRevE.72.046211}
\begin{eqnarray}
    g(\omega) =
\begin{cases} 
\frac{1}{2 \sigma} & |\omega| \leq \sigma, \\
0 & |\omega| > \sigma,
\end{cases} \label{eq: g omega unif}
\end{eqnarray}
with $2\sigma \geq 0$ being the width of the distribution. Here, for a fixed width $\sigma$, the quantity $r_1^\mathrm{st}$ shows a subcritical bifurcation (a first-order phase transition) from $r_1^\mathrm{st} = 0$ to $r_1^\mathrm{st}  \neq 0$ as $K_1$ is increased beyond a critical value $K_1^\mathrm{c}$ given by~\cite{PhysRevE.72.046211}
\begin{eqnarray}
    K_1^\mathrm{c} = \frac{4 \sigma}{\pi}. \label{eq: Kc model 3}
\end{eqnarray}
 
\subsection{Noisy Kuramoto model with first and second-harmonic interactions and identical frequencies \label{subsec: noisy-harmonic-biharmonic Bare}}

\begin{figure}[htbp!]
    \centering
    \includegraphics[width=7 cm]{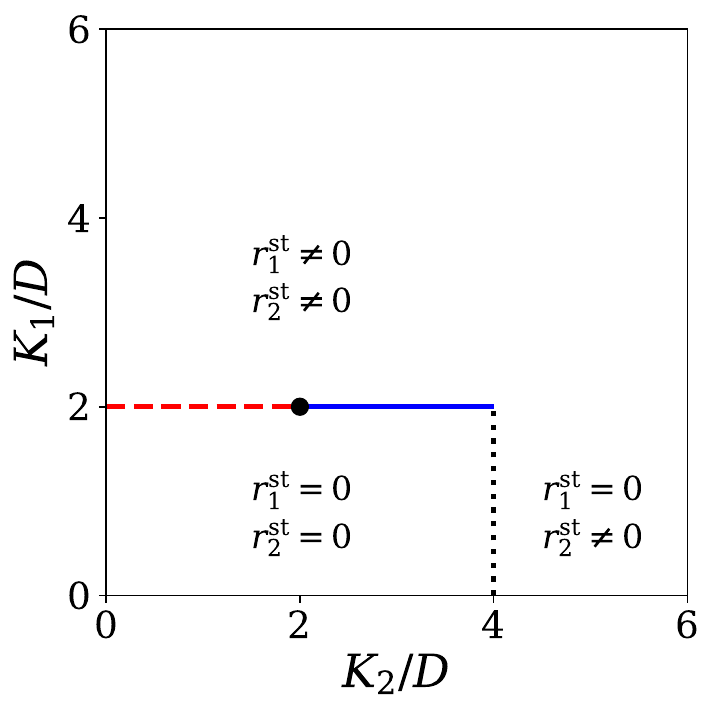}
    \caption{Phase diagram of the model~\eqref{eq: Kuramoto case IV}: The horizontal line at $K_1=2D$ shows the transition points where $r_1^\mathrm{st}$ shows a transition upon changing $K_1$ at a fixed $K_2$, with the nature being continuous for $K_2\leq 2D$ (red dashed line) and first-order for $2D<K_2\leq 4D$ (blue solid line). The vertical dotted line at $K_2 = 4D$ shows the transition points where $r_2^\mathrm{st}$ shows a transition upon changing $K_2$ at a fixed $K_1$, with the nature being continuous everywhere.}
    \label{fig:bi-phdiag}
\end{figure}

In this case, $F(q) = K_1 \sin{q} + K_2 \sin{2q}$, while we have $g(\omega)=\delta(\omega)$. The evolution equation of the $j$th oscillator becomes~\cite{Vlasov_2015, majumder2025finitesizefluctuationsstochasticcoupled}
\begin{align}
    \frac{d \theta_j}{dt} &= \frac{K_1}{N} \sum_{k = 1}^N \sin{(\theta_k-\theta_j)}+ \frac{K_2}{N} \sum_{k = 1}^N \sin{[2(\theta_k-\theta_j)]}\nonumber\\
    &+ \sqrt{2D}\, \eta_j(t). \label{eq: Kuramoto case IV}
\end{align}
To characterize all the phases for this model, we would need both $z_1(t) = r_1(t) e^{i\psi_1(t)}$ and $z_2(t) = r_2(t) e^{i\psi_2(t)}$. The phase diagram in the $(K_1,K_2)$-plane for a fixed $D$ is shown in Fig.~\ref{fig:bi-phdiag}. Note that from the definition of $r_1$ and $r_2$, a non-zero $r_1^\mathrm{st}$ necessarily implies non-zero $r_2^\mathrm{st}$. However, a non-zero $r_2^\mathrm{st}$ corresponds to a state that may or may not have a non-zero $r_1^\mathrm{st}$~\cite{Pikovsky_2014}.

\section{Continuum Limit of the Kuramoto Model\label{con lim}}

Before moving on to the discussion on resetting, let us first discuss the mathematical framework used to describe the bare dynamics of the Kuramoto model in the thermodynamic limit $N \to \infty$. For simplicity, we will consider the interaction to have only the first and second harmonic terms, i.e., $K_1,K_2\neq0$ and $K_{l>2}=0$ in Eq.~\eqref{eq: Generalized Kuramoto}. We may imagine partitioning the system into two subsystems: subsystem-$\mathrm{r}$ with oscillators labeled $j=1,2,\ldots,n$ and subsystem-$\mathrm{nr}$ with oscillators labeled $j=n+1,n+2,\ldots,N$, with $f\equiv n/N$ being the fraction of the total number of oscillators that are in subsystem-$\mathrm{r}$. Equation~\eqref{eq: Generalized Kuramoto} describing the dynamics of any arbitrary oscillator at angle $\theta_j$ and with frequency $\omega_j$ may be written as
\begin{align}
    &\frac{d\theta_j}{dt}= \omega_j +\frac{K_1f}{n}\sum_{k=1}^n\sin{(\theta_k-\theta_j)}\nonumber\\
    &+\frac{K_1\bar{f}}{N-n}\sum_{k=n+1}^N\sin{(\theta_k-\theta_j)}+\frac{K_2f}{n}\sum_{k=1}^n\sin[2(\theta_k-\theta_j)]\nonumber\\
    &+\frac{K_2\bar{f}}{N-n}\sum_{k=n+1}^N\sin[2(\theta_k-\theta_j)]+\sqrt{2D}\, \eta_j(t),\label{eq: describe well two sub}
\end{align}
where $\bar{f}\equiv 1-f$. Here, the index $j$ may refer to an oscillator from either the subsystem-$\mathrm{r}$ or the subsystem-$\mathrm{nr}$.

Using Eq.~\eqref{eq: order parameter any order}, we may now define the order parameters for each of the subsystems separately as $z_{1,\mathrm{r}} \equiv r_{1,\mathrm{r}} e^{i\psi_{1,\mathrm{r}}} \equiv (1/n) \sum_{j = 1}^n e^{i\theta_j}$,~$z_{2,\mathrm{r}} \equiv r_{2,\mathrm{r}} e^{i\psi_{2,\mathrm{r}}} \equiv (1/n) \sum_{j = 1}^n e^{i2\theta_j}$,~$z_{1,\mathrm{nr}} \equiv r_{1,\mathrm{nr}} e^{i\psi_{1,\mathrm{nr}}} \equiv [1/(N-n)] \sum_{j = n+1}^N e^{i\theta_j}$, and $z_{2,\mathrm{nr}} \equiv r_{2,\mathrm{nr}} e^{i\psi_{2,\mathrm{nr}}} \equiv [1/(N-n)] \sum_{j = n+1}^N e^{i2\theta_j}$. Then, we may rewrite Eq.~\eqref{eq: describe well two sub} as
\begin{align}
    \frac{d\theta_j}{dt} &= \omega_j +K_1f\,r_{1,\mathrm{r}} \sin{(\psi_{1,\mathrm{r}}-\theta_j)}\nonumber\\
    &+K_1\bar{f}\,r_{1,\mathrm{nr}} \sin{(\psi_{1,\mathrm{nr}}-\theta_j)}+K_2f\,r_{2,\mathrm{r}} \sin{(\psi_{2,\mathrm{r}}-2\theta_j)}\nonumber\\
    &+K_2\bar{f}\,r_{2,\mathrm{nr}} \sin{(\psi_{2,\mathrm{nr}}-2\theta_j)}+\sqrt{2D}\, \eta_j(t).\label{eq: describe well two sub 2}
\end{align}
Note that the dynamics of each oscillator is influenced by both the subsystems. Moreover, the influence of the interaction on the dynamics of each oscillator at any time instant depends on the interaction strengths $K_1,K_2$ as well as on the value of the order parameters at that instant.

We now focus on the thermodynamic limit. Let $(\theta_{\mathrm{r}},\omega_\mathrm{r})$ be respectively the angle and the frequency of an oscillator from the subsystem-$\mathrm{r}$, and $(\theta_{\mathrm{nr}},\omega_\mathrm{nr})$ be that of an oscillator from the subsystem-$\mathrm{nr}$. 
The state of the total system may be described in terms of a joint probability distribution $P(\theta_\mathrm{r},\omega_\mathrm{r},\theta_\mathrm{nr},\omega_\mathrm{nr},t)$, normalized as
\begin{align}
\int_{-\infty}^{+\infty}d\omega_\mathrm{r}\,g(\omega_\mathrm{r})\int_{-\infty}^{+\infty}d\omega_\mathrm{nr}\,g(\omega_\mathrm{nr})\int_0^{2\pi}d\theta_\mathrm{r}\int_0^{2\pi}d\theta_\mathrm{nr}\nonumber\\
\times P(\theta_{\mathrm{r}},\omega_\mathrm{r},\theta_{\mathrm{n}r},\omega_\mathrm{nr},t) = 1~\forall~t.
\label{eq:P-normalization}
\end{align}
The time evolution of this distribution is given by the Fokker-Planck equation, which reads as~\cite{risken1996fokker}
\begin{align}
   \frac{\partial P}{\partial t} =D \left[\frac{\partial^2 P}{\partial \theta^2_\mathrm{r}} +  \frac{\partial^2 P}{\partial \theta^2_\mathrm{nr}}\right] -\left[ \frac{\partial \left(P h_\mathrm{r} \right)}{\partial \theta_\mathrm{r}}+ \frac{\partial \left(P h_\mathrm{nr} \right)}{\partial \theta_\mathrm{nr}} \right] \label{eq: FP Kuramoto bare},
\end{align}
where we have
\begin{widetext}
\begin{align}
   & h_\mathrm{x} \equiv \omega_\mathrm{x} +K_1f \int d\theta'_\mathrm{r}d\omega'_\mathrm{r} g(\omega_\mathrm{r}) P(\theta'_\mathrm{r},\omega'_\mathrm{r} ,t| \theta_\mathrm{x} ,\omega_\mathrm{x})\sin(\theta'_\mathrm{r}-\theta_\mathrm{x})+K_1\bar{f} \int d\theta'_\mathrm{nr}d\omega'_\mathrm{nr} g(\omega_{\mathrm{nr}})P(\theta'_\mathrm{nr},\omega'_\mathrm{nr} ,t|  \theta_\mathrm{x},\omega_\mathrm{x}) \sin(\theta'_\mathrm{nr}-\theta_\mathrm{x}) \nonumber\\
    &+K_2f \int d\theta'_\mathrm{r}d\omega'_\mathrm{r} g(\omega_\mathrm{r}) P(\theta'_\mathrm{r},\omega'_\mathrm{r} ,t| \theta_\mathrm{x} ,\omega_\mathrm{x})\sin{[2(\theta'_\mathrm{r}-\theta_\mathrm{x})]}+K_2\bar{f} \int d\theta'_\mathrm{nr}d\omega'_\mathrm{nr} g(\omega_{\mathrm{nr}})P(\theta'_\mathrm{nr},\omega'_\mathrm{nr} ,t|  \theta_\mathrm{x},\omega_\mathrm{x}) \sin{[2(\theta'_\mathrm{nr}-\theta_\mathrm{x})]}, \label{eq: hx K1 K2 bare}
\end{align}
\end{widetext}
with $\mathrm{x} \equiv \mathrm{r}, \mathrm{nr}$. The first two bracketed terms on the rhs of Eq.~\eqref{eq: FP Kuramoto bare} are respectively the usual diffusion term due to Gaussian noise and the drift term due to inter-oscillator interactions.

In Eq.~\eqref{eq: hx K1 K2 bare} with $\mathrm{x}=\mathrm{r}$, the first and the third integral term represent the interaction of an oscillator from the subsystem-$\mathrm{r}$ that has a given angle $\theta_\mathrm{r}$ with the rest of the oscillators from the same subsystem, while the second and the fourth integral refer to its interaction with all the oscillators from the subsystem-$\mathrm{nr}$. A similar explanation holds for $\mathrm{x}=\mathrm{nr}$ in Eq.~\eqref{eq: hx K1 K2 bare}. The conditional probabilities are given by
\begin{equation}
    P(\theta'_\mathrm{y},\omega'_\mathrm{y} ,t| \theta_\mathrm{x} ,\omega_\mathrm{x}) = \frac{\mathbf{P}(\theta_\mathrm{x} ,\omega_\mathrm{x},\theta'_\mathrm{y},\omega'_\mathrm{y},t)}{P(\theta_\mathrm{x} ,\omega_\mathrm{x},t)}, \label{eq: con prob}
\end{equation}
where $\mathrm{x},\mathrm{y}$ can both be either $\mathrm{r}$ or $\mathrm{nr}$, and we have $P(\theta_\mathrm{x} ,\omega_\mathrm{x},t) = \int d\theta'_\mathrm{y} d\omega'_\mathrm{y}\,g(\omega'_\mathrm{y})\,\mathbf{P}(\theta_\mathrm{x} ,\omega_\mathrm{x},\theta'_\mathrm{y},\omega'_\mathrm{y},t)$. For example, when $\mathrm{x}=\mathrm{r}$ and $\mathrm{y}=\mathrm{nr}$,   $\mathbf{P}(\theta_\mathrm{x} ,\omega_\mathrm{x},\theta'_\mathrm{y},\omega'_\mathrm{y},t)$ reduces to $P(\theta_\mathrm{r},\omega_\mathrm{r},\theta_\mathrm{nr},\omega_\mathrm{nr},t)$. The probability distribution $\mathbf{P}(\theta_\mathrm{x} ,\omega_\mathrm{x},\theta'_\mathrm{y},\omega'_\mathrm{y},t)$ may be written as a marginal of a suitable joint distribution, e.g., in the case of $\mathrm{x} = \mathrm{r}$ and $\mathrm{y} = \mathrm{r}$, we may write $\mathbf{P}(\theta_\mathrm{r} ,\omega_\mathrm{r},\theta'_\mathrm{r},\omega'_\mathrm{r},t) = \int d\theta_\mathrm{nr}d\omega_\mathrm{nr}d\theta'_\mathrm{nr}d\omega'_\mathrm{nr}\,g(\omega_\mathrm{nr})g(\omega'_\mathrm{nr})$ $\mathbb{P}(\theta_\mathrm{r},\omega_\mathrm{r}, \theta_\mathrm{nr}, \omega_\mathrm{nr},\theta'_\mathrm{r},\omega'_\mathrm{r},\theta'_\mathrm{nr},\omega'_\mathrm{nr},t)$.

Since our model given by Eq.~\eqref{eq: describe well two sub} is a mean-field model, we invoke a mean-field approximation, implying the following factorization property of the joint probability density 
\begin{align}
    \nonumber & \mathbb{P}(\theta_\mathrm{r},\omega_\mathrm{r}, \theta_\mathrm{nr}, \omega_\mathrm{nr},\theta'_\mathrm{r},\omega'_\mathrm{r},\theta'_\mathrm{nr},\omega'_\mathrm{nr},t)= \\
    & \hspace{1cm}P(\theta_\mathrm{r}, \omega_\mathrm{r},\theta_\mathrm{nr},\omega_\mathrm{nr},t)P(\theta'_\mathrm{r}, \omega'_\mathrm{r},\theta'_\mathrm{nr},\omega'_\mathrm{nr},t), \label{eq: approx 8 pt to 4 pt}
\end{align}
implying
\begin{eqnarray}
    P(\theta'_\mathrm{y},\omega'_\mathrm{y} | \theta_\mathrm{x} ,\omega_\mathrm{x}) = P(\theta'_\mathrm{y},\omega'_\mathrm{y} ). \label{eq: con approx bare}
\end{eqnarray}
The mean-field approximation encodes the fact that oscillators in different frequency groups are independent in the sense that the probability that one finds an oscillator in the subsystem-$\mathrm{x}$ with phase $\theta_\mathrm{x}$ and frequency $\omega_\mathrm{x}$ is independent of the probability of finding an oscillator in the subsystem-$\mathrm{y}$ with phase $\theta'_\mathrm{y}$ and frequency $\omega'_\mathrm{y}$. Using this mean-field approximation, we may express $h_\mathrm{x}$ in terms of the order parameter as follows
\begin{align}
    h_\mathrm{x} &= \omega_\mathrm{x}+K_1f\,r_{1,\mathrm{r}} \sin{(\psi_{1,\mathrm{r}}-\theta_\mathrm{x})}\nonumber \\
    &+K_1\bar{f}\,r_{1,\mathrm{nr}} \sin{(\psi_{1,\mathrm{nr}}-\theta_\mathrm{x})}\nonumber\\
    &+K_2f\,r_{2,\mathrm{r}} \sin{(\psi_{2,\mathrm{r}}-2\theta_\mathrm{x})}+K_2\bar{f}\,r_{2,\mathrm{nr}} \sin{(\psi_{2,\mathrm{nr}}-2\theta_\mathrm{x})},
\end{align}
where we have
\begin{align}
    \nonumber &z_{1,\mathrm{r}}= r_{1,\mathrm{r}}e^{i\psi_{1,\mathrm{r}}} \\
     &= \int d\theta_\mathrm{r}d \omega_\mathrm{r}d\theta_\mathrm{nr}d\omega_\mathrm{nr} e^{i \theta_\mathrm{r}} g(\omega_\mathrm{r})g(\omega_\mathrm{nr})P(\theta_\mathrm{r}, \omega_\mathrm{r},\theta_\mathrm{nr},\omega_\mathrm{nr},t), \label{eq: zx1}
\end{align}
and similarly, $\nonumber z_{1,\mathrm{nr}}= r_{1,\mathrm{nr}}e^{i\psi_{1,\mathrm{nr}}}$ is defined as an average of $e^{i \theta_\mathrm{nr}}$, $\nonumber z_{2,\mathrm{r}}=r_{2,\mathrm{r}}e^{i\psi_{2,\mathrm{r}}}$ is defined as an average of $e^{i 2\theta_\mathrm{r}}$, $\nonumber z_{2,\mathrm{nr}}= r_{2,\mathrm{nr}}e^{i\psi_{2,\mathrm{nr}}}$ is defined as an average of $e^{i 2\theta_\mathrm{nr}}$.

\section{\label{sec:level3}Subsystem Resetting Protocol}

In the backdrop of the stationary-state results summarized in the Sec.~\ref{sec:level2}, we are interested in the following broad question in this work: Given that under the bare dynamics, the different variants of the Kuramoto system reach a stationary state, what modification thereof, if any, is brought about by the protocol of  subsystem resetting?

Let us discuss in more detail the context and the protocol of subsystem resetting. From the discussions in Sec.~\ref{sec:level2}, we know that Kuramoto-type oscillator systems exhibit a stationary-state phase transition from an incoherent to a synchronized phase as the interaction strength is tuned. In particular, when the coupling constant $K_1$ exceeds a critical value $K_1^\mathrm{c}$, the system develops macroscopic coherence among the angles, whereas for $K_1<K_1^\mathrm{c}$, the oscillators desynchronize in the long-time limit. Keeping this in mind, the situation we are interested in within the framework of the Kuramoto model is the following: Consider a system of coupled oscillators that is initially prepared in a fully synchronized configuration, i.e., all oscillator angles are equal at the initial time. The subsequent evolution of the system depends on the value of the coupling constant(s). If $K_1>K_1^\mathrm{c}$, the dynamics naturally drives the system toward a synchronized stationary state. By contrast, if 
$K_1\leq K^\mathrm{c}_1$, the system progressively loses coherence and evolves toward an incoherent state. We repeatedly interrupt this natural dynamics at random times and reset the angles the oscillators of the subsystem-$\mathrm{r}$ to a fixed configuration (reset configuration). We do not interrupt the dynamics of the rest of the oscillators belonging to subsystem-$\mathrm{nr}$. The system evolves following the bare dynamics between two subsequent reset events. In the remainder of the paper, we will call the subsystem-$\mathrm{r}$ as the reset subsystem and the subsystem-$\mathrm{nr}$ as the non-reset subsystem (we had anticipated this nomenclature in the preceding section while assigning the labels r and nr to the two subsystems).

The time interval between two reset events is chosen from an exponential distribution with rate parameter $\lambda>0$. During each reset event, we reset to zero the angles of a given fraction $\alpha$ of the oscillators of the reset subsystem, while that for the rest of the oscillators of the reset subsystem are reset to $\pi$. Using the definition from Eq.~\eqref{eq: order parameter any order}, we obtain that the magnitude of the $l=1$ order parameter for the reset configuration is $r_0 \equiv |2\alpha-1|$, while the same for $l=2$ is unity. Thus, at every reset event, $r_{1,\mathrm{r}}$ is reset to $r_0$ and $r_{2,\mathrm{r}}$ is reset to unity. By changing $\alpha$, we can change the amount of synchronization $r_0$ of the reset configuration. The question we want to address concerns if and how the phase diagram of the bare model gets modified under such a protocol. In particular, we focus on the manipulation of the phase diagram of the order parameter $r_{1,\mathrm{nr}}$ through our protocol of subsystem resetting. As we will reveal, one can move or even quite remarkably eliminate phase transitions of the bare model, simply by a suitable choice of the three parameters characterizing the resetting protocol, namely, $f,\lambda,r_0$. 

In the absence of resetting, reset and non-reset subsystems behave identically under bare dynamics. Let at any given parameter values $K_1,K_2$, the stationary-state value of both the order parameters $r_{1,\mathrm{r}}$ and $r_{1,\mathrm{nr}}$ equal $r_{1}^\mathrm{B}$ under the bare dynamics. In presence of resetting, we expect that if we reset the order parameter $r_{1,\mathrm{r}}$ to a value $r_0>r_{1}^\mathrm{B}$, the stationary-state value of $r_{1,\mathrm{nr}}$, denoted by $r^\mathrm{st}_{1,\mathrm{nr}}$, will also be greater that $r_{1}^\mathrm{B}$. Similarly, if $r_0<r_1^\mathrm{B}$, we will have $r^\mathrm{st}_{1,\mathrm{nr}}<r_1^\mathrm{B}$. We can argue this by focusing on the dynamics of the oscillators from the non-reset subsystem that follows Eq.~\eqref{eq: describe well two sub 2}.
To maintain the stationary state, desynchronizing effects originating from frequency disorder and the noise are balanced by the ordering/synchronizing effects coming from the interaction terms that depend on the four order parameters $r_{1,\mathrm{r}}, r_{1,\mathrm{nr}}, r_{2,\mathrm{r}}, r_{2,\mathrm{nr}}$. If $r_0>r^\mathrm{B}_1$, the interaction between the reset and the non-reset oscillators increases the ordering effect, resulting in a new stationary state with higher synchrony than the bare model~(see Eq.~\eqref{eq: describe well two sub 2}). Similarly, if $r_0<r^\mathrm{B}_1$, the interaction between the reset and the non-reset oscillators decreases the ordering effect, resulting in lower synchrony than the bare model~(see Eq.~\eqref{eq: describe well two sub 2}). In Figs.~\ref{fig: model 1},~\ref{fig: model 2},~and~\ref{fig: model 4}, we show that our results support our expectation.

\section{Analysis for First-Harmonic Interaction}
\label{sec:first-harmonic}
In this section, we describe our analytical formalism for studying the effects of subsystem resetting in the set-up of Eq.~\eqref{eq: Generalized Kuramoto}
 with $K_1\ne 0,~K_{l \ge2}=0$. The analysis for the general case is given in Sec.~\ref{sec: app to gen}.
 \subsection{General Theory \label{sec: harmonic theory}}

From the protocol of subsystem resetting discussed in Sec.~\ref{sec:level3}, we may modify the Fokker-Planck equation given in Eq.~\eqref{eq: FP Kuramoto bare} as follows:
\begin{eqnarray}
    \nonumber \frac{\partial P}{\partial t} =&& D \left[\frac{\partial^2 P}{\partial \theta^2_\mathrm{r}} +  \frac{\partial^2 P}{\partial \theta^2_\mathrm{nr}}\right] -\left[ \frac{\partial \left(P h_\mathrm{r} \right)}{\partial \theta_\mathrm{r}}+ \frac{\partial \left(P h_\mathrm{nr} \right)}{\partial \theta_\mathrm{nr}} \right] -\lambda P\\
    &&+ \lambda \left[ \alpha \delta(\theta_\mathrm{r}) + (1-\alpha) \delta(\theta_\mathrm{r}-\pi)\right]\nonumber\\
    && \times\int_{-\infty}^{+\infty} d \omega'_\mathrm{r} g(\omega'_\mathrm{r})\int_0^{2\pi} d \theta'_\mathrm{r} P(\theta'_\mathrm{r},\theta_\mathrm{nr},\omega'_\mathrm{r},\omega_\mathrm{nr},t), \label{eq: FP Kuramoto Finite Resetting} 
\end{eqnarray}
where we have $h_\mathrm{x}$ with $\mathrm{x} \equiv \mathrm{r}, \mathrm{nr}$ given in Eq.~\eqref{eq: hx K1 K2 bare}, with $K_2=0$. The last two terms in Eq.~\eqref{eq: FP Kuramoto Finite Resetting} account for probability loss and gain due to resetting at rate $\lambda$. In the absence of the $K_2$ term, the only relevant order parameters are $z_{1,\mathrm{r}} \equiv r_{1,\mathrm{r}}e^{i\psi_{1,\mathrm{r}}} $ and $z_{1,\mathrm{nr}} \equiv r_{1,\mathrm{nr}}e^{i\psi_{1,\mathrm{nr}}} $. Hence, until the end of this section, we will drop the subscript $1$ for brevity.

In the stationary state (st), we have $\partial P/\partial t = 0$ in Eq.~\eqref{eq: FP Kuramoto Finite Resetting}. To proceed, we use the approximation that in the stationary state, we have 
\begin{eqnarray}
    \nonumber && \mathbb{P}_\mathrm{st}(\theta_\mathrm{r},\omega_\mathrm{r}, \theta_\mathrm{nr}, \omega_\mathrm{nr},\theta'_\mathrm{r},\omega'_\mathrm{r},\theta'_\mathrm{nr},\omega'_\mathrm{nr}) \approx \\
    && \hspace{1cm}P_\mathrm{st}(\theta_\mathrm{r}, \omega_\mathrm{r},\theta_\mathrm{nr},\omega_\mathrm{nr})P_\mathrm{st}(\theta'_\mathrm{r}, \omega'_\mathrm{r},\theta'_\mathrm{nr},\omega'_\mathrm{nr}).\label{eq:approx_reset_nonreset}
\end{eqnarray}
Note that a similar approximation was invoked earlier for the bare dynamics, see Eq.~\eqref{eq: approx 8 pt to 4 pt}, and we are now invoking it for the dynamics in presence of resetting. At this stage, it is pertinent to assess the validity of the approximation in Eq.~\eqref{eq:approx_reset_nonreset} across different regimes of the reset rate $\lambda$.  First, in absence of resetting, i.e., at  $\lambda=0$,  the factorization approximation in Eq.~\eqref{eq:approx_reset_nonreset} holds exactly in the limit $N\to \infty$ because of the mean-field nature of all the models considered in this paper: in a mean-field system, oscillators interact only through the global order parameter, so that any two oscillators become statistically independent as $N \to \infty$. Note that for $\lambda=0$, the subscripts r and nr become redundant. Next, in the case of resetting at infinite rate, i.e., as $\lambda \to \infty$, the reset subsystem remains effectively frozen in time at the reset configuration~\cite{acharya2025manipulating,PhysRevE.109.064137}. The non-reset subsystem then evolves in a static mean-field sourced entirely by $r_0$. The time scale separation between the dynamics of the two subsystems then ensures that the factorization continues to hold exactly in this limit as well. On the basis of the above, we expect the factorization approximation to hold for small and large $\lambda$. For intermediate reset rates, the shift in the behavior of the order parameters of the non-reset subsystem is bounded between those in the two limits, $\lambda=0$ and $\lambda\to \infty$. In this regime, the factorization may not hold a priori. However, we will show that one can invoke it as an approximation to get a very good estimate of the order parameters of the non-reset subsystem, which we will validate by comparing with simulation results.

Now, using the approximation~\eqref{eq:approx_reset_nonreset}, Eq.~\eqref{eq: con prob} becomes
\begin{eqnarray}
    P_\mathrm{st}(\theta'_\mathrm{y},\omega'_\mathrm{y} | \theta_\mathrm{x} ,\omega_\mathrm{x}) \approx P_\mathrm{st}(\theta'_\mathrm{y},\omega'_\mathrm{y} ) \label{eq: con approx}
\end{eqnarray}
for any $\mathrm{x},\mathrm{y}$, where $P_\mathrm{st}(\theta'_\mathrm{r},\omega'_\mathrm{r} )$ and $P_\mathrm{st}(\theta'_\mathrm{nr},\omega'_\mathrm{nr} )$ are just marginals of  $P_\mathrm{st}(\theta'_\mathrm{r}, \omega'_\mathrm{r},\theta'_\mathrm{nr},\omega'_\mathrm{nr})$. Approximation~\eqref{eq: con approx} along with Eq.~\eqref{eq: hx K1 K2 bare} when substituted in Eq.~\eqref{eq: FP Kuramoto Finite Resetting} with the stationary-state condition $\partial P/\partial t=0$ makes it a closed equation for $P_\mathrm{st}(\theta_{r},\omega_\mathrm{r},\theta_{nr},\omega_\mathrm{nr})$, which helps to solve the problem in the stationary state. This approximation in turn simplifies Eq.~\eqref{eq: hx K1 K2 bare} in the stationary state to give
\begin{equation}
     h_\mathrm{x} = \omega_\mathrm{x} +K_1f r^\mathrm{st}_\mathrm{r}\sin(\psi^\mathrm{st}_\mathrm{r}-\theta_\mathrm{x}) +K_1\bar{f} r^\mathrm{st}_\mathrm{nr}\sin(\psi^\mathrm{st}_\mathrm{nr}-\theta_\mathrm{x}), \label{eq: hx st}
\end{equation}
where the stationary-state values of the order parameters are given by
\begin{align}
    \nonumber z^\mathrm{st}_\mathrm{r} &= r^\mathrm{st}_\mathrm{r}e^{i\psi^\mathrm{st}_\mathrm{r}} \\
     &= \int d\theta_\mathrm{r}d \omega_\mathrm{r}d\theta_\mathrm{nr}d\omega_\mathrm{nr} e^{i \theta_\mathrm{r}} g(\omega_\mathrm{r})g(\omega_\mathrm{nr})P_\mathrm{st}(\theta_\mathrm{r}, \omega_\mathrm{r},\theta_\mathrm{nr},\omega_\mathrm{nr}), \label{eq: zx}
\end{align}
and similarly, $\nonumber z^\mathrm{st}_\mathrm{nr}= r^\mathrm{st}_\mathrm{nr}e^{i\psi^\mathrm{st}_\mathrm{nr}}$ is defined as an average of $e^{i \theta_\mathrm{nr}}$. 

Next, we use the fact that $P_\mathrm{st}(\theta_\mathrm{r},\omega_\mathrm{r},\theta_\mathrm{nr},\omega_\mathrm{nr})$ is $2\pi$-periodic in both $\theta_\mathrm{r}$ and $\theta_\mathrm{nr}$. This allows us to expand it into a two-dimensional Fourier series, which reads as
\begin{align}
    \nonumber &P_\mathrm{st}(\theta_\mathrm{r},\omega_\mathrm{r},\theta_\mathrm{nr},\omega_\mathrm{nr}) \nonumber \\
    &= \sum_{l = -\infty}^{\infty}\sum_{m = -\infty}^{\infty} \mathcal{P}_{l,m}(\omega_\mathrm{r},\omega_\mathrm{nr}) e^{i l \theta_\mathrm{r}+i m\theta_\mathrm{nr}}. \label{eq: fourier}
\end{align}
Now, $P_\mathrm{st}(\theta_\mathrm{r},\omega_\mathrm{r},\theta_\mathrm{nr},\omega_\mathrm{nr})$ being real and normalized, see Eq.~\eqref{eq:P-normalization}, we get $\left(\mathcal{P}_{l,m}\right)^{*} = \mathcal{P}_{-l,-m}$ and $\mathcal{P}_{0,0} = 1/(4 \pi^2)$, respectively. Using Eq.~\eqref{eq: fourier} in Eq.~\eqref{eq: FP Kuramoto Finite Resetting} along with Eq.~\eqref{eq: hx st} and comparing the coefficients $e^{i l \theta_\mathrm{r}+i m\theta_\mathrm{nr}}$ from the various terms in the equation, we get the following relation between the various $\mathcal{P}_{l,m}(\omega_\mathrm{r},\omega_\mathrm{nr})$'s:
\begin{align}
     &\left[(l^2+m^2)D+i(l \omega_\mathrm{r}+m\omega_\mathrm{nr})+\lambda\right] \mathcal{P}_{l,m} \nonumber \\
     &+\gamma\left(l\mathcal{P}_{l+1,m}+m\mathcal{P}_{l,m+1}\right)-\gamma^{*}\left(l\mathcal{P}_{l-1,m}+m\mathcal{P}_{l,m-1}\right) \nonumber \\
     & =  \lambda\left[ \alpha  + (-1)^l(1-\alpha) \right] \mathcal{P}_{0,m}, \label{eq: Fourier Relation Kuramoto SubReset Sup}
\end{align}
where we have defined
\begin{eqnarray}
    \gamma \equiv \left[\frac{ K_1f r^\mathrm{st}_\mathrm{r}e^{i \psi^\mathrm{st}_\mathrm{r}}+K_1\bar{f} r^\mathrm{st}_\mathrm{nr} e^{i \psi^\mathrm{st}_\mathrm{nr}} }{2}\right].\label{eq: gamma}
\end{eqnarray}
Furthermore, using the Fourier expansion given in Eq.~\eqref{eq: fourier} in the definition of the order parameter given in Eq,~\eqref{eq: zx}, we obtain
\begin{align}
     z^\mathrm{st}_\mathrm{r}  &=  4 \pi^2 \int_{-\infty}^{+\infty} d \omega_\mathrm{r}d \omega_\mathrm{nr} g(\omega_\mathrm{r}) g(\omega_\mathrm{nr}) \mathcal{P}_{-1,0}(\omega_\mathrm{r},\omega_\mathrm{nr}), \label{order fourier sup1}\\
     z^\mathrm{st}_\mathrm{nr}& =  4 \pi^2 \int_{-\infty}^{+\infty} d \omega_\mathrm{r} d \omega_\mathrm{nr}g(\omega_\mathrm{r}) g(\omega_\mathrm{nr}) \mathcal{P}_{0,-1}(\omega_\mathrm{r},\omega_\mathrm{nr}). \label{order fourier sup2}
\end{align}
Before proceeding, note that we may also define the Kuramoto-Daido order parameters, see Eq.~\eqref{eq: order parameter any order}, for both reset and non-reset subsystems. One has in the stationary state that $z^\mathrm{st}_{l,\mathrm{x}} \equiv  r^\mathrm{st}_{l,\mathrm{x}}e^{il\psi^\mathrm{st}_{l,\mathrm{x}}}$, defined as the average of $e^{i l\theta_\mathrm{x}}$ with $\mathrm{x=r,nr}$. Using Eq.~\eqref{eq: fourier}, we get 
\begin{align}
     z^\mathrm{st}_{l,\mathrm{r} } &=  4 \pi^2 \int_{-\infty}^{+\infty} d \omega_\mathrm{r}d \omega_\mathrm{nr} g(\omega_\mathrm{r}) g(\omega_\mathrm{nr}) \mathcal{P}_{-l,0}(\omega_\mathrm{r},\omega_\mathrm{nr}), \label{lorder fourier sup1}\\
     z^\mathrm{st}_{l,\mathrm{nr}}& =  4 \pi^2 \int_{-\infty}^{+\infty} d \omega_\mathrm{r} d \omega_\mathrm{nr}g(\omega_\mathrm{r}) g(\omega_\mathrm{nr}) \mathcal{P}_{0,-l}(\omega_\mathrm{r},\omega_\mathrm{nr}). \label{lorder fourier sup2}
\end{align}

Let us now focus on obtaining the stationary-state order parameters $z^\mathrm{st}_\mathrm{r}$ and $z^\mathrm{st}_\mathrm{nr}$.  
To this end, Eqs.~\eqref{order fourier sup1}~and~\eqref{order fourier sup2} imply that we need to find only the quantities $\mathcal{P}_{-1,0}$ and $\mathcal{P}_{0,-1}$. Since $\mathcal{P}_{-1,0} = \left(\mathcal{P}_{1,0}\right)^{*}$ and $\mathcal{P}_{0,-1} = \left(\mathcal{P}_{0,1}\right)^{*}$, we will focus on obtaining the quantities $\mathcal{P}_{1,0}$ and $\mathcal{P}_{0,1}$. Putting successively $m=0$ and $l=0$ in Eq.~\eqref{eq: Fourier Relation Kuramoto SubReset Sup}, we obtain respectively that
\begin{align}
    &\left[l^2D+il \omega_\mathrm{r}+\lambda\right] \mathcal{P}_{l,0}+l \gamma \mathcal{P}_{l+1,0}-l \gamma^{*}\mathcal{P}_{l-1,0}\nonumber\\
    &= \frac{\lambda}{4\pi^2}\left[ \alpha  + (-1)^l(1-\alpha) \right] 
  \label{eq: Fourier Relation l-axis Kuramoto SubReset Sup} ,\\
  &\left[m^2D+im \omega_\mathrm{nr}\right] \mathcal{P}_{0,m}+m \gamma \mathcal{P}_{0,m+1}-m \gamma^{*}\mathcal{P}_{0,m-1} = 0 
  \label{eq: Fourier Relation m-axis Kuramoto SubReset Sup} .
\end{align}

We first consider Eq.~\eqref{eq: Fourier Relation l-axis Kuramoto SubReset Sup}. Clearly, for $l=0$, Eq.~\eqref{eq: Fourier Relation l-axis Kuramoto SubReset Sup} reproduces the known result $\mathcal{P}_{0,0}  = 1/(4 \pi^2)$, whereas for $l>0$, it couples three consecutive Fourier components for each $l$: $\mathcal{P}_{l-1,0}, \mathcal{P}_{l,0}$ and $\mathcal{P}_{l+1,0}$. Hence, each $\mathcal{P}_{l,0} $ may be expressed as a linear combination of $\mathcal{P}_{l-1,0}$ and $\mathcal{P}_{l-2,0}$, for all $l>2$. For example, $\mathcal{P}_{2,0} $ may be expressed as a linear combination of $\mathcal{P}_{1,0}$ and $\mathcal{P}_{0,0}$. Since we already know the expression for $\mathcal{P}_{0,0} $, we may express $\mathcal{P}_{2,0}$ solely as a linear function of $\mathcal{P}_{1,0}$. Moving onto the next $l$ value, $\mathcal{P}_{3,0}$ may be expressed as a linear combination of $\mathcal{P}_{2,0}$ and $\mathcal{P}_{1,0}$ using Eq.~\eqref{eq: Fourier Relation l-axis Kuramoto SubReset Sup}. Since it follows from the above that $\mathcal{P}_{1,0}$ may be expressed solely as a linear function of $\mathcal{P}_{2,0}$, we may further express $\mathcal{P}_{3,0}$ solely as a linear function of $\mathcal{P}_{2,0}$. Proceeding this way, we observe that we may express each $\mathcal{P}_{l,0}$ as a linear function of $\mathcal{P}_{l-1,0}$, for $l\geq2$. Motivated by this argument, we make the following ansatz
\begin{eqnarray}
    \mathcal{P}_{l,0} = \Gamma_{l} \mathcal{P}_{l-1,0} + \Delta_{l},~l\geq2,\label{eq: l-axis ansatz Kuramoto SubReset Sup}
\end{eqnarray}
which, when used in Eq.~\eqref{eq: Fourier Relation l-axis Kuramoto SubReset Sup}, gives
\begin{eqnarray}
    \Gamma_l &=& \frac{l \gamma^{*}}{\left(l^2D+il \omega_\mathrm{r}+\lambda\right)+l \gamma \Gamma_{l+1}},\label{eq: gamma l}\\
    \Delta_l  &=& \frac{\frac{\lambda}{4\pi^2}\left[ \alpha  + (-1)^l(1-\alpha) \right]-l \gamma  \Delta_{l+1}}{\left(l^2D+il \omega_\mathrm{r}+\lambda\right)+l \gamma \Gamma_{l+1}},\label{eq: delta l}
\end{eqnarray}
both valid for $l\geq2$. Specifically, for $l=2$, we have
\begin{align}
    \mathcal{P}_{2,0} = \Gamma_{2} \mathcal{P}_{1,0} + \Delta_2. \label{eq: p 2}
\end{align}

Now, putting $l=1$ in Eq.~\eqref{eq: Fourier Relation l-axis Kuramoto SubReset Sup}, we obtain
\begin{align}
    \left(D+i \omega_\mathrm{r}+\lambda\right) \mathcal{P}_{1,0}+ \gamma \mathcal{P}_{2,0}- \gamma^{*}\mathcal{P}_{0,0}= \frac{\lambda}{4\pi^2}\left( 2\alpha -1 \right). \label{eq: p1 p2 p0}
\end{align}
Putting Eq.~\eqref{eq: p 2} into Eq.~\eqref{eq: p1 p2 p0}, we obtain $\mathcal{P}_{1,0}$ as
\begin{equation}
    \mathcal{P}_{1,0} = \Gamma_1 \mathcal{P}_{0,0} + \Delta_1,
\end{equation}
where $\Gamma_1$ and $\Delta_1$ follow the exact same form as given in Eqs.~\eqref{eq: gamma l}~and~\eqref{eq: delta l}, respectively. Thus, Eqs.~\eqref{eq: l-axis ansatz Kuramoto SubReset Sup},~\eqref{eq: gamma l},
~and~\eqref{eq: delta l} are valid even for $l\geq1$. Then, putting $l=1$ in Eq.~\eqref{eq: l-axis ansatz Kuramoto SubReset Sup}, we obtain
\begin{eqnarray}
    \mathcal{P}_{1,0}(\omega_\mathrm{r}) 
    &=& \frac{\Gamma_1 (\omega_\mathrm{r}) }{4 \pi^2} + \Delta_1(\omega_\mathrm{r}),
    \label{eq:P10-1}
\end{eqnarray}
where both $\Gamma_1$ and $\Delta_1$ have forms of infinite continued fraction. For example, $\Gamma_1$ is given by
\begin{align}
    \Gamma_1(\omega_\mathrm{r}) &= \frac{ \gamma^{*}}{\left(D+i \omega_\mathrm{r}+\lambda\right)+ \gamma\left[ \frac{2 \gamma^{*}}{\left(4 D+ 2 i \omega_\mathrm{r}+\lambda\right)+2 \gamma \left[\ddots\right]}\right]}. \label{eq: Gamma Cont Frac}
\end{align}
 Now, $0\leq|z^\mathrm{st}_{l,\mathrm{r}}|\leq 1$ being determined by $\mathcal{P}_{-l,0}(\omega_\mathrm{r},\omega_\mathrm{nr})$ (see Eq.~\eqref{lorder fourier sup1}), the latter quantities cannot diverge for arbitrary values of $\omega_\mathrm{r},\omega_\mathrm{nr}$. This further restricts that $\Gamma_l(\omega_\mathrm{r})$ and $\Delta_l(\omega_\mathrm{r})$ must converge to finite values for general values of $\omega_\mathrm{r}$. 

For further computation, let us discuss how to approximate the infinite continued fractions for $\Gamma_1$ and $\Delta_1$. Suppose the sequence $\{\Gamma_1, \Gamma_2,\ldots,\Gamma_{l-1},\Gamma_{l},\Gamma_{l+1},\ldots\}$ is convergent. Then, there exists a large-enough value of $l$, say $L$, such that we may put $\Gamma_{l+1}= \Gamma_{l}~\forall ~l\geq L$, up to a desired precision. Using this in Eq.~\eqref{eq: gamma l}, we obtain a quadratic equation for $\Gamma_l~\forall~l\geq L$. The root of this equation will provide the convergent value of the sequence, which reads as
\begin{align}
 \Gamma_{l,\pm} &= \frac{1}{2l\gamma} \bigl[-\left(l^2D+il \omega_\mathrm{r}+\lambda\right)\nonumber\\
 &\pm\sqrt{\left(l^2D+il \omega_\mathrm{r}+\lambda\right)^2+4l^2|\gamma|^2}~\bigr].
\end{align}
Now, from Eq.~\eqref{eq: gamma}, we have $|\gamma|<K_1$. Furthermore, considering $D$, $\omega_\mathrm{r}$ and $\lambda$ to be finite, we observe that $\Gamma_{l,-}$ diverges as $l\to\infty$ for arbitrary values of $\omega_\mathrm{r}$. Hence, the negative root cannot be the large-$l$ expression for $\Gamma_l$. On the other hand, for large $l$, the quantity $\Gamma_{l,+}$ becomes
\begin{eqnarray}
    \Gamma_{l,+} \approx \frac{l|\gamma|^2}{2\gamma\left(l^2D+il \omega_\mathrm{r}+\lambda\right)}.
\end{eqnarray}
In the limit $l \to \infty$, we obtain the convergent value as
\begin{align}
    \Gamma_{l\to\infty,+}=\begin{cases}
        0,~~~~~~~~\mathrm{if}~~D\neq0,\\
    \frac{\gamma^{*}}{2\omega_\mathrm{r}},~~~~\mathrm{if}~~D=0.
    \end{cases}\label{eq: gamma infty}
\end{align}
Hence, for numerical computations, we first consider a large enough $L$ such that $\Gamma_L= 
\Gamma_{l\to\infty,+}$, to our desired precision. Now, we may express $\Gamma_1$ as
\begin{eqnarray}
    \Gamma_1(\omega_\mathrm{r}) = \frac{\gamma^{*}}{a_1+\frac{2|\gamma|^2}{a_2+\frac{\ddots}{a_L+L\gamma\Gamma_{L}}}}, \label{eq: towards gamma1 approximation}
\end{eqnarray}
where we have defined $a_l \equiv \left(l^2D+il \omega_\mathrm{r}+\lambda\right) $. Equation~\eqref{eq: towards gamma1 approximation} is still an exact expression of $\Gamma_1(\omega_\mathrm{r})$. We now approximate Eq.~\eqref{eq: towards gamma1 approximation} by replacing $\Gamma_L$ by $\Gamma_{l\to\infty,+}$. The resulting expression is what we use for all further numerical computations.

In a similar way as above, we may approximate $\Delta_1$ for numerical computations. Assuming the sequence $\{\Delta_1, \Delta_2,\ldots,\Delta_{l-1},\Delta_{l},\Delta_{l+1},\ldots\}$ to be convergent, there exists a large-enough value of $l$, say $L'$, such that we may put $\Delta_{l+1}= \Delta_{l}~\forall ~l\geq L'$, up to a desired precision. Thus, $\forall~l \geq \bar{L} \equiv \mathrm{max}(L,L')$, we may replace $\Delta_{l+1} = \Delta_{l}$ and $\Gamma_{l} = \Gamma_{l\to\infty,+}$ in Eq.~\eqref{eq: delta l} and obtain
\begin{align}
    &\Delta_{l}\left[1+\frac{\gamma}{lD+i\omega_\mathrm{r}+\gamma \Gamma_{l\to\infty,+}}\right]= \frac{\frac{\lambda}{4\pi^2}\left[ \alpha  + (-1)^{l} (1-\alpha) \right]}{l^2D+il\omega_\mathrm{r}+l\gamma \Gamma_{l\to\infty,+}},
\end{align}
which immediately gives
\begin{eqnarray}
    \Delta_{l\to\infty} = 0. \label{eq: Delta l to infty}
\end{eqnarray}
Thus, similar to the case of $\Gamma_1(\omega_\mathrm{r})$, here also we approximate $\Delta_{\bar{L}}$ by $\Delta_{l\to\infty}$ in the expression of $\Delta_1$ and use that expression for all further numerical computations.

Using the fact that $(\mathcal{P}_{-1,0}) = (\mathcal{P}_{1,0})^{*}$ and Eq.~\eqref{eq:P10-1} in Eq.~\eqref{order fourier sup1}, we finally obtain
\begin{align}
r^\mathrm{st}_\mathrm{r}e^{i\psi^\mathrm{st}_\mathrm{r}} = 4 \pi^2 \int_{-\infty}^{+\infty} d \omega_\mathrm{r} g(\omega_\mathrm{r}) \left[\frac{\Gamma^{*}_1 (\omega_\mathrm{r}) }{4 \pi^2} + \Delta^{*}_1(\omega_\mathrm{r}) \right].\label{eq; z st reset subsystem}
\end{align}

We now focus on Eq.~\eqref{eq: Fourier Relation m-axis Kuramoto SubReset Sup}. Following the same line of argument as invoked following Eq.~\eqref{eq: Fourier Relation m-axis Kuramoto SubReset Sup}, we make an ansatz similar to Eq.~\eqref{eq: l-axis ansatz Kuramoto SubReset Sup}:
\begin{eqnarray}
    \mathcal{P}_{0,m+1} = \Lambda_{m+1}\mathcal{P}_{0,m}+\Pi_ {m+1} \label{eq: p 0 m+1}.
\end{eqnarray}
Equation~\eqref{eq: Fourier Relation m-axis Kuramoto SubReset Sup} gives
\begin{eqnarray}
    \Lambda_m &=& \frac{\gamma^{*}}{mD+i\omega_\mathrm{nr}+ \gamma \Lambda_{m+1}},\label{eq: lambda l}\\
    \Pi_m  &=& -\frac{ \gamma  \Pi_{m+1}}{mD+i\omega_\mathrm{nr}+ \gamma \Lambda_{m+1}}.\label{eq: pi l}
\end{eqnarray}
As before, since we have $0\leq|z^\mathrm{st}_{m,\mathrm{nr}}|\leq 1~\forall~m$, the quantities $\mathcal{P}_{0,-m}(\omega_\mathrm{r},\omega_\mathrm{nr})$ cannot diverge for arbitrary values of $\omega_\mathrm{r},\omega_\mathrm{nr}$. This further restricts that $\Lambda_m(\omega_\mathrm{nr})$ and $\Pi_m(\omega_\mathrm{nr})$ must converge to a finite value for general values of $\omega_\mathrm{nr}$. Using a similar argument as for the case of $\Gamma_l$ and $\Delta_l$, we obtain
\begin{align}
    \Lambda_{m\to\infty,+}=\begin{cases}
        0,~~~~~~~~\mathrm{if}~~D\neq0,\\
    \frac{\gamma^{*}}{2\omega_\mathrm{nr}},~~~~\mathrm{if}~~D=0,
    \end{cases}\label{eq: lambda infty}
\end{align}
and
\begin{eqnarray}
    \Pi_{m\to\infty} = 0. \label{eq: Pi m inf}
\end{eqnarray}
Using Eq.~\eqref{eq: Pi m inf} in Eq.~\eqref{eq: pi l}, we obtain
\begin{eqnarray}
    \Pi_m(\omega_\mathrm{nr}) = 0~\forall~m. \label{eq: pi m = 0}
\end{eqnarray}
Then, using the fact that $(\mathcal{P}_{0,-1}) = (\mathcal{P}_{0,1})^{*}$, we obtain on using the expression of $\mathcal{P}_{0,-1}$ in Eq.~\eqref{order fourier sup2} that
\begin{align}
r^\mathrm{st}_\mathrm{nr}e^{i\psi^\mathrm{st}_\mathrm{nr}} = \int_{-\infty}^{+\infty} d \omega_\mathrm{nr} g(\omega_\mathrm{nr}) \Lambda^{*}_1 (\omega_\mathrm{nr}).\label{eq; z st non-reset subsystem}
\end{align}
Then, using Eqs.~\eqref{eq; z st reset subsystem}~and~\eqref{eq; z st non-reset subsystem} in Eq.~\eqref{eq: gamma}, we obtain
\begin{eqnarray}
    \gamma = \frac{K_1}{2} \int_{-\infty}^{+\infty} d \omega g(\omega)\left[f\Gamma_1^{*}(\omega) + \bar{f}\Lambda_1^*(\omega)+4\pi^2f\Delta_1^{*}(\omega)\right],\nonumber\\ \label{eq: transcendental gamma}
\end{eqnarray}
where $\Gamma_1(\omega),\Lambda_1(\omega),$~and~$\Delta_1(\omega)$ are all function of $\gamma$ and $\gamma^{*}$ as well. Solving the self-consistency equation~\eqref{eq: transcendental gamma} numerically, we obtain the solution for $\gamma$. Putting it back into Eq.~\eqref{eq; z st non-reset subsystem}, we may obtain the stationary-state value of the order parameter of the non-reset subsystem.

\subsubsection{Stationary-State Distribution}
\begin{figure*}[htbp!]
\includegraphics[width=1\linewidth]{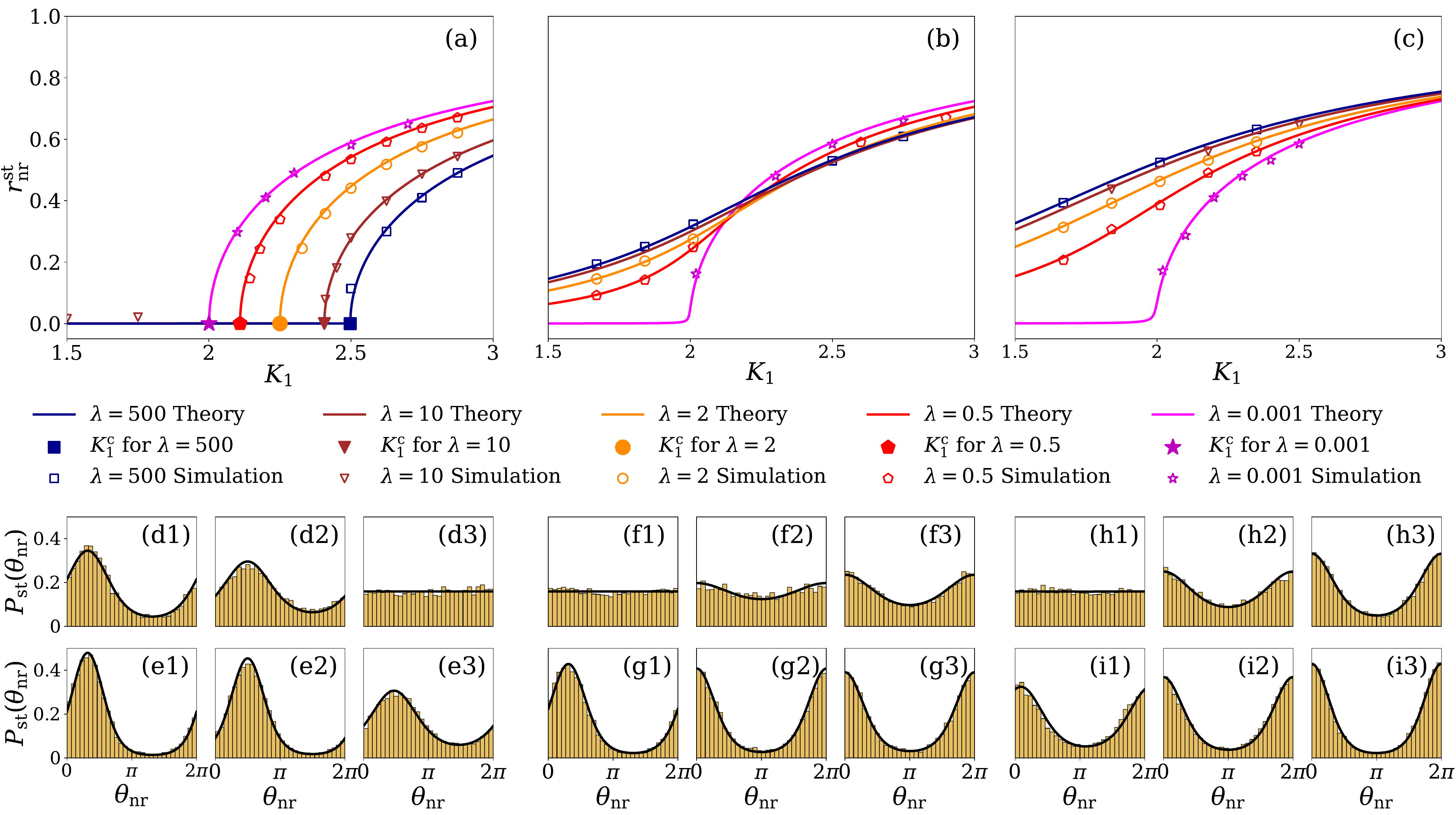}
   \caption{Results for the noisy Kuramoto model with first-harmonic interaction and identical frequencies ~(Sec.~\ref{subsec: Iso-noisy-harmonic Bare}): Agreement between theory (solid lines) and simulations (unfilled markers) in $r^\mathrm{st}_\mathrm{nr}$ versus $K_1$ for $f=0.2$ is shown  for reset configuration $r_0=0.0$ (panel (a)), $r_0=0.4$ (panel (b)), and $r_0=1.0$ (panel (c)). In (a), the filled markers denote the theoretically-obtained transition points $K_1^\mathrm{c}(\lambda)$, Eq.~\eqref{eq: critical K model 1 a}. Agreement between the theoretically-obtained stationary-state distribution (black lines) of the phase-angles of the oscillators from the non-reset subsystem (obtained from Eq.~\eqref{eq: eq: p of theta nr4}) and numerically-obtained histogram is shown in (d) -- (i). In each of these plots, $f$ is chosen to be $0.2$. For $r_0=0$, the plots are shown in (d) for $K_1=2.25$ and in (e) for $K_1=2.7$. For $r_0=0.4$, the plots are shown in (f) for $K_1=1.75$ and in (g) for $K_1=2.5$. For $r_0=1.0$, the plots are shown in (h) for $K_1=1.75$ and in (i) for $K_1=2.2$. In figures (d) -- (i), subplot (1) corresponds to $\lambda = 0.001$, (2) to $\lambda = 0.5$, and (3) to $\lambda = 500$.  In all simulations reported in the paper, the dynamics is integrated in time using a combination of the fourth-order Runge-Kutta method and the Euler–Maruyama method~\cite{Debraj}, with integration time step chosen to be 0.0005 for $\lambda=500$ and $0.001$ for rest of the $\lambda$ values. The system size is $N=10^4$.}

    \label{fig: model 1}
\end{figure*}

Once we obtain the solution of $\gamma$ from Eq.~\eqref{eq: transcendental gamma}, we may compute the distribution of the oscillator angles ($\theta_\mathrm{nr}$) of the non-reset subsystem in the stationary state. We start from the definition
\begin{eqnarray}
    P_\mathrm{st}(\theta_\mathrm{nr}) = \int_{-\infty}^{+\infty}d\omega_\mathrm{r}g(\omega_\mathrm{r})&&\int_{-\infty}^{+\infty}d\omega_\mathrm{nr}g(\omega_\mathrm{nr})\int_0^{2\pi}d\theta_\mathrm{r}\nonumber\\
&&\times P_\mathrm{st}(\theta_{\mathrm{r}},\omega_\mathrm{r},\theta_{\mathrm{n}r},\omega_\mathrm{nr}). \label{eq: p of theta nr}
\end{eqnarray}
Using the Fourier expansion of $P_\mathrm{st}(\theta_\mathrm{r}, \omega_\mathrm{r},\theta_\mathrm{nr},\omega_\mathrm{nr})$ given by Eq.~\eqref{eq: fourier} in Eq.~\eqref{eq: p of theta nr}, we obtain
\begin{eqnarray}
    P_\mathrm{st}(\theta_\mathrm{nr})=2\pi\sum_{m=-\infty}^{\infty}e^{im\theta_\mathrm{nr}}\int_{-\infty}^{+\infty}&&d\omega_\mathrm{r}d\omega_\mathrm{nr} g(\omega_\mathrm{r})g(\omega_\mathrm{nr})\nonumber\\
    &&\times \mathcal{P}_{0,m}(\omega_\mathrm{r},\omega_\mathrm{r}). \label{eq: p of theta nr2}
\end{eqnarray}
We now use Eqs.~\eqref{eq: p 0 m+1}~and~\eqref{eq: lambda l} along with Eq.~\eqref{eq: pi m = 0} in Eq.~\eqref{eq: p of theta nr2} and obtain
\begin{eqnarray}
    P_\mathrm{st}(\theta_\mathrm{nr})=&&\left[\frac{1}{2\pi}\sum_{m=1}^{\infty}e^{im\theta_\mathrm{nr}}\int_{-\infty}^{+\infty}d\omega_\mathrm{nr} g(\omega_\mathrm{nr}) \prod_{j=1}^{m}\Lambda_j(\omega_\mathrm{nr})\right.\nonumber\\
    && + \mathrm{c.c.} \Bigg]+ \frac{1}{2\pi} .\label{eq: eq: p of theta nr3}
\end{eqnarray}
Using the convergence of $\Lambda_m$, we now assume that for $m\geq\tilde{L}$, we may put $\Lambda_{m+1} = \Lambda_m =\Lambda_{m\to\infty,+}$ up to a desired precision. Using this, we may rewrite Eq.~\eqref{eq: eq: p of theta nr3} as
\begin{eqnarray}
    P_\mathrm{st}(\theta_\mathrm{nr})&&=\left[\frac{1}{2\pi}\sum_{m=1}^{\tilde{L}-1}e^{im\theta_\mathrm{nr}}\int_{-\infty}^{+\infty}d\omega_\mathrm{nr} g(\omega_\mathrm{nr}) \prod_{j=1}^{m}\Lambda_j\right.\nonumber \\
    && +\frac{e^{i\tilde{L}\theta_{\mathrm{nr}}}\Lambda_\infty}{2\pi\left(1-e^{i\theta_\mathrm{nr}}\Lambda_\infty\right)}\int_{-\infty}^{+\infty}d\omega_\mathrm{nr} g(\omega_\mathrm{nr}) \prod_{j=1}^{\tilde{L}-1}\Lambda_j \nonumber\\
    && + \mathrm{c.c.} \Bigg]+ \frac{1}{2\pi} .\label{eq: eq: p of theta nr4}
\end{eqnarray}
For brevity, we use $\Lambda_\infty$ to denote $\Lambda_{m\to \infty,+}$, whose expression is given in Eq.~\eqref{eq: lambda infty}. Equation~\eqref{eq: eq: p of theta nr4} provides the final expression for $P_\mathrm{st}(\theta_\mathrm{nr})$. In the case of $D\neq0$, we have $\Lambda_\infty = 0$, which makes the second term inside the bracket of Eq.~\eqref{eq: eq: p of theta nr4} and its complex conjugate to vanish. Note that  $P_\mathrm{st}(\theta_\mathrm{nr})$ contains the resetting rate $\lambda$ as a parameter. Therefore, by numerically computing $P_\mathrm{st}(\theta_\mathrm{nr})$ from Eq.~\eqref{eq: eq: p of theta nr4} for different values of $\lambda$, we may study how the stationary distribution of the non-reset subsystem changes as we increase the resetting rate, as shown in Fig.~\ref{fig: model 1} (panels (d1)--(i3)).

\subsubsection{Transition Points}~\label{sec:critical_points_harmonic}

Let us now obtain the transition point of the order-disorder transition in presence of resetting. Note that Eq.~\eqref{eq: transcendental gamma} has the form of a self-consistent equation $\gamma=\mathcal{F}(\gamma)$. We are interested in the following: If $\gamma=0$ is a solution of Eq.~\eqref{eq: transcendental gamma}, does the equation also admit a $\gamma\neq0$ solution? Assuming there is only one $\gamma\neq0$ solution possible in the physically-meaningful range $0<|\gamma|\leq K_1/2$, its existence depends on the nature of the function $\mathcal{F}(\gamma)$ near $\gamma=0$. More precisely, upon changing the parameters appearing in $\mathcal{F}(\gamma)$, when its slope at $\gamma=0$ crosses unity, the equation will admit a nonzero solution. In order to obtain the slope of $\mathcal{F}(\gamma)$ at $\gamma=0$, we need to Taylor expand $\mathcal{F}(\gamma)$ around that point, i.e., consider the small-$\gamma$ expansion of $\mathcal{F}(\gamma)$. Replacing $\mathcal{F}$ by its Taylor expansion in $\gamma=\mathcal{F}(\gamma)$, if we obtain that $\gamma=0$ is a solution, this validates our initial assumption of $\gamma=0$ as a valid solution. We start by expanding the expressions of $\Gamma_1^*$, $\Lambda_1^*$ , and $\Delta_1^*$ around $\gamma=0$, and obtain for $\Gamma_1$ that
\begin{align}
    \Gamma_1(\omega_\mathrm{r}) &= \frac{\gamma^{*}}{a_1}\left[1-\frac{1}{a_1}\frac{2|\gamma|^2}{a_2+2\gamma\left[\frac{3\gamma^{*}}{a_3+3\gamma\left[\ddots\right]}\right]}\right],\label{eq: Gamma Cont Frac1}
\end{align}
where, we recall that $a_l = (l^2D+il\omega_\mathrm{r} +\lambda )$. A further Taylor expansion of Eq.~\eqref{eq: Gamma Cont Frac1} yields
\begin{align}
    \Gamma_1(\omega_\mathrm{r}) &=  \frac{\gamma^{*}}{a_1}-\frac{2\gamma^{*}|\gamma|^2}{a_1^2a_2}+\mathcal{O}(|\gamma|^5)
.\label{eq: Gamma Cont Frac2}
\end{align}

In a similar way, an expansion of $\Lambda_1(\omega_\mathrm{nr})$ gives
\begin{eqnarray}
    \Lambda_1(\omega_\mathrm{nr}) = \frac{\gamma^{*}}{c_1}-\frac{2\gamma^{*}|\gamma|^2}{c_1^2c_2}+\mathcal{O}(|\gamma|^5),\label{eq: lambda 1 expression final}
\end{eqnarray}
where we have defined $c_l \equiv (l^2D+il\omega_\mathrm{nr} )$. We now focus on $\Delta_1$. The denominator of $\Delta_1$ (see Eqs.~\eqref{eq: gamma l} and~\eqref{eq: delta l}) is the same as that of $\Gamma_1$. Hence, we may write using the expansion of $\Gamma_1$ that
\begin{eqnarray}
    \Delta_1(\omega_\mathrm{r}) = (b_1-\gamma\Delta_2)\frac{1}{a_1}\left[1-\frac{2|\gamma|^2}{a_1a_2}+\mathcal{O}\left(|\gamma|^4\right)\right], \label{eq: Delta 1 expansion sup1}
\end{eqnarray}
where we have defined $b_l \equiv (\lambda/(4\pi^2))\left[ \alpha  + (-1)^l(1-\alpha) \right]$. Now, $\Delta_2$ may be expanded as
\begin{eqnarray}
    \Delta_2(\omega_\mathrm{r}) = \left(b_2-2\gamma\Delta_3\right)\frac{1}{a_2}\left[1-\frac{6|\gamma|^2}{a_2a_3}+\mathcal{O}\left(|\gamma|^4\right)\right],
\end{eqnarray}
while $\Delta_3$ may be expanded as
\begin{eqnarray}
    \Delta_3 (\omega_\mathrm{r})= \left(b_3-3\gamma\Delta_4\right)\left[\frac{1}{a_3}+\mathcal{O}\left(|\gamma|^2\right)\right],
\end{eqnarray}
and $\Delta_4$ may be expanded as
\begin{eqnarray}
    \Delta_4 (\omega_\mathrm{r})= \frac{b_4}{a_4} +\mathcal{O}\left(|\gamma|\right).
\end{eqnarray}

Then, putting all of them back into Eq.~\eqref{eq: Delta 1 expansion sup1}, we get
\begin{eqnarray}
    &&\Delta_1(\omega_\mathrm{r})  =\frac{b_1}{a_1} -\frac{b_2}{a_1a_2}\gamma+2\left(\frac{b_3}{a_1a_2a_3}\gamma^2-\frac{b_1}{a_1^2a_2}|\gamma|^2\right)\nonumber \\
    &&+\left[\left(\frac{6b_2}{a_1a_2^2a_3}+\frac{2b_2}{a^2_1a^2_2}\right)\gamma|\gamma|^2-\frac{6b_4}{a_1a_2a_3a_4}\gamma^3\right]+\ldots . \label{eq: delta 1 expression final}
\end{eqnarray}
Now, we write $\gamma = \delta e^{i\Phi}$, with $\delta \equiv |\gamma|$ being a small quantity. Putting Eqs.~\eqref{eq: Gamma Cont Frac2},~\eqref{eq: lambda 1 expression final}~and~\eqref{eq: delta 1 expression final} back into Eq.~\eqref{eq: transcendental gamma} , we obtain
\begin{eqnarray}
    \mathcal{A} + \mathcal{B}\delta+\mathcal{C}\delta^2 + \ldots=0, \label{eq: stability equation}
\end{eqnarray}
where
\begin{widetext}
\begin{align}
\mathcal{A} &\equiv \frac{\lambda(2\alpha-1)K_1f}{2} \int_{-\infty}^{+\infty} d \omega \frac{g(\omega)}{D-i\omega+\lambda},\label{eq: mathcal A}\\
    \mathcal{B}&\equiv \frac{K_1}{2} \int_{-\infty}^{+\infty} d \omega g(\omega)\left[\frac{fe^{i\Phi}}{D-i\omega+\lambda}+\frac{(1-f)e^{i\Phi}}{D-i\omega} -\frac{\lambda fe^{-i\Phi}}{(D-i\omega+\lambda)(4D-i2\omega+\lambda)} \right]-e^{i\Phi},\label{eq: mathcal B}\\
\mathcal{C} &\equiv \lambda(2\alpha-1)K_1f \int_{-\infty}^{+\infty} d \omega g(\omega) \Bigg[ \frac{e^{-i2\Phi}}{(D-i\omega+\lambda)(4D-i2\omega+\lambda)(9D-i3\omega+\lambda)}-\left.\frac{1}{(D-i\omega+\lambda)^2(4D-i2\omega+\lambda)}\right]. \label{eq: mathcal C}
\end{align}
\end{widetext}

Near the transition point, $\delta$ is small. Hence, only the first few terms in Eq.~\eqref{eq: stability equation} are important in determining the value of $\delta$. As we go more into the synchronized phase, the value of $\delta$ increases, and we need to consider higher-order terms in Eq.~\eqref{eq: stability equation}. The fact that the transition point is signaled by having the slope of $\mathcal{F}(\gamma)$ at $\gamma=0$ equal to unity translates to having $\mathcal{B}=0$.

Before moving forward, let us discuss the results presented in Eqs.~\eqref{eq: stability equation},~\eqref{eq: mathcal A},~\eqref{eq: mathcal B},~and~\eqref{eq: mathcal C}. In the non-resetting case, clearly $\lambda = 0$, which immediately gives $\mathcal{A} = \mathcal{C} = 0$. It then follows that Eq.~\eqref{eq: stability equation} admits a solution $\gamma=0$, implying an order-disorder transition, which is consistent with the discussion in Sec.~\ref{sec:level2}. Furthermore, Eq.~\eqref{eq: mathcal B} simplifies to
\begin{eqnarray}
    \mathcal{B}_{\lambda=0} = \left[\frac{K_1}{2}\int_{-\infty}^{+\infty} d\omega \frac{g(\omega)}{D-i\omega}-1\right]e^{i\Phi}. \label{eq:single_harmonic_criticiality_noreset}
\end{eqnarray}
Thus, the solution of the equation
\begin{equation}
    \mathcal{B}_{\lambda=0} = 0,
\end{equation}
will give us the transition points. For model~\ref{subsec: Iso-noisy-harmonic Bare}, putting $g(\omega) = \delta(\omega)$, we immediately obtain $K_1^\mathrm{c} =2D$, which agrees with Eq.~\eqref{eq: Kc model 1}. For model~\eqref{subsec: noiseless-harmonic-lorentzian Bare}, converting the integral~\eqref{eq:single_harmonic_criticiality_noreset} into a contour integral in the complex-$\omega$ plane and evaluating it using the theorem of residues, one immediately obtains the transition point to be $K_1^\mathrm{c} = 2/[\pi g(\omega_0)]$, which agrees with Eq.~\eqref{eq: Kc general g omega harmonic }.

In the presence of resetting, if we reset to an incoherent state ($r_0=0$), we have $\alpha = 1/2$, which again gives $\mathcal{A} = \mathcal{C} = 0$. Thus, Eq.~\eqref{eq: stability equation} admits a solution $\gamma=0$, indicating in this case that the system shows an order-disorder transition even in the presence of resetting, although the transition points now depend on the resetting rate $\lambda$.

While resetting to a partially synchronized or a fully synchronized configuration, we have $\alpha \neq 1/2$. Hence, the quantity $\mathcal{A}$ becomes non-zero in general, indicating that $\gamma=0$ is not a solution of Eq.~\eqref{eq: stability equation}. This, in turn, implies that the system does not show an order-disorder transition. 

Let us remark that the formalism presented in this section until now hold in the very general set-up of Eq.~\eqref{eq: Generalized Kuramoto} with $K_1\ne 0,~K_{l \ge2}=0$: one may choose any distribution $g(\omega)$, and our results will apply equally well to these choices. For illustrative purposes, we now use our theory to present explicit results for a few representative cases. 

\subsection{Application to Model~\ref{subsec: Iso-noisy-harmonic Bare} \label{subsec: app 1 critical}}

In this case, we have the frequency distribution as $g(\omega) = \delta(\omega).$ Thus Eqs.~\eqref{eq: mathcal A},~\eqref{eq: mathcal B},~and~\eqref{eq: mathcal C} reduce to
\begin{eqnarray}
    \mathcal{A} &=& \frac{\lambda(2\alpha-1)K_1f}{2(D+\lambda)} ,\label{eq: mathcal A model 1}\\
    \mathcal{B}&=& \frac{K_1}{2} \left[\frac{fe^{i\Phi}}{D+\lambda}+\frac{\bar{f}e^{i\Phi}}{D} -\frac{\lambda fe^{-i\Phi}}{(D+\lambda)(4D+\lambda)} \right]-e^{i\Phi}, \nonumber\\\label{eq: mathcal B model 1}\\
    \mathcal{C} &=& \frac{\lambda(2\alpha-1)K_1f}{(D+\lambda)(4D+\lambda)}  \Bigg[ \frac{e^{-i2\Phi}}{(9D+\lambda)}-\frac{1}{(D+\lambda)}\Bigg], \label{eq: mathcal C model 1}
\end{eqnarray}
where recall that $\bar{f} = (1-f)$.

Let us now focus on the case of resetting to an incoherent state, i.e., $\alpha = 1/2$. This immediately gives from Eqs.~\eqref{eq: mathcal A model 1}~and~\eqref{eq: mathcal C model 1} that $\mathcal{A} = \mathcal{C} =0$. Hence, $\delta = 0$ becomes a solution of Eq.~\eqref{eq: stability equation}, indicating the presence of an order-disorder transition. The transition points may be obtained from the condition
\begin{equation}
    \mathcal{B} = 0. \label{eq: B = 0 for model 1}
\end{equation}
Since $\mathcal{B}$ is a complex quantity, it is useful to express Eq.~\eqref{eq: B = 0 for model 1} in terms of its real and imaginary parts. Multiplying both sides of Eq.~\eqref{eq: B = 0 for model 1} by $\delta$ and defining $\gamma \equiv \delta e^{i\Phi} = \delta \cos{\Phi} + i \delta \sin{\Phi} \equiv \gamma_\mathrm{R}+i\gamma_\mathrm{I}$, we obtain
\begin{eqnarray}
    \mathbf{C}^{-}\gamma_\mathrm{R} &=& 0,  \label{eq: gamma real model 1}\\
 \mathbf{C}^{+}\gamma_\mathrm{I} &=& 0 \label{eq: gamma imaginary model 1},
\end{eqnarray}
where we have defined
\begin{equation}
    \mathbf{C}^{\pm} \equiv \frac{K_1}{2} \left[\frac{f}{D+\lambda}+\frac{\bar{f}}{D} \pm\frac{\lambda f}{(D+\lambda)(4D+\lambda)} \right]-1.
\end{equation}

For $K_1$ smaller than a critical value $K_1^\mathrm{c}(\lambda)$, we have both $\mathbf{C}^{\pm}\neq 0$. Hence, the only solution that Eqs.~\eqref{eq: gamma real model 1}
~and~\eqref{eq: gamma imaginary model 1} can have is $\gamma_\mathrm{R} = \gamma_\mathrm{I} = 0$, indicating that the system is in the incoherent state. As we increase $K_1$ keeping $D$ and $\lambda$ fixed, the quantity $\mathbf{C}^{+}$ becomes zero at
\begin{equation}
    K_1 = K_1^\mathrm{c}(\lambda) \equiv \frac{2D(D+\lambda)(4D+\lambda)}{(1-f)\lambda^2+D(5-3f)\lambda+4D^2},\label{eq: critical K model 1 a}
\end{equation}
whereas $\mathbf{C^{-}}$ remains nonzero. Hence, right after the transition point, $\gamma_\mathrm{I}$ becomes non-zero, whereas $\gamma_\mathrm{R}$ remains zero, indicating $\Phi = \pi/2$ at the transition point. Furthermore, taking the limit $\lambda \to \infty$ in Eq.~\eqref{eq: critical K model 1 a}, we obtain
\begin{equation}
    K_1^{\mathrm{c}}(\lambda\to \infty) = \frac{2D}{1-f}.
\end{equation}
This expression provides, for a fixed fraction of the total system being reset, the maximum change that we can induce in the value of the transition point. Agreement between theory and simulations for this model is shown in Fig.~\ref{fig: model 1}. 

Let us now summarize the main features of the $r_\mathrm{nr}^\mathrm{st}$ versus $K_1$ plots in Fig.~\ref{fig: model 1}: (i) When resetting to an incoherent state $r_0=0$, the continuous phase transition of the bare dynamics is preserved (panel (a)). By contrast, for $r_0\ne 0$, the bare-model phase transition becomes a crossover (panels (b) and (c)). In panel (a), the transition point shifts monotonically to the right (i.e., the order-disorder transition takes place at a higher value of $K_1$) as one implements resetting over a faster time scale, that is, with increasing reset rate $\lambda$. We observe from panels (b) and (c) that for $K_1<K_1^c$ (where $K_1^c$ is the bare-model transition point), one obtains enhanced synchrony with the increase of $\lambda$. On the other hand, provided $0<r_0<1$, one has for $K_1>K_1^c$ that the amount of synchrony decreases with increasing $\lambda$. For $r_0=1$, again, one has enhanced synchrony with increasing $\lambda$. These features may be understood as follows: as explained in Sec.~\ref{sec:level3}, resetting at a given value of the coupling parameter(s) drives the order parameter of the non-reset subsystem to a value that lies between the reset value $r_0$ and the corresponding stationary value for the bare model. For $K_1<K_1^c$, the latter value is zero, and so one definitely has $r_0>0$; with increasing $\lambda$, when one has more resets, the value of $r_\mathrm{nr}^\mathrm{st}$ is drawn more towards the value $r_0$, thus becoming increasingly larger in magnitude. For $K_1>K_1^c$, the bare-model stationary value is nonzero, and so one has $r_0$ either (1) greater, or (2) lesser than the stationary value, unless $r_0=1$ when evidently only the scenario (1) applies. When (1) applies, with increasing $\lambda$, the value of $r_\mathrm{nr}^\mathrm{st}$ becomes increasingly larger in magnitude with increase of $\lambda$. When (2) applies, instead, the value of $r_\mathrm{nr}^\mathrm{st}$, being increasingly drawn to the value $r_0$, becomes increasingly smaller in magnitude with the increase of $\lambda$. These observations explain the aforementioned features of panels (b) and (c). On the basis of the above, we conclude that subsystem resetting serves as a mechanism to shift, suppress, or enhance synchronization by anchoring the order parameter to the reset configuration. 

\subsection{Application to Model~\ref{subsec: noiseless-harmonic-lorentzian Bare} }

Here, Eqs.~\eqref{eq: mathcal A},~\eqref{eq: mathcal B},~and~\eqref{eq: mathcal C} yields
\begin{eqnarray}
    \mathcal{A} &=& \frac{\lambda(2\alpha-1)K_1f}{2(\sigma+\lambda)} ,\label{eq: mathcal A model 2}\\
    \mathcal{B}&=& \frac{K_1}{2} \left[\frac{fe^{i\Phi}}{\sigma+\lambda}+\frac{\bar{f}e^{i\Phi}}{\sigma} -\frac{\lambda fe^{-i\Phi}}{(\sigma+\lambda)(2\sigma+\lambda)} \right]-e^{i\Phi}, \nonumber \\\label{eq: mathcal B model 2}\\
    \mathcal{C} &=& \frac{\lambda(2\alpha-1)K_1f}{(\sigma+\lambda)(2\sigma+\lambda)}  \Bigg[ \frac{e^{-i2\Phi}}{(3\sigma+\lambda)}-\frac{1}{(\sigma+\lambda)}\Bigg]. \label{eq: mathcal C model 2}
\end{eqnarray}
We now use a similar argument as used in the previous Sec.~\ref{subsec: app 1 critical} to obtain the transition point as
\begin{align}
    K_1 = K_1^\mathrm{c}(\lambda) \equiv \frac{2\sigma(\sigma+\lambda)(2\sigma+\lambda)}{(1-f)\lambda^2+(3-f)\sigma\lambda+2\sigma^2},\label{eq: critical K model 2}
\end{align}
and $\Phi = \pi/2$ at the transition point. Furthermore, taking the limit $\lambda \to \infty$ in Eq.~\eqref{eq: critical K model 2}, we obtain
\begin{equation}
    K_1^{\mathrm{c}}(\lambda\to \infty) = \frac{2\sigma}{1-f},
\end{equation}
which gives for a fixed fraction of the total system being reset the maximum change that can be induced in the value of the transition point.

For this particular model, the case of resetting to a fully-synchronized state was studied in Ref.~\cite{PhysRevE.109.064137} using the celebrated Ott-Antonsen ansatz.~\cite{Ott_2008}. This powerful ansatz was introduced to obtain a low-dimensional description for the noiseless Kuramoto model with harmonic interaction and Lorentzian frequency disorder. However, for more general models such as the current work, the applicability of the method used in~Ref.~\cite{PhysRevE.109.064137} is limited. In Appendix~\ref{app: A}, we show that our method reproduces the results obtained in Ref.~\cite{PhysRevE.109.064137}.

\begin{figure*}[htbp!]
\includegraphics[width=1\linewidth]{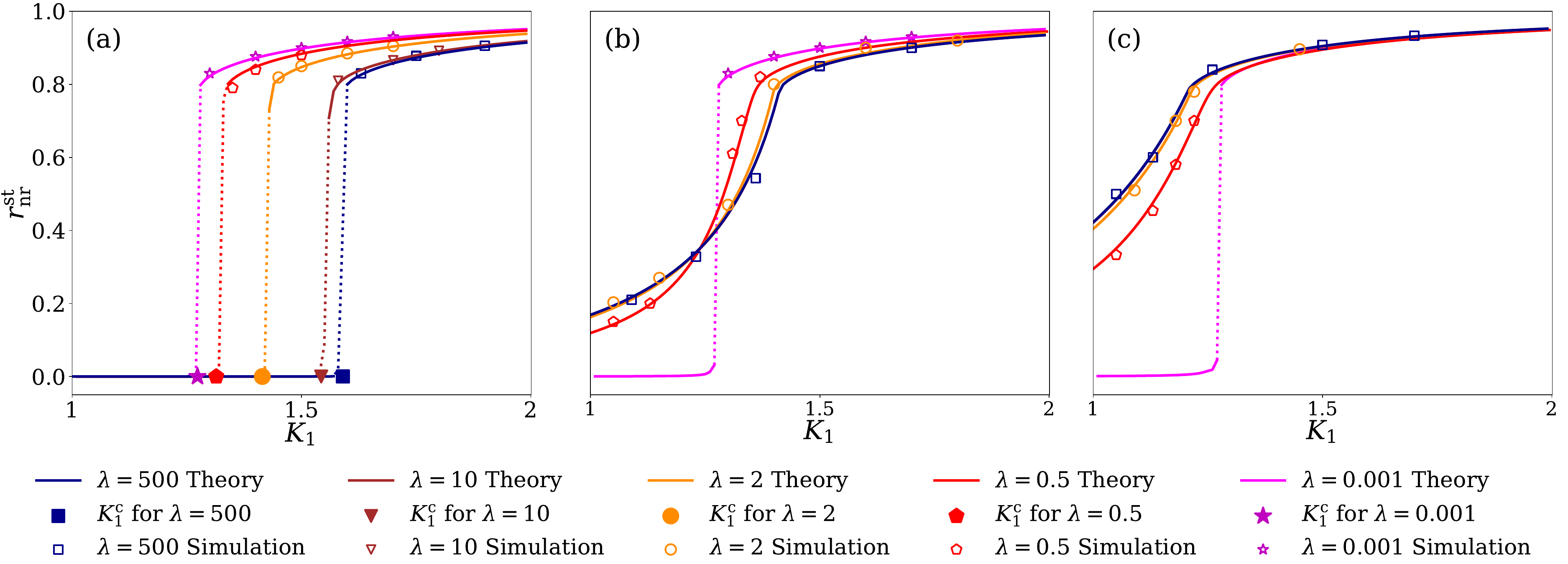}
\caption{Results for the noiseless Kuramoto model with first-harmonic interaction and with uniformly-distributed frequencies (Sec.~\ref{subsec: noiseless-harmonic-uniform Bare}): Agreement between theory (solid lines) and simulations (unfilled markers) in $r^\mathrm{st}_\mathrm{nr}$ versus $K_1$ for $f=0.2$ is shown for reset configuration $r_0=0.0$ (panel (a)), $r_0=0.4$ (panel (b)), and $r_0=1.0$ (panel (c)). In (a), the filled markers denote the theoretically-obtained transition points $K_1^\mathrm{c}(\lambda)$, Eq.~\eqref{eq: critical K model 3}. The system size is $N=10^4$.}
    \label{fig: model 2}
\end{figure*}

\subsection{Application to Model~\ref{subsec: noiseless-harmonic-uniform Bare} }

In this case, $g(\omega)$ is defined in Eq.~\eqref{eq: g omega unif}. Using this in Eqs.~\eqref{eq: mathcal A},~\eqref{eq: mathcal B}~and~\eqref{eq: mathcal C}, we obtain
\begin{align}
    \mathcal{A} &= \frac{\lambda(2\alpha-1)K_1f}{2\sigma} \tan^{-1}\left(\frac{\sigma}{\lambda}\right) ,\label{eq: mathcal A model 3}\\
    \mathcal{B} &= \frac{K_1}{2\sigma} \bigg[2f\cos{\Phi}\tan^{-1}\left(\frac{\sigma}{\lambda}\right)-fe^{-i\Phi}\tan^{-1}\left(\frac{2\sigma}{\lambda}\right)\nonumber\\
    &+\frac{\pi \bar{f}}{2}e^{i\Phi}\bigg]-e^{i\Phi},\label{eq: mathcal B model 3}\\
    \mathcal{C}&=\frac{(2\alpha-1)K_1 f}{2\lambda \sigma}\bigg[\frac{2\lambda \sigma}{\lambda^2+\sigma^2}+\Big(4+e^{-i2\Phi}\Big)\tan^{-1}\left(\frac{\sigma}{\lambda}\right) \nonumber\\
    &-4\Big(1+e^{-i2\Phi}\Big)\tan^{-1}\left(\frac{2\sigma}{\lambda}\right)+3e^{-i2\Phi}\tan^{-1}\left(\frac{3\sigma}{\lambda}\right)\bigg].\label{eq: mathcal C model 3}
\end{align}

Let us now focus on the case of resetting to an incoherent state, i.e., $\alpha = 1/2$. This immediately gives from Eqs~\eqref{eq: mathcal A model 3}~and~\eqref{eq: mathcal C model 3} that $\mathcal{A} = \mathcal{C} =0$. Hence, $\delta = 0$ becomes a solution of Eq.~\eqref{eq: stability equation}, indicating the presence of an order-disorder transition. We now follow the argument used in Sec.~\ref{subsec: app 1 critical} to obtain the transition points from the condition
\begin{equation}
    \mathcal{B} = 0. \label{eq: B = 0 for model 3}
\end{equation}
Since $\mathcal{B}$ is a complex quantity, expressing  Eq.~\eqref{eq: B = 0 for model 3} in terms of its real and imaginary parts, multiplying both sides of the equation by $\delta$, and defining $\gamma \equiv \delta e^{i\Phi} = \delta \cos{\Phi} + i \delta \sin{\Phi} \equiv \gamma_\mathrm{R}+i\gamma_\mathrm{I}$, we obtain
\begin{eqnarray}
    \mathbf{C}^{-}\gamma_\mathrm{R} &=& 0,  \label{eq: gamma real model 3}\\
 \mathbf{C}^{+}\gamma_\mathrm{I} &=& 0 \label{eq: gamma imaginary model 3},
\end{eqnarray}
where we have defined
\begin{align}
    \mathbf{C}^{+} &= \frac{K_1}{4\sigma}\bigg[2f\tan^{-1}\left(\frac{2\sigma}{\lambda}\right)+\pi\bar{f}\bigg]-1,\\
    \mathbf{C}^-&=\mathbf{C}^{+}+\frac{K_1f}{\sigma}\bigg[\tan^{-1}\left(\frac{\sigma}{\lambda}\right)-\tan^{-1}\left(\frac{2\sigma}{\lambda}\right)\bigg].
\end{align}
Since for arbitrary positive $x>0$, we have $(\tan^{-1}x-\tan^{-1}2x)<0$, we conclude that $\mathbf{C}^-<\mathbf{C}^+$ for all parameter range. For $K_1<K_1^\mathrm{c}(\lambda)$, we have both $\mathbf{C}^{\pm}\neq 0$. Hence, the only solution that Eqs.~\eqref{eq: gamma real model 3}
~and~\eqref{eq: gamma imaginary model 3} can have is $\gamma_\mathrm{R} = \gamma_\mathrm{I} = 0$, indicating that the system is in the incoherent state. As we increase $K_1$ keeping $\lambda$ fixed, the quantity $\mathbf{C}^{+}$ becomes zero at
\begin{equation}
    K_1 = K_1^\mathrm{c}(\lambda) \equiv \frac{4\sigma}{(1-f)\pi+2f\tan^{-1}(2\sigma/\lambda)},\label{eq: critical K model 3}
\end{equation}
whereas $\mathbf{C^{-}}$ remains nonzero. Hence, right after the transition point, $\gamma_\mathrm{I}$ becomes non-zero, whereas $\gamma_\mathrm{R}$ remains zero, indicating $\Phi = \pi/2$ at the transition point. Furthermore, taking the limit $\lambda \to \infty$ in Eq.~\eqref{eq: critical K model 3}, we obtain
\begin{equation}
    K_1^{\mathrm{c}}(\lambda\to \infty) = \frac{4\sigma}{(1-f)\pi}.
\end{equation}
Agreement between theory and simulations for this model is shown in Fig.~\ref{fig: model 2}. From the figure, we see that the features summarized above for the behavior of $r_\mathrm{nr}^\mathrm{st}$ in Fig.~\ref{fig: model 1} continue to hold here. Indeed, the phase transition of the bare model is retained on including resetting effects as long as the reset configuration is fully disordered ($r_0=0$) and is otherwise (i.e., with $r_0\ne 0$) converted into a crossover. Moreover, the behavior seen in Fig.~\ref{fig: model 1} in the $r_\mathrm{nr}^\mathrm{st}$ versus $K$ plots for $r_0\ne 0$ also applies in the current situation. 

\section{Analysis for general interaction \label{sec: app to gen}}

We now extend the analysis done in Sec.~\ref{sec: harmonic theory} for the general model as defined in Eq.~\eqref{eq: Generalized Kuramoto}. Our starting point is Eq.~\eqref{eq: FP Kuramoto Finite Resetting}, with Eq.~\eqref{eq: hx K1 K2 bare} being modified to
\begin{align}
     &h_\mathrm{x}  \equiv\omega_\mathrm{x}\nonumber \\
    &+f \sum_{l=1}^{\infty}K_l\int d\theta'_\mathrm{r}d\omega'_\mathrm{r} g(\omega_\mathrm{r}') P(\theta'_\mathrm{r},\omega'_\mathrm{r} ,t| \theta_\mathrm{x} ,\omega_\mathrm{x})\sin\left[l(\theta'_\mathrm{r}-\theta_\mathrm{x})\right] \nonumber \\
    & +\bar{f}\sum_{l=1}^{\infty}K_l \int d\theta'_\mathrm{nr}d\omega'_\mathrm{nr} g(\omega_\mathrm{nr}') P(\theta'_\mathrm{nr},\omega'_\mathrm{nr} ,t|  \theta_\mathrm{x},\omega_\mathrm{x}) \nonumber\\
    &~\qquad\times\sin\left[l(\theta'_\mathrm{nr}-\theta_\mathrm{x})\right]. \label{eq: hx general}
\end{align}
In the stationary state, using Eq.~\eqref{eq: con approx}, we may rewrite Eq.~\eqref{eq: hx general} as
\begin{eqnarray}
    h_\mathrm{x} = \omega_\mathrm{x}&& +f \sum_{l=1}^{\infty}K_lr^\mathrm{st}_{l,\mathrm{r}} \sin\left[l\left(\psi^\mathrm{st}_{l,\mathrm{r}} -\theta_\mathrm{x}\right)\right]\nonumber \\
    &&+\bar{f} \sum_{l=1}^{\infty}K_lr^\mathrm{st}_{l,\mathrm{nr}} \sin\left[l\left(\psi^\mathrm{st}_{l,\mathrm{nr}}-\theta_\mathrm{x}\right)\right], \label{eq: hx st general}
\end{eqnarray}
where the stationary-state values of the order parameters may be defined as in Eq.~\eqref{eq: zx}. Using the Fourier expansion of $P_\mathrm{st}(\theta_\mathrm{r}, \omega_\mathrm{r},\theta_\mathrm{nr},\omega_\mathrm{nr},t)$ given by Eq.~\eqref{eq: fourier} in Eq.~\eqref{eq: FP Kuramoto Finite Resetting} along with Eq.~\eqref{eq: hx st general} and comparing the coefficients $e^{i l \theta_\mathrm{r}+i m\theta_\mathrm{nr}}$ from the various terms in the equation, we get the following relation between the various $\mathcal{P}_{l,m}(\omega_\mathrm{r},\omega_\mathrm{nr})$'s:
\begin{align}
     &\left[(l^2+m^2)D+i(l \omega_\mathrm{r}+m\omega_\mathrm{nr})+\lambda\right] \mathcal{P}_{l,m} +\sum_{k=1}^{\infty}\gamma_k\left(l\mathcal{P}_{l+k,m}\right.\nonumber \\
     &\left.+m\mathcal{P}_{l,m+k}\right)-\sum_{k=1}^{\infty}\gamma^{*}_k\left(l\mathcal{P}_{l-k,m}+m\mathcal{P}_{l,m-k}\right) \nonumber \\
     & =  \lambda\left[ \alpha  + (-1)^l(1-\alpha) \right] \mathcal{P}_{0,m}, \label{eq: Fourier Relation Kuramoto SubReset Sup general}
\end{align}
where we have defined
\begin{eqnarray}
    \gamma_k \equiv \left[\frac{ K_kf r^\mathrm{st}_{k,\mathrm{r}}e^{ik \psi^\mathrm{st}_{k,\mathrm{r}}}+K_k\bar{f} r^\mathrm{st}_{k,\mathrm{nr}} e^{i k\psi^\mathrm{st}_{k,\mathrm{nr}}} }{2}\right].\label{eq: gamma general}
\end{eqnarray}

Following Eqs.~\eqref{lorder fourier sup1}~and~\eqref{lorder fourier sup2}, it is clear that to obtain the stationary-state values of the order parameters $z^\mathrm{st}_{l,\mathrm{x}} = r^\mathrm{st}_{l,\mathrm{x}}e^{il\psi^\mathrm{st}_{l,\mathrm{x}}}$, we need to find only the quantities $\mathcal{P}_{-l,0}$ and $\mathcal{P}_{0,-l}$. Since $\mathcal{P}_{-l,0} = \left(\mathcal{P}_{l,0}\right)^{*}$ and $\mathcal{P}_{0,-l} = \left(\mathcal{P}_{0,l}\right)^{*}$, we will focus on obtaining the quantities $\mathcal{P}_{l,0}$ and $\mathcal{P}_{0,l}$. Putting successively $m=0$ and $l=0$ in Eq.~\eqref{eq: Fourier Relation Kuramoto SubReset Sup general}, we obtain respectively that
\begin{align}
    &\left[l^2D+il \omega_\mathrm{r}+\lambda\right] \mathcal{P}_{l,0}+l \sum_{k=1}^{\infty} \left(\gamma_k \mathcal{P}_{l+k,0}-\gamma^{*}_k\mathcal{P}_{l-k,0}\right)\nonumber\\
    &= \frac{\lambda}{4\pi^2}\left[ \alpha  + (-1)^l(1-\alpha) \right] 
  \label{eq: Fourier Relation l-axis Kuramoto SubReset Sup general1} ,\\
  &\left[m^2D+im \omega_\mathrm{nr}\right] \mathcal{P}_{0,m}+m\sum_{k=1}^{\infty} \left(\gamma_k \mathcal{P}_{0,m+k}- \gamma^{*}_k\mathcal{P}_{0,m-k}\right)\nonumber\\
  &= 0 
  \label{eq: Fourier Relation m-axis Kuramoto SubReset Sup general1} .
\end{align}
To proceed further, let us assume that $K_l = 0~\forall~l>M$. Hence, we have $\gamma_k = 0~\forall~k>M$, which reduces Eqs.~\eqref{eq: Fourier Relation l-axis Kuramoto SubReset Sup general1}~and~\eqref{eq: Fourier Relation m-axis Kuramoto SubReset Sup general1} to
\begin{align}
    &\left[l^2D+il \omega_\mathrm{r}+\lambda\right] \mathcal{P}_{l,0}+l \sum_{k=1}^{M} \left(\gamma_k \mathcal{P}_{l+k,0}-\gamma^{*}_k\mathcal{P}_{l-k,0}\right)\nonumber\\
    &= \frac{\lambda}{4\pi^2}\left[ \alpha  + (-1)^l(1-\alpha) \right] ,
  \label{eq: Fourier Relation l-axis Kuramoto SubReset Sup general} \\
  &\left[m^2D+im \omega_\mathrm{nr}\right] \mathcal{P}_{0,m}+m\sum_{k=1}^{M} \left(\gamma_k \mathcal{P}_{0,m+k}- \gamma^{*}_k\mathcal{P}_{0,m-k}\right)\nonumber\\
  &= 0 \label{eq: Fourier Relation m-axis Kuramoto SubReset Sup general} .
\end{align}

Let us now focus on Eq.~\eqref{eq: Fourier Relation l-axis Kuramoto SubReset Sup general}. Following the same line of argument as invoked following Eq.~\eqref{eq: Fourier Relation m-axis Kuramoto SubReset Sup}, we make an ansatz similar to Eq.~\eqref{eq: l-axis ansatz Kuramoto SubReset Sup}:
\begin{eqnarray}
    \mathcal{P}_{l,0} &=& \sum_{k=1}^M \Gamma_{k,l}\mathcal{P}_{l-k,0} + \Delta_l, \label{eq: l ansatz general}\\
    \mathcal{P}_{0,m} &=& \sum_{k=1}^M \Lambda_{k,m}\mathcal{P}_{0,m-k} + \Pi_m. \label{eq: m ansatz general}
\end{eqnarray}
Putting Eqs.~\eqref{eq: l ansatz general}~and~\eqref{eq: m ansatz general} back into Eq.~\eqref{eq: Fourier Relation m-axis Kuramoto SubReset Sup general}, we obtain the recursion relations for $\Gamma_{k,l},~\Delta_l,~\Lambda_{k,m}$,~and~$\Pi_m$. Using these recursion relations, we may express each of the $\Gamma_{k,l},~\Delta_l,~\Lambda_{k,m}$,~and~$\Pi_m$ in a continued fraction form. We put these expressions into Eqs.~\eqref{eq: l ansatz general}~and~\eqref{eq: m ansatz general} for $l,m = 1,2,\ldots,M$ and get equations for $\mathcal{P}_{M,0},\mathcal{P}_{M-1,0},\ldots,\mathcal{P}_{1,0}$ and $\mathcal{P}_{0,M},\mathcal{P}_{0,M-1},\ldots,\mathcal{P}_{0,1}$. Our goal is to solve these $2M$ equations to get expressions of $\mathcal{P}_{M,0},\mathcal{P}_{M-1,0},\ldots,\mathcal{P}_{1,0}$ and $\mathcal{P}_{0,M},\mathcal{P}_{0,M-1},\ldots,\mathcal{P}_{0,1}$ solely using $\Gamma_{k,l},~\Delta_l,~\Lambda_{k,m}$,~and~$\Pi_m$. 
It is yet not doable, since these $2M$ equations are not closed with respect to $\mathcal{P}_{M,0},\mathcal{P}_{M-1,0},\ldots,\mathcal{P}_{1,0}$ and $\mathcal{P}_{0,M},\mathcal{P}_{0,M-1},\ldots,\mathcal{P}_{0,1}$. For example, the expression for $\mathcal{P}_{0,M-1}$ contains $\mathcal{P}_{0,-1}$, the expression for $\mathcal{P}_{0,M-2}$ contains $\mathcal{P}_{0,-1}$ and $\mathcal{P}_{0,-2}$. Going like this, the expression for $\mathcal{P}_{0,1}$ contains $\mathcal{P}_{0,-1},\mathcal{P}_{0,-2},\ldots,\mathcal{P}_{0,-(M-1)}$. Similar is the situation for $\mathcal{P}_{l,0}$'s with $l = 1,2,\ldots, M$. Thus, we have $2M$ equations and a total of $4M-2$ unknowns: $\mathcal{P}_{M,0},\mathcal{P}_{M-1,0},\ldots,\mathcal{P}_{1,0}$ and $\mathcal{P}_{0,M},\mathcal{P}_{0,M-1},\ldots,\mathcal{P}_{0,1}$ and $\mathcal{P}_{-1,0},\mathcal{P}_{-2,0},\ldots,\mathcal{P}_{-(M-1),0}$ and $\mathcal{P}_{0,-1},\mathcal{P}_{0,-2},\ldots,\mathcal{P}_{0,-(M-1)}$. Now, taking complex conjugate of the aforementioned $2M$ equations, we obtain another set of $2M$ equations. With these $4M$ equations along with the fact that $\mathcal{P}_{0,- l} = \left(\mathcal{P}_{0,l}\right)^{*}$ and $\mathcal{P}_{- m,0} = \left(\mathcal{P}_{m,0}\right)^{*}$, we finally express $\mathcal{P}_{M,0},\mathcal{P}_{M-1,0},\ldots,\mathcal{P}_{1,0}$ and $\mathcal{P}_{0,M},\mathcal{P}_{0,M-1},\ldots,\mathcal{P}_{0,1}$ solely in terms of $\Gamma_{k,l},~\Delta_l,~\Lambda_{k,m}$,~and~$\Pi_m$. Then putting these expressions into Eqs.~\eqref{lorder fourier sup1}~and~\eqref{lorder fourier sup2} and solving them simultaneously, we finally obtain the stationary-state values of the order parameters.

For $M=2$, this method is worked out in the next section.

\subsection{Application to Model~\ref{subsec: noisy-harmonic-biharmonic Bare}}
For the Kuramoto model with first- and second-harmonic interaction, we are interested in finding the values of the quantities $r^\mathrm{st}_\mathrm{1,r},r^\mathrm{st}_\mathrm{2,r},r^\mathrm{st}_\mathrm{1,nr}$~and~$r^\mathrm{st}_\mathrm{2,nr}$. Hence, following Eqs.~\eqref{lorder fourier sup1}~and~\eqref{lorder fourier sup2}, our quantities of interest are $\mathcal{P}_{-1,0}, \mathcal{P}_{-2,0}, \mathcal{P}_{0,-1}$~and~$\mathcal{P}_{0,-2}$. Since $\mathcal{P}_{-l,0} = \left(\mathcal{P}_{l,0}\right)^{*}$, we will focus on finding $\mathcal{P}_{1,0}, \mathcal{P}_{2,0}, \mathcal{P}_{0,1}$~and~$\mathcal{P}_{0,2}$ 

Let us first focus on finding $\mathcal{P}_{1,0}$~and~$\mathcal{P}_{2,0}$. For the case under consideration, we have $M=2$. Putting $M=2$ in Eq.~\eqref{eq: l ansatz general}, we obtain
\begin{eqnarray}
    \mathcal{P}_{l,0} = \Gamma_{1,l}\mathcal{P}_{l-1,0}+\Gamma_{2,l}\mathcal{P}_{l-2,0}+\Delta_{l}. \label{eq: l=2 ansatz general}
\end{eqnarray}
Then, putting the expansion given by Eq.~\eqref{eq: l=2 ansatz general} for $\mathcal{P}_{l+2}$ and $\mathcal{P}_{l+1}$ into Eq.~\eqref{eq: Fourier Relation l-axis Kuramoto SubReset Sup general} with $M=2$, we obtain
\begin{align}
    \Gamma_{1,l} &= \frac{l\gamma^{*}_1-l\Gamma_{2,l+1}\left(\gamma_1+\gamma_2\Gamma_{1,l+2}\right)}{a_l + l\gamma_2\Gamma_{2,l+2}+l\Gamma_{1,l+1}\left(\gamma_1+\gamma_2\Gamma_{1,l+2}\right)}, \label{eq: gamma 1 l}\\
    \Gamma_{2,l} &= \frac{l\gamma^{*}_2}{a_l + l\gamma_2\Gamma_{2,l+2}+l\Gamma_{1,l+1}\left(\gamma_1+\gamma_2\Gamma_{1,l+2}\right)},\label{eq: gamma 2 l}\\
    \Delta_{l} &= \frac{b_l-l\gamma_2\Delta_{l+2}-l\Delta_{l+1}\left(\gamma_1+\gamma_2\Gamma_{1,l+2}\right)}{a_l + l\gamma_2\Gamma_{2,l+2}+l\Gamma_{1,l+1}\left(\gamma_1+\gamma_2\Gamma_{1,l+2}\right)}\label{eq: Delta l},
\end{align}
where recall that $a_l = \left(l^2D+il\omega_\mathrm{r}+\lambda\right)$ and $b_l = (\lambda/(4\pi^2))\left[ \alpha  + (-1)^l(1-\alpha) \right]$. Clearly, for $K_2=0$, we have $\gamma_2 = 0$ from Eq.~\eqref{eq: gamma general}. Putting this in Eq~\eqref{eq: gamma 2 l}, we obtain $\Gamma_{2,l} = 0~\forall~l$. Using these in Eqs.~\eqref{eq: gamma 1 l}~and~\eqref{eq: Delta l}, we get back Eqs.~\eqref{eq: gamma l}~and~\eqref{eq: delta l}, thus proving our consistency. Now, for $l=1$ and $l=2$, we have from Eq.~\eqref{eq: l=2 ansatz general} that
\begin{eqnarray}
    \mathcal{P}_{1,0} &=& \frac{\Gamma_{1,1}}{4\pi^2} + \Gamma_{2,1} \mathcal{P}_{-1,0} + \Delta_{1}, \label{eq: p 1 0}\\
    \mathcal{P}_{2,0} &=&  \Gamma_{1,2} \mathcal{P}_{1,0} +\frac{\Gamma_{2,2}}{4\pi^2} + \Delta_{2}\label{eq: p 2 0}.
\end{eqnarray}
In Eq.~\eqref{eq: p 1 0}, we may replace $\mathcal{P}_{-1,0} = \left(\mathcal{P}_{1,0}\right)^{*}$ and take complex conjugate of Eq.~\eqref{eq: p 1 0} to solve $\mathcal{P}_{1,0}$ in terms of $\Gamma$'s and $\Delta$'s, which reads as
\begin{align}
    \mathcal{P}_{1,0} = \frac{\Gamma_{1,1}+\Gamma^{*}_{1,1}\Gamma_{2,1} + 4\pi^2\left(\Delta_1+\Delta^{*}_1\Gamma_{2,1}\right)}{4\pi^2\left(1-\Gamma^{*}_{2,1}\Gamma_{2,1}\right)}, \label{eq: p 1 0 final}
\end{align}
where $\Gamma_{1,1},\Gamma_{2,1},\Delta_1$ are given by Eqs.~\eqref{eq: gamma 1 l},~\eqref{eq: gamma 2 l},~and~\eqref{eq: Delta l}. Putting the expression of $\mathcal{P}_{1,0}$ from Eq.~\eqref{eq: p 1 0 final} into Eq.~\eqref{eq: p 2 0}, we obtain $\mathcal{P}_{2,0}$. 

Now, we have to find the large-$l$ behavior of the quantities $\Gamma_{1,1},\Gamma_{2,1},\Delta_1$ in order to approximate them for numerical computations. Using a similar argument as in the paragraph following Eq.~\eqref{eq: Gamma Cont Frac}, here also we conclude that $\Gamma_{1,l},\Gamma_{2,l}, $ and $\Delta_l$ must converge. Hence, there exists a value of $l$, say $l=L$, such that $\Gamma_{1,l+1}=\Gamma_{1,l}$, $\Gamma_{2,l+1}=\Gamma_{2,l}$ and $\Delta_{l+1}=\Delta_{l}$, to our desired precision for all $l\geq L$. Hence, for $l\geq L$, we obtain
\begin{eqnarray}
   && l \gamma_2\Gamma_{1,l}^3 + l\gamma_1 \Gamma_{1,l}^2+2l\gamma_2 \Gamma_{1,l}\Gamma_{2,l}+a_l\Gamma_{1,l}+l\gamma_1\Gamma_{2,l} = l\gamma_1^{*}, \nonumber \\\label{eq: large l equation 1}\\ 
&&l \gamma_2\Gamma_{2,l}^2+l \gamma_2\Gamma_{1,l}^2\Gamma_{2,l}+l \gamma_1\Gamma_{1,l}\Gamma_{2,l}+a_l \Gamma_{2,l} = l\gamma_2^{*}. \label{eq: large l equation 2}
\end{eqnarray}
If the noise strength $D\neq 0$, then for large $l$, the quantity $a_l$ behaves as $a_l \sim l^2$. Hence, the most dominant term on the left-hand side of Eqs.~\eqref{eq: large l equation 1}~and~\eqref{eq: large l equation 2} is the term containing $a_l$. Then, for large $l$, we may write
\begin{eqnarray}
    \Gamma_{1,l} = \frac{l\gamma_1^{*}}{l^2D+il\omega_\mathrm{r}+\lambda},~~\Gamma_{2,l}=\frac{l\gamma_2^{*}}{l^2D+il\omega_\mathrm{r}+\lambda},
\end{eqnarray}
which immediately gives
\begin{eqnarray}
    \Gamma_{1,l \to \infty} = \Gamma_{2,l \to \infty} = 0,~\mathrm{if}~D\neq 0. \label{eq:Gamma_1l_and_Gamma_2l_convergence}
\end{eqnarray}
For $D=0$, we may write $a_l \approx i\omega_\mathrm{r}$ for large $l$. Hence, Eqs.~\eqref{eq: large l equation 1}~and~\eqref{eq: large l equation 2} become
\begin{eqnarray}
   && \gamma_2\Gamma_{1,l}^3 + \gamma_1 \Gamma_{1,l}^2+2\gamma_2 \Gamma_{1,l}\Gamma_{2,l}+i\omega_\mathrm{r}\Gamma_{1,l}+\gamma_1\Gamma_{2,l} = \gamma_1^{*}, \nonumber \\ \label{eq: large l equation 1 D = 0}\\
&&\gamma_2\Gamma_{2,l}^2+ \gamma_2\Gamma_{1,l}^2\Gamma_{2,l}+ \gamma_1\Gamma_{1,l}\Gamma_{2,l}+i\omega_\mathrm{r} \Gamma_{2,l} = \gamma_2^{*}. \label{eq: large l equation 2 D = 0}
\end{eqnarray}
Finding the roots of Eqs.~\eqref{eq: large l equation 1 D = 0}~and~\eqref{eq: large l equation 2 D = 0}, we obtain expressions for $\Gamma_{1,l\to \infty}$ and $\Gamma_{2\to \infty}$ in the case of $D=0$. Using these expressions for $\Gamma_{1,l\to \infty}$ and $\Gamma_{2,l\to\infty}$ , and following a similar argument as given in the paragraph following Eq.~\eqref{eq: towards gamma1 approximation}, here also we approximate $\Gamma_{1,l}$ , and $\Gamma_{2,l}$ for further numerical computations. Using a similar argument as done in the case of obtaining Eq.~\eqref{eq: Delta l to infty}, here also we obtain $\Delta_{l\to\infty} = 0$, and we use this fact to approximate $\Delta_1$ and $\Delta_2$.

Using the fact that $(\mathcal{P}_{-1,0}) = (\mathcal{P}_{1,0})^{*}$ and $(\mathcal{P}_{-2,0}) = (\mathcal{P}_{2,0})^{*}$, we obtain on using the expression of $\mathcal{P}_{-1,0}$ and $\mathcal{P}_{-2,0}$ in Eq.~\eqref{lorder fourier sup1} that
\begin{align}
r^\mathrm{st}_\mathrm{1,r}e^{i\psi^\mathrm{st}_\mathrm{1,r}} =& \int_{-\infty}^{+\infty} d \omega_\mathrm{r} g(\omega_\mathrm{r})\Bigg[ \nonumber \\
&\left. \frac{\Gamma^{*}_{1,1}+\Gamma_{1,1}\Gamma^{*}_{2,1} + 4\pi^2\left(\Delta^{*}_1+\Delta_1\Gamma^{*}_{2,1}\right)}{\left(1-\Gamma^{*}_{2,1}\Gamma_{2,1}\right)}\right] ,\label{eq; z st non-reset subsystem L2 1}\\
r^\mathrm{st}_\mathrm{2,r}e^{i2\psi^\mathrm{st}_\mathrm{2,r}} =& \int_{-\infty}^{+\infty} d \omega_\mathrm{r} g(\omega_\mathrm{r})\Bigg[ \Gamma^{*}_{2,2}+4\pi^2 \Delta^{*}_2\nonumber \\
&\left. +~\Gamma^{*}_{1,2} \frac{\Gamma^{*}_{1,1}+\Gamma_{1,1}\Gamma^{*}_{2,1} + 4\pi^2\left(\Delta^{*}_1+\Delta_1\Gamma^{*}_{2,1}\right)}{\left(1-\Gamma^{*}_{2,1}\Gamma_{2,1}\right)}\right].\label{eq; z st non-reset subsystem L2 2}
\end{align}

We now focus on finding $\mathcal{P}_{0,1}$~and~$\mathcal{P}_{0,2}$. Putting $M=2$ into Eq.~\eqref{eq: m ansatz general}, we obtain
\begin{eqnarray}
    \mathcal{P}_{0,m} = \Lambda_{1,m}\mathcal{P}_{0,m-1}+\Lambda_{2,m}\mathcal{P}_{0,m-2}+\Pi_{m}. \label{eq: m=2 ansatz general}
\end{eqnarray}
Thus, putting the expansion given by Eq.~\eqref{eq: m=2 ansatz general} for $\mathcal{P}_{0,m+2}$ and $\mathcal{P}_{0,m+1}$ into Eq.~\eqref{eq: Fourier Relation m-axis Kuramoto SubReset Sup general} with $M=2$, we obtain
\begin{align}
    \Lambda_{1,m} &= \frac{m\gamma^{*}_1-m\Lambda_{2,m+1}\left(\gamma_1+\gamma_2\Lambda_{1,m+2}\right)}{c_m + m\gamma_2\Lambda_{2,m+2}+m\Lambda_{1,m+1}\left(\gamma_1+\gamma_2\Lambda_{1,m+2}\right)}, \label{eq: lambda 1 m}\\
    \Lambda_{2,m} &= \frac{m\gamma^{*}_2}{c_m + m\gamma_2\Lambda_{2,m+2}+m\Lambda_{1,m+1}\left(\gamma_1+\gamma_2\Lambda_{1,m+2}\right)},\label{eq: lambda 2 m}\\
    \Pi_{m} &= -\frac{m\gamma_2\Pi_{m+2}+m\Pi_{m+1}\left(\gamma_1+\gamma_2\Lambda_{1,m+2}\right)}{c_m + m\gamma_2\Lambda_{2,m+2}+m\Lambda_{1,m+1}\left(\gamma_1+\gamma_2\Lambda_{1,m+2}\right)}\label{eq: Pi m},
\end{align}
where we have $c_m = \left(m^2D+im\omega_\mathrm{nr}\right)$. Using a similar argument as for the case of $\Gamma_{1,l},\Gamma_{2,l}$ and $\Delta_l$, we obtain
\begin{eqnarray}
    \Lambda_{1,m\to\infty,+} =\Lambda_{2,m\to\infty,+}= 0,~\mathrm{if}~~D\neq0,\label{eq:Lambda_1l_and_Lambda_2l_convergence}
\end{eqnarray}
and
\begin{eqnarray}
    \Pi_{m\to\infty} = 0. \label{eq: Pi m inf M2}
\end{eqnarray}
Using Eq.~\eqref{eq: Pi m inf M2} in  Eq.~\eqref{eq: Pi m}, we obtain
\begin{eqnarray}
    \Pi_m(\omega_\mathrm{nr}) = 0~\forall~m. \label{eq: pi m = 0 bi harmonic}
\end{eqnarray}
Now, for $m=1$ and $m=2$, we have from Eq.~\eqref{eq: m=2 ansatz general} that
\begin{align}
    &\mathcal{P}_{0,1} =\frac{\Lambda_{1,1}}{4\pi^2} + \Lambda_{2,1} \mathcal{P}_{0,-1}, \label{eq: p 0 1}\\
    &\mathcal{P}_{0,2} =  \Lambda_{1,2} \mathcal{P}_{0,1} +\frac{\Lambda_{2,2}}{4\pi^2}\label{eq: p 0 2}.
\end{align}

Using the fact that $(\mathcal{P}_{0,-1}) = (\mathcal{P}_{0,1})^{*}$, we obtain on using the expression of $\mathcal{P}_{0,-1}$ in Eq.~\eqref{lorder fourier sup2} that
\begin{align}
r^\mathrm{st}_\mathrm{1,nr}e^{i\psi^\mathrm{st}_\mathrm{1,nr}} =& \int_{-\infty}^{+\infty} d \omega_\mathrm{nr} g(\omega_\mathrm{nr})\left(\frac{\Lambda^{*}_{1,1}+\Lambda_{1,1}\Lambda^{*}_{2,1} }{1-\Lambda^{*}_{2,1}\Lambda_{2,1}}\right) ,\label{eq; z st non-reset subsystem M2 1}\\
r^\mathrm{st}_\mathrm{2,nr}e^{i2\psi^\mathrm{st}_\mathrm{2,nr}}=&  \int_{-\infty}^{+\infty} d \omega_\mathrm{nr} g(\omega_\mathrm{nr})\left[\Lambda^{*}_{1,2} \left(\frac{\Lambda^{*}_{1,1}+\Lambda_{1,1}\Lambda^{*}_{2,1} }{1-\Lambda^{*}_{2,1}\Lambda_{2,1}}\right) \right. \nonumber \\
&+\Lambda^{*}_{2,2}\Bigg].\label{eq; z st non-reset subsystem M2 2}
\end{align}

Solving Eq.~\eqref{eq; z st non-reset subsystem L2 1},~\eqref{eq; z st non-reset subsystem L2 2},~\eqref{eq; z st non-reset subsystem M2 1},~and~\eqref{eq; z st non-reset subsystem M2 2} simultaneously, we obtain the stationary-state order parameters of the non-reset subsystem.

\subsubsection{Transition Points of Model~\ref{subsec: noisy-harmonic-biharmonic Bare}}

\begin{figure*}[htbp!]
 \includegraphics[width=1\linewidth]{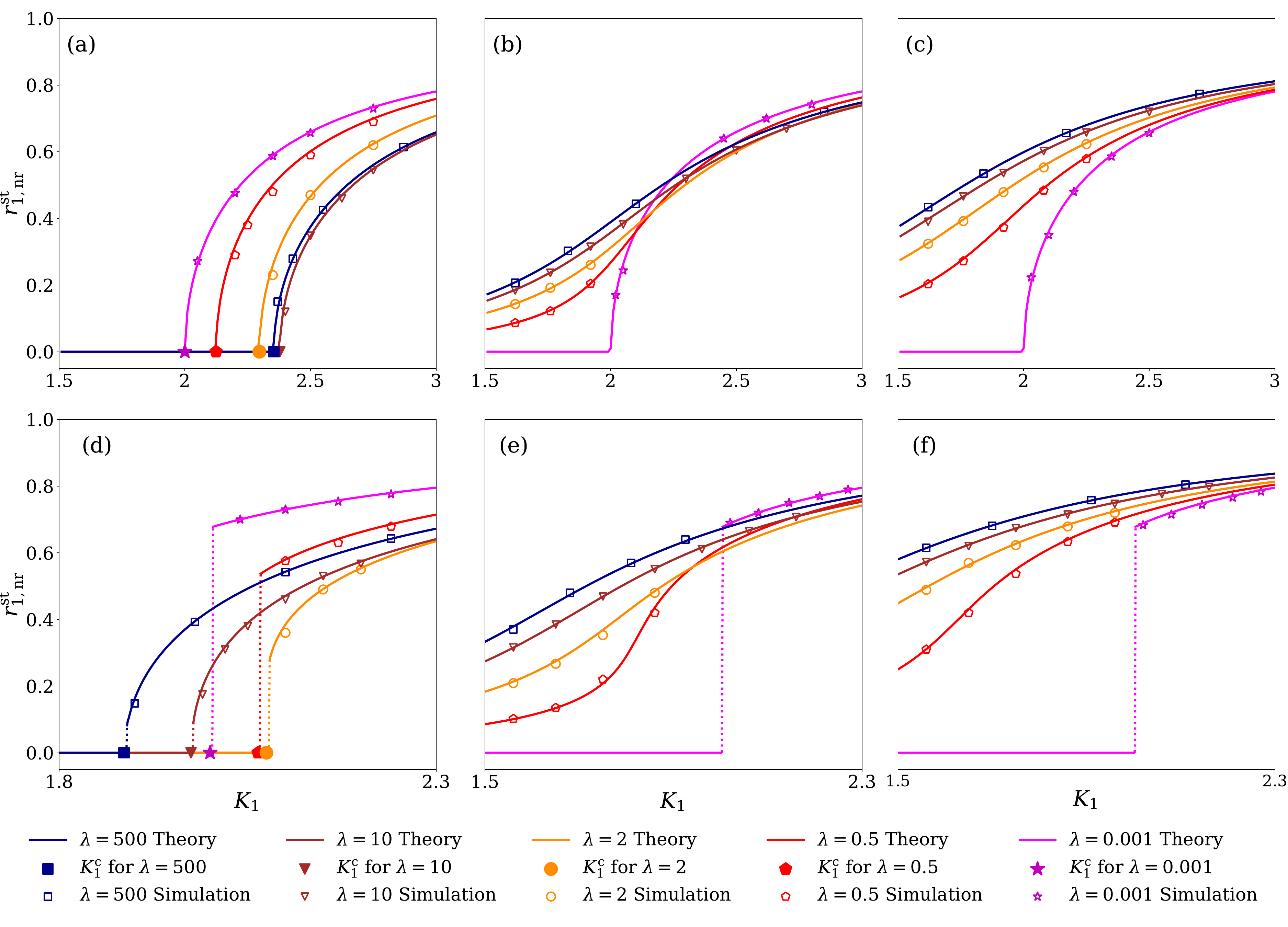}
       \caption{Results for noisy Kuramoto model with first and second harmonic interaction and identical frequencies~(Sec.~\ref{subsec: noisy-harmonic-biharmonic Bare}): Agreement between theory (solid lines) and simulations (open markers) for the stationary order parameter of the non-reset subsystem, $r_{1,\mathrm{nr}}^{\mathrm{st}}$, as a function of $K_1$ for $f=0.2, D=1.0$. Panels (a)–(c) correspond to $K_2=1.0<2D$, while panels (d)–(f) correspond to $K_2 = 3.0>2D$. In each case, the reset configuration is $r_0=0.0$ [(a),(d)], $r_0=0.4$ [(b),(e)], and $r_0=1.0$ [(c),(f)]. Filled markers indicate the theoretical transition points $K_1^{\mathrm c}$ given by Eq.~\eqref{eq:K1c_general_model_II_C}. The system size is $N=5\times 10^3$.} 
    \label{fig: model 4}
\end{figure*}

We now move on to obtaining the transition points for the order-disorder transition corresponding to the two order parameters $z^\mathrm{st}_{1,\mathrm{nr}}$ and $z^\mathrm{st}_{2,\mathrm{nr}}$ in the presence of subsystem resetting for the model~\ref{subsec: noisy-harmonic-biharmonic Bare}. Before delving into the problem, let us first understand the situation physically. As it turns out that there is no transition in $z^\mathrm{st}_{1,\mathrm{nr}}$ for $\alpha \neq 1/2$ (see Fig.~\ref{fig: model 4} obtained by numerically solving Eqs.~\eqref{eq; z st non-reset subsystem L2 1},~\eqref{eq; z st non-reset subsystem L2 2},~\eqref{eq; z st non-reset subsystem M2 1},~and~\eqref{eq; z st non-reset subsystem M2 2} and verified by simulations), we will exclusively focus on the $\alpha = 1/2$ case here. Following Sec.~\ref{sec:level3}, in the case of $\alpha = 1/2$, the reset configuration is chosen such that the angles of half of the oscillators belonging to the reset subsystem are reset to $0$, and those of the other half are reset to $\pi$. Considering $n_r = f N$, the number of oscillators in the reset subsystem, we may calculate the synchronization order parameters ($z_{0,k};~k=1,2$) of the reset configuration, which read as
\begin{align}
&z_{0,1} = \frac{1}{n_r}\Big[\tfrac{n_r}{2} e^{i \cdot 0} + \tfrac{n_r}{2} e^{i \pi}\Big] = 0,\nonumber\\
&\label{eq:resetting_order_parameters}\\
&z_{0,2} = \frac{1}{n_r}\Big[\tfrac{n_r}{2} e^{i \cdot 0} + \tfrac{n_r}{2}  e^{i 2\pi}\Big] = 1.\nonumber
\end{align}
Hence, under this reset protocol, the order-parameter $r_{1,\mathrm{r}}= |z_{1,\mathrm{r}}|$ is reset to $0$ and $r_{2,\mathrm{r}}=|z_{2,\mathrm{r}}|$ is reset to unity. From our earlier results on first-harmonic interaction discussed in this work as well as from  Refs.~\cite{PhysRevE.109.064137, acharya2025manipulating}, we know that in case of resetting to a fully-synchronized state, i.e., when $r_{1,\mathrm{r}}$ is reset to unity, the transition in $r^\mathrm{st}_{1,\mathrm{nr}}$ in the bare dynamics is replaced by a crossover, thereby making $r^\mathrm{st}_{1,\mathrm{nr}}=0$ not a solution in the stationary state. We may expect that for the case at hand, $r^\mathrm{st}_{2,\mathrm{r}}=0$ will not be a feasible solution, and therefore, $r^\mathrm{st}_{2,\mathrm{nr}}$ will not show any transition, although $r^\mathrm{st}_{1,\mathrm{nr}}$ will show transition.

To validate the last statement mathematically, let us first assume that there is an order-disorder transition existing for both the order parameters $r^\mathrm{st}_{1,\mathrm{nr}}$ and $r^\mathrm{st}_{2,\mathrm{nr}}$. Hence $\gamma_1=0=\gamma_2$ should be a solution of the self-consistent equations of the form $\gamma_k = \mathcal{F}_k(\gamma_1, \gamma_2);~k=1,2$ given in Eq.~\eqref{eq: gamma general}. If this assumption is true, we may perform a Taylor expansion of $\mathcal{F}_k$ around the point $\gamma_1=0=\gamma_2$. Using this Taylor series expansion into the equation $\gamma_k = \mathcal{F}_k(\gamma_1, \gamma_2)$, we should consistently obtain $\gamma_1=0=\gamma_2$ as a solution. Following these steps, we will unveil that only $\gamma_1=0$ but not $\gamma_2=0$ is a solution. This proves the absence of an order-disorder transition in $r^\mathrm{st}_{2,\mathrm{nr}}$.

We now go into the details of the aforementioned computation. According to Eq.~\eqref{eq: gamma general} along with Eqs.~\eqref{eq; z st non-reset subsystem L2 1} and \eqref{eq; z st non-reset subsystem M2 1}, evaluating $\gamma_1$ requires the expansion of $\Gamma_{1,1}$, $\Gamma_{2,1}$, $\Delta_1$ corresponding to the reset subsystem and $\Lambda_{1,1}$, and $\Lambda_{2,1}$ corresponding to the non-reset subsystem. Now, from Eq.~\eqref{eq: gamma 1 l}, retaining only the leading-order terms, we find 
$\Gamma_{1,1} = \gamma_1/a_1$ and $\Gamma_{2,1} =\gamma_2/a_1$. Similarly, one obtains 
$\Lambda_{1,1} = \frac{\gamma_1}{c_1}$, $\Lambda_{2,1} = \frac{\gamma_2}{c_1}$, and $\Delta_1 = \frac{b_1}{a_1}- \frac{b_2 \gamma_1}{a_1 a_2}- \frac{b_3 \gamma_2}{a_1 a_3}$. By substituting these forms into Eq.~\eqref{eq: gamma general}, we obtain $\mathcal{A}_1 + \mathcal{B}_1(\gamma_1,\gamma_2,\gamma_1^*,\gamma_2^*) + \dots=0$, with the coefficients given by
\begin{align}
\mathcal{A}_1 &= \frac{2 \pi^2 K_1 f b_1}{a_1^*}, \label{eq:coefficient_A_gamma1} \\
\mathcal{B}_1 &= \frac{K_1}{2} \Bigg[\left( \frac{f}{a_1^*} + \frac{\bar{f}}{c_1^*} \right)\gamma_1
- 4\pi^2 f \left(\frac{b_2 \gamma_1^*}{a_1^* a_2^*}+\frac{b_3 \gamma_2^*}{a_1^* a_3^*}\right)
\nonumber\\
&\quad~+\frac{4\pi^2 f b_1}{|a_1|^2}\gamma_2
\Bigg]. \label{eq:linear_gamma1_equation}
\end{align}
Now, for $\alpha = 1/2$, the quantity $b_1$ vanishes, leading to $\mathcal{A}_1 = 0$, and allowing for an incoherent solution $\gamma_1 = 0$. The threshold for the order-disorder transition in $r^\mathrm{st}_{1,\mathrm{nr}}$ is then determined by the condition $\mathcal{B}_1 = 0$. Next, according to Eq.~\eqref{eq: gamma general} along with Eqs.~\eqref{eq; z st non-reset subsystem L2 2} and \eqref{eq; z st non-reset subsystem M2 2}, evaluating $\gamma_2$ requires the expressions of $\Gamma_{1,1}$, $\Gamma_{1,2}$, $\Gamma_{2,1}$, $\Gamma_{2,2}$, $\Delta_2$ corresponding to the reset subsystem and $\Lambda_{1,1}$, $\Lambda_{1,2}$, $\Lambda_{2,1}$ and $\Lambda_{2,2}$ corresponding to the non-reset subsystem. One may eventually obtain $ \mathcal{A}_2 + \mathcal{B}_2(\gamma_1,\gamma_2,\gamma_1^*,\gamma_2^*) + \dots=0$, with
\begin{align}
\mathcal{A}_2 &= \frac{2 \pi^2 K_2 f b_2}{ a_2^*}, \label{eq:coefficient_A_gamm2} \\
\mathcal{B}_2 &= K_2 \biggl[\left( \frac{f}{a_2^*} + \frac{\bar{f}}{c_2^*} \right)\gamma_2
+ 4\pi^2 f \left(\frac{b_1 \gamma_1}{a_1^* a_2^*}-\frac{b_3 \gamma_1^*}{a_2^* a_3^*}\right)
\nonumber\\
&\quad~-\frac{4\pi^2 f b_4}{a_2^* a_4^*}\gamma_2^*\biggr]. \label{eq:linear_gamma2_equation}
\end{align}
In contrast to Eq.~\eqref{eq:coefficient_A_gamma1}, here, the presence of a non-vanishing $\mathcal{A}_2$ term (here, $b_2 \neq 0$ for $\alpha=1/2$) indicates that $\gamma_2 = 0$ is no longer a valid solution. This implies that for any finite $\lambda$, the quantity $\gamma_2$ can never be zero, implying that $r^\mathrm{st}_{2,\mathrm{nr}}$ does not show an order-disorder transition for $\lambda\neq0$. Thus we have proved what we had set out to. Hence, we need to do the Taylor series expansion of $\mathcal{F}_k(\gamma_1, \gamma_2);~k=1,2$ around the point $\gamma_1=0$ and $\gamma_2 = \gamma_2^{(0)}\neq0$, where $\gamma_2^{(0)}$ will be obtained self-consistently. Note that, in the case of $\lambda=0$, putting $g(\omega) = \delta(\omega)$, and solving Eqs.~\eqref{eq:linear_gamma1_equation} and ~\eqref{eq:linear_gamma2_equation} yield the transition points of the bare model, which match with the previously-presented results, Fig.~\ref{fig:bi-phdiag}.

In the presence of resetting, as discussed in the preceding paragraphs, 
transition in $\gamma_1$ may arise for the cases $K_1 \neq 0,\, K_2 = 0$ and $K_1 \neq 0,\, K_2 \neq 0$. For $K_2 = 0$, Eq.~\eqref{eq: gamma general} implies $\gamma_2=0$, while using $\gamma_1 = \delta e^{i\Phi}$, Eq.~\eqref{eq:linear_gamma1_equation} reduces exactly to 
Eq.~\eqref{eq: mathcal B}. This is expected, since in this limit, the system 
effectively reduces to the single-harmonic case discussed in 
Sec.~\ref{sec:critical_points_harmonic}. 

We now focus on obtaining the transition point of the transition in $\gamma_1$ in the case of $K_2\neq0$. We have from our earlier discussion that $\gamma_2$ cannot be zero in the stationary state. Hence, to obtain the transition point of $\gamma_1$, we first need to compute the value of $\gamma_2$ at the transition point of $\gamma_1$. Inside the incoherent phase as well as at the transition point, we have $\gamma_1=0$. Putting this condition into Eqs.~\eqref{eq; z st non-reset subsystem L2 2} and \eqref{eq; z st non-reset subsystem M2 2} and using them in the definition of $\gamma_2$ in Eq.~\eqref{eq: gamma general}, we obtain a self-consistent relation of $\gamma_2$ which is valid in the incoherent region of $\gamma_1$ as well as at the transition point. Note that $K_1$ dependency of $\gamma_2$ comes only through $\gamma_1$ (see Eq.~\eqref{eq: gamma general}). Since $\gamma_1=0$ in the incoherent phase as well as at the transition point, $\gamma_2$ becomes independent of $K_1$, and remains constant (equal to say $\gamma_2^{(0)}$). Let us now find the self-consistent relation of $\gamma_2^{(0)}$. To avoid confusion, we will denote the continued fractions at $\gamma_1=0$ as $\Gamma_{1,l}^{(0)},~ \Gamma_{2,l}^{(0)},~ \Delta_l^{(0)}, ~\Lambda_{1,l}^{(0)},~\Lambda_{l,l}^{(0)} $. Putting $\gamma_1=0$ in Eq.~\eqref{eq: gamma 1 l}, we obtain that
\begin{align}
    \Gamma_{1,l}^{(0)} = -\frac{l\gamma_2\Gamma_{2,l+1}^{(0)}\Gamma_{1,l+2}^{(0)}}{a_l + l\gamma_2^{(0)}\big(\Gamma_{2,l+2}^{(0)}+\Gamma_{1,l+1}^{(0)}\Gamma_{1,l+2}^{(0)}\big)}, \label{eq:Gamma_1l_0}
\end{align}
which upon using Eq.~\eqref{eq:Gamma_1l_and_Gamma_2l_convergence} gives $\Gamma_{1,l}^{(0)}=0~~\forall~l$. Similarly, putting $\gamma_1=0$ in Eq.~\eqref{eq: lambda 1 m} along with Eq.~\eqref{eq:Lambda_1l_and_Lambda_2l_convergence}, we obtain $\Lambda_{1,l}^{(0)}=0~~\forall~l$. Focusing on Eqs.~\eqref{eq: gamma 2 l},~\eqref{eq: Delta l},~\eqref{eq: lambda 2 m} and putting $\gamma_1=0$, we obtain

\begin{align}
    \Gamma_{2,l}^{(0)} &= \frac{l\big[\gamma_2^{(0)}\big]^*}{a_l + l\gamma_2^{(0)}\Gamma_{2,l+2}^{(0)}},~~~~    \Delta_{l}^{(0)}= \frac{b_l-l\gamma_2^{(0)}\Delta_{l+2}^{(0)}}{a_l + l\gamma_2^{(0)}\Gamma_{2,l+2}^{(0)}},\label{eq:gamma_2_0_cf}\\
\Lambda_{2,m}^{(0)} &= \frac{m\big[\gamma^{(0)}_2\big]^{*}}{c_m + m\gamma_2^{(0)}\Lambda_{2,m+2}^{(0)}}.
\end{align}
Putting them into Eqs.~\eqref{eq; z st non-reset subsystem L2 2}~and~\eqref{eq; z st non-reset subsystem M2 2} along with the definition given in Eq.~\eqref{eq: gamma general}, we obtain the self-consistent equation for $\gamma_2^{(0)}$, which reads as
\begin{align}
    \gamma_2^{(0)} =
    \frac{K_2 f}{2}\left\{\big[\Gamma_{2,2}^{(0)} \big]^{*}+ 4\pi^2\big[\Delta_2^{(0)}\big]^*\right\}
    + \frac{K_2\bar{f}}{2}\,\big[\Lambda_{2,2}^{(0)}\big]^*.
    \label{eq:gamma2_background}
\end{align}
For $\alpha=1/2$, $b_l$ vanishes for odd $l$ and equals
$\lambda/(4\pi^2)$ for even $l$. The former fact implies that $\Delta_l^{(0)}=0$ for odd $l$. However,
$\Delta_l^{(0)}$ for even $l$ is driven by
 nonzero $b_l$ for every even $l$ and is non-trivially coupled through
$\gamma_2^{(0)}$, as
\begin{align}
    \Delta_2^{(0)} = \frac{b_2 - 2\gamma_2^{(0)}\Delta_4^{(0)}}{A_2^{(0)}},
    \quad
    \Delta_4^{(0)} = \frac{b_4 - 4\gamma_2^{(0)}\Delta_6^{(0)}}{A_4^{(0)}},
    \quad \dots \label{eq:delta_even_l_gamma_2_0}
\end{align}
Solving Eq.~\eqref{eq:gamma2_background}, we  obtain $\gamma_2^{(0)}$.

After computing $\gamma_2^{(0)}$, we move on to compute the transition point of $\gamma_1$. From Eqs~\eqref{eq; z st non-reset subsystem L2 1},~\eqref{eq; z st non-reset subsystem M2 1} along with the expressions of the continued fraction, it is clear that we may write Eq.~\eqref{eq: gamma general}
 for $\gamma_1$ as
\begin{align}
    \gamma_1 = \mathcal{F}_1(\gamma_1,\gamma_2)
    \label{eq:F_map1}.
\end{align}
When $\gamma_1=0$, we have $\gamma_2=\gamma_2^{(0)}$. However, in the synchronized phase, when $\gamma_1$ is nonzero, $\gamma_2$ becomes dependent on $\gamma_1$ and starts deviating from $\gamma_2^{(0)}$. Hence, we may write $\gamma_2 = \gamma_2^{(0)}+\delta\gamma_2$ and put it into Eq.~\eqref{eq:F_map1}. Close to the transition point, $\gamma_1$ is a small quantity. In Appendix~\ref{app: B}, we show that in this region, we have $\delta\gamma_2\sim\mathcal{O}(|\gamma_1|^2)$. Since the transition point depends on the behavior of $\mathcal{F}_1$ to linear order in $\gamma_1$, we may replace $\gamma_2$ by $\gamma_2^{(0)}$ in Eq.~\eqref{eq:F_map1} near the transition point (see the discussion in Sec.~\ref{sec:critical_points_harmonic}).

Now, Eq.~\eqref{eq:F_map1} is similar to the form of equations described in Eq.~\ref{sec:critical_points_harmonic}. Here, we do not have an explicit expression for $\mathcal{F}_1$ due to the presence of $\gamma_2^{(0)}$. We now adopt the following method to calculate the transition points. Since $\gamma_1 \in \mathbb{C}$, writing $\gamma_1=\gamma_\mathrm{R}+i \gamma_\mathrm{I}$, doing a Taylor series expansion of $\mathcal{F}_1$ around $\gamma_1=0$, and keeping up to first order terms, we obtain that
\begin{align}
    \begin{pmatrix} \gamma_\mathrm{R} \\ \gamma_\mathrm{I} \end{pmatrix}
    =
    \frac{K_1}{2}\,\mathbf{J}\,
    \begin{pmatrix} \gamma_\mathrm{R} \\ \gamma_\mathrm{I} \end{pmatrix},
    \label{eq:linearised_iteration}
\end{align}
where $\mathbf{J}$ is the $2\times 2$ Jacobian matrix given by derivatives of
$\mathcal{F}_1$ with respect to $(\gamma_\mathrm{R},\gamma_\mathrm{I})$ and evaluated at $\gamma_1=0$ :
\begin{align}
    \mathbf{J} =
    \begin{pmatrix}
        \partial_{\gamma_\mathrm{R}}\,\mathrm{Re}[\mathcal{F}_1] &
        \partial_{\gamma_\mathrm{I}}\,\mathrm{Re}[\mathcal{F}_1] \\[4pt]
        \partial_{\gamma_\mathrm{R}}\,\mathrm{Im}[\mathcal{F}_1] &
        \partial_{\gamma_\mathrm{I}}\,\mathrm{Im}[\mathcal{F}_1]
    \end{pmatrix}_{\!\gamma_1=0}.
    \label{eq:jacobian}
\end{align}
As we tune $K_1$ keeping other parameters fixed, $\gamma_1$ becomes non-zero when $K_1$ crosses $K_1^\mathrm{c}$. To obtain a nonzero solution of $\gamma_1$ from Eq.~\eqref{eq:linearised_iteration}, we must have
\begin{align}
    \mathrm{det}\bigg[\frac{K^\mathrm{c}_1}{2}\,\mathbf{J}-\mathbb{I}\bigg]=0, \label{eq: con crit}
\end{align}
where $\mathbb{I}$ is a $2\times2$ identity matrix. Now let the eigenvalues of the matrix $\mathbf{J}$ be $\mu_+(\mathbf{J})$ and $\mu_{-}(\mathbf{J})$ with $\mu_+(\mathbf{J})>\mu_{-}(\mathbf{J})$. As we increase $K_1$, at $K_1=K_{1,+} = 2/\mu_+(\mathbf{J})$, one eigenvalue of the matrix $(K^\mathrm{c}_1\mathbf{J}/2-\mathbb{I})$ becomes zero, thereby satisfying Eq.~\eqref{eq: con crit}. From this point onward, $\gamma_1 \neq 0$ becomes a valid solution. If we further increase $K_1$, Eq.~\eqref{eq: con crit} will again be satisfied at $K_1=K_{1,-}=2/\mu_{-}(\mathbf{J})$. Since $K_{1,+}<K_{1,-}$, the system shows phase transition at
\begin{align}
    K_1^c= K_{1+}^c= \frac{2}{\mu_+(\mathbf{J})},\label{eq:K1c_general_model_II_C}
\end{align}
where $\mu_+(\mathbf{J})$ denotes the largest eigenvalue of $\mathbf{J}$. The transition points in Fig~\ref{fig: model 4} are obtained using Eq.~\eqref{eq:K1c_general_model_II_C}.

We now summarize the main features of Fig~\ref{fig: model 4}. Similar to Figs.~\ref{fig: model 1} and~\ref{fig: model 2}, the phase transition of the bare model, be it first-order or continuous, is retained on including resetting effects as long as the reset configuration is fully disordered ($r_0=0$) and is otherwise (i.e., with $r_0\ne 0$) turned into a crossover. Moreover, the behavior seen in Figs.~\ref{fig: model 1} and~\ref{fig: model 2} in the $r_\mathrm{nr}^\mathrm{st}$ versus $K$ plots for $r_0\ne 0$ and with increase of $\lambda$ continue to hold here. 

\begin{figure}
    \centering
\includegraphics[width=7 cm]{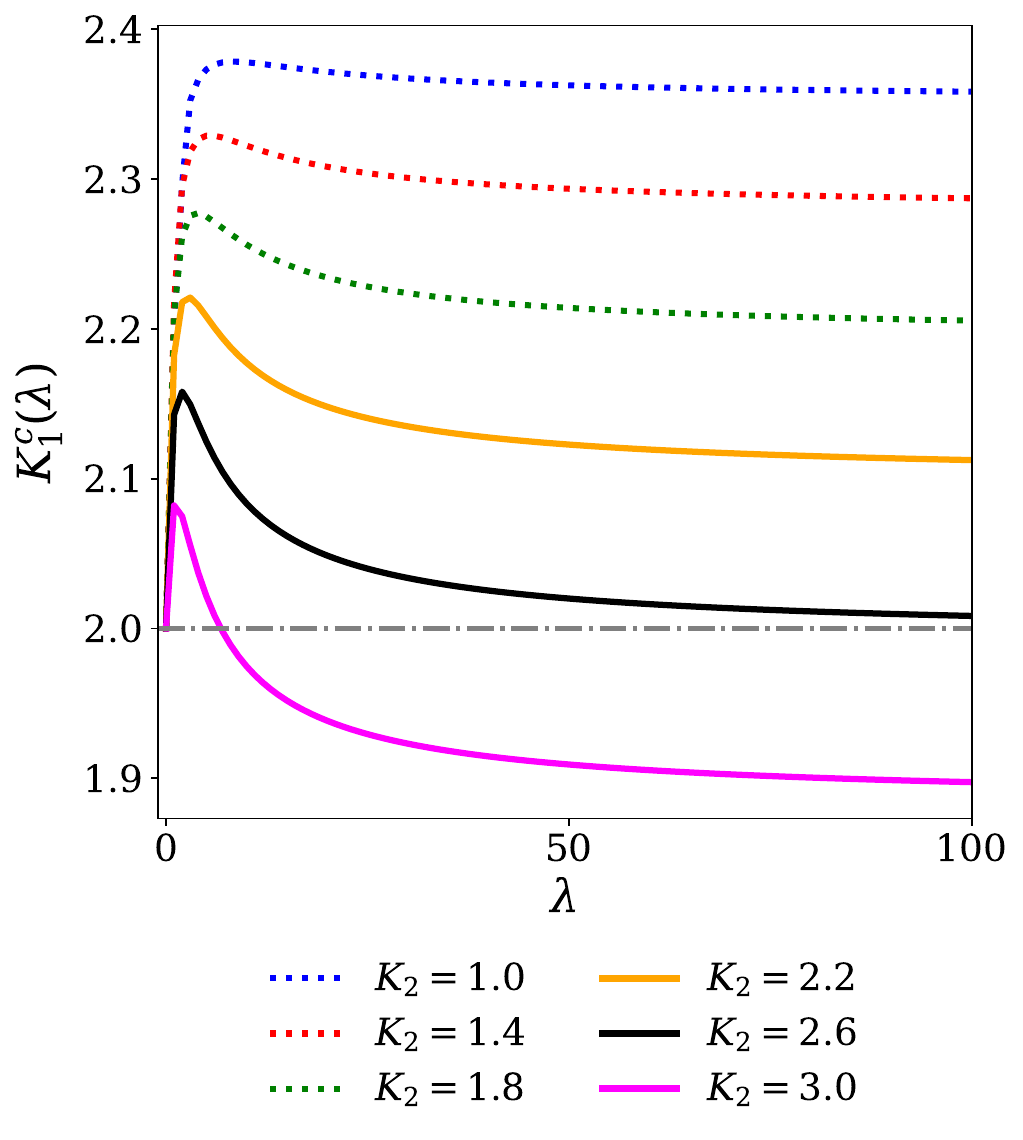}
    \caption{Noisy Kuramoto model with first and second harmonic interaction and identical frequencies~(Sec.~\ref{subsec: noisy-harmonic-biharmonic Bare}): ~$K_1^{c}(\lambda)$ as a function of $\lambda$ for fixed values of $K_2$, obtained using Eq.~\eqref{eq:K1c_general_model_II_C}. The dash-dotted line indicates $K_1^c (\lambda=0)=2D$, see Fig.~\ref{fig:bi-phdiag}. The dotted lines correspond to values of $K_2$ for which the bare model shows a continuous transition in $r_{1,\mathrm{nr}}^{\mathrm{st}}$, while the solid lines correspond to the values of $K_2$ for which the bare model shows a first-order transition in $r_{1,\mathrm{nr}}^{\mathrm{st}}$.}
    \label{fig: reentrant_window}
\end{figure}

The question then is: does the presence of the second-harmonic interaction add any new feature? The answer is in the affirmative and points to a remarkable effect of a re-entrant phase transition. With respect to panels (a) and (d) in Fig.~\ref{fig: model 4}, we see that in contrast to the corresponding plots in Figs.~\ref{fig: model 1} and~\ref{fig: model 2} (compare with panel (a) in each of these figures), we observe the following: With increase of $\lambda$, the transition point changes non-monotonically as a function of $\lambda$. Indeed, with increase of $\lambda$, the transition point shifts initially to the right and eventually to the left. To illustrate this effect more clearly, we show in Fig.~\ref{fig: reentrant_window} the transition point $K_1^c(\lambda)$ as a function of $\lambda$ for multiple values of $K_2$, namely, three for which the bare model shows a first-order transition (the values $K_2=2.2, 2.4, 3.0$) and the three for which it shows a continuous transition (the values $K_2=1.0, 1.4, 1.8$). The re-entrant nature is evident from the following feature: at a fixed $K_1$ satisfying $K_1^c(\lambda=0) < K_1 <  [K_1^c(\lambda)]_\mathrm{max}$, where $[K_1^c(\lambda)]_\mathrm{max}$ is the maximum value of $K_1^c(\lambda)$ attained over the whole range of $\lambda$, the system with increase of $\lambda$ goes successively from a disordered to an ordered and back to a disordered phase. The effect gets more pronounced as $K_2$ is increased. 

We now discuss the origin of the re-entrant behavior. As noted after Eq.~\eqref{eq:resetting_order_parameters}, the reset configuration is chosen such that $z_{1,\mathrm{r}}=0$ and $z_{2,\mathrm{r}}=1$. Thus, dynamically, the reset subsystem tends to suppress synchronization of the non-reset subsystem through the first harmonic interaction modulated by the coupling parameter $K_1$, while promoting synchronization through the second harmonic interaction modulated by the coupling parameter $K_2$ (see Eq.~\eqref{eq: describe well two sub 2}). These competing effects generate opposing influences during the resetting dynamics. For small resetting rates, the dominance of the effect of the first harmonic interaction between the reset and the non-reset subsystem over the second harmonic makes the system difficult to synchronize, leading to an increase of $K_1^\mathrm{c}(\lambda)$ with $\lambda$. On the other hand, when $\lambda$ is high, the second-harmonic interaction dominates, leading to a decrease of $K_1^\mathrm{c}(\lambda)$ with $\lambda$. Since the re-entrant effect originates from $z_{2,\mathrm{r}}$ being reset to $1$, which acts through the second-harmonic coupling $K_2$, the non-monotonicity of $K_1^\mathrm{c}(\lambda)$ becomes more pronounced with increasing $K_2$, as can be seen in Fig.~\ref{fig: reentrant_window}.

Re-entrant transitions generically refer to situations where an ordered phase exists only within a finite parameter window, and which disappears on varying a control parameter in either direction. For example, in the rotor synchronization model~\cite{Komarov_2014}, increasing coupling can induce synchronization, then suppress it, and later restore it, due to the competing effects of noise, inertia, and frequency dispersion that yield multiple self-consistent states. Similarly, in disordered solids~\cite{PhysRevB.27.2902}, a crystalline phase may exist only at intermediate temperatures, being destabilized at low temperatures by quenched disorder (via dislocation unbinding) and at high temperatures by thermal fluctuations. More generally, re-entrant behavior across statistical physics arises from competing mechanisms that favour and disrupt order in different regimes, leading to non-monotonic phase boundaries.

\section{Conclusion}
We have developed a general framework to study subsystem resetting in interacting many-body systems, focusing on Kuramoto-type models with both noisy and noiseless dynamics. Using a continued-fraction approach, we derived self-consistent relations for the stationary-state order parameter of the non-reset subsystem and demonstrated its applicability across models with different interaction structures. Our results show that subsystem resetting provides a powerful and flexible control mechanism for collective behavior, enabling systematic tuning of synchronization transitions, shifting of transition points, and the emergence of nontrivial features such as re-entrant behavior and phase-boundary restructuring. In contrast to global resetting, subsystem resetting preserves memory effects and leads to qualitatively new nonequilibrium phenomena. These findings establish subsystem resetting as a versatile tool for controlling interacting systems and open avenues for exploring its effects in more complex settings, including networks and experimentally-relevant platforms~\cite{Ginot_2026}. 

From a theoretical point of view, it would be interesting to ask how the effects of subsystem resetting depend on the dimension of the degree of freedom of the Kuramoto oscillators~\cite{SarthakODDdDifferent,ddimentional_Rupak}. Moreover, in reality, most of the synchronizing systems are made up of a finite number of interacting units. Hence, the natural extension would be to develop a finite-size theory of subsystem resetting to broaden its applicability~\cite{majumder2025finitesizefluctuationsstochasticcoupled}. Another direction to pursue is to consider models beyond the ones treated here, and investigate how the continued-fraction approach discussed in this work applies to more general nonlinear oscillator systems with complex coupling structures, and to network-coupled oscillators~\cite{RODRIGUES20161,PhysRevLett.96.240602,PhysRevLett.97.100601,PhysRevE.79.011102}.

\section{acknowledgments}
This work was supported by the Department of Atomic Energy, Government of India, under Project Identification Number RTI-4012. The computations were carried out on the computing clusters at the Department of Theoretical Physics, TIFR, Mumbai. We also thank Ajay Salve and Kapil Ghadiali for their computational support. 

\appendix

\section{Agreement with previous results \label{app: A}}

For the model discussed in Sec.~\ref{subsec: noiseless-harmonic-lorentzian Bare}, Ref~\cite{PhysRevE.109.064137} reported the time evolution equation of the average order parameter in the presence of subsystem resetting by using the Ott-Antonsen~\cite{Ott_2008, Ott_2009} ansatz. In this section, we will show that we can obtain the same equation using the general theory discussed in Sec.~\ref{sec: harmonic theory}, in particular, from Eq.~\eqref{eq: FP Kuramoto Finite Resetting}. While deriving, we will consider arbitrary noise strength $D$, but while making a comparison with Ref.~\cite{PhysRevE.109.064137}, we will put $D=0$. We start by taking the time derivative of Eq.~\eqref{eq: zx} for $\mathrm{x} = \mathrm{r}$, which gives
\begin{eqnarray}
    \frac{d z_\mathrm{r}}{dt}  &=& \int d\theta_\mathrm{r}d \omega_\mathrm{r}d\theta_\mathrm{nr}d\omega_\mathrm{nr} e^{i \theta_\mathrm{r}} g(\omega_\mathrm{r})g(\omega_\mathrm{nr})\frac{\partial P}{\partial t}. \label{eq: d zr dt}
\end{eqnarray}

We now use Eq.~\eqref{eq: FP Kuramoto Finite Resetting} in Eq.~\eqref{eq: d zr dt} to compute the rhs. Let us compute term by term. The first term in Eq.~\eqref{eq: FP Kuramoto Finite Resetting} gives
\begin{align}
    &\int d\theta_\mathrm{r}d \omega_\mathrm{r}d\theta_\mathrm{nr}d\omega_\mathrm{nr} e^{i \theta_\mathrm{r}} g(\omega_\mathrm{r})g(\omega_\mathrm{nr})\left[D\frac{\partial^2 P}{\partial \theta_\mathrm{r}^2}\right] \nonumber\\
    &=-D\int d\theta_\mathrm{r}d \omega_\mathrm{r}d\theta_\mathrm{nr}d\omega_\mathrm{nr} e^{i \theta_\mathrm{r}} g(\omega_\mathrm{r})g(\omega_\mathrm{nr})P (\theta_\mathrm{r},\theta_\mathrm{nr},\omega_\mathrm{r},\omega_\mathrm{nr},t)\nonumber\\
    &= -Dz_\mathrm{r}.
\end{align}
Here we have used twice integration by parts with respect to the variable $\theta_\mathrm{r}$. The boundary terms are zero due to the $2\pi$-periodicity in $\theta_\mathrm{r}$. The second term in Eq.~\eqref{eq: FP Kuramoto Finite Resetting} gives
\begin{align}
    \int d\theta_\mathrm{r}d \omega_\mathrm{r}d\theta_\mathrm{nr}d\omega_\mathrm{nr} e^{i \theta_\mathrm{r}} g(\omega_\mathrm{r})g(\omega_\mathrm{nr})\left[D\frac{\partial^2 P}{\partial \theta_\mathrm{nr}^2}\right] =0,
\end{align}
where we have used the $2\pi$-periodicity property of $\partial^2 P/\partial \theta_\mathrm{nr}^2$ in the variable $\theta_\mathrm{nr}$. The third term in Eq.~\eqref{eq: FP Kuramoto Finite Resetting} gives that
\begin{align}
    &\int d\theta_\mathrm{r}d \omega_\mathrm{r}d\theta_\mathrm{nr}d\omega_\mathrm{nr} e^{i \theta_\mathrm{r}} g(\omega_\mathrm{r})g(\omega_\mathrm{nr})\left[-\frac{\partial (Ph_\mathrm{r})}{\partial \theta_\mathrm{r}}\right] \nonumber\\
   &= i\int d\theta_\mathrm{r}d \omega_\mathrm{r}d\theta_\mathrm{nr}d\omega_\mathrm{nr} e^{i \theta_\mathrm{r}} g(\omega_\mathrm{r})g(\omega_\mathrm{nr})Ph_\mathrm{r} \nonumber\\
   &= i\int d\theta_\mathrm{r}d \omega_\mathrm{r}d\theta_\mathrm{nr}d\omega_\mathrm{nr} e^{i \theta_\mathrm{r}} g(\omega_\mathrm{r})g(\omega_\mathrm{nr})P\bigg[\omega_\mathrm{r} \nonumber\\
   &+K_1f \int d\theta'_\mathrm{r}d\omega'_\mathrm{r} g(\omega'_\mathrm{r}) P(\theta'_\mathrm{r},\omega'_\mathrm{r} ,t| \theta_\mathrm{r} ,\omega_\mathrm{r})\sin(\theta'_\mathrm{r}-\theta_\mathrm{r}) \nonumber\\
    & +K_1\bar{f} \int d\theta'_\mathrm{nr}d\omega'_\mathrm{nr} g(\omega'_\mathrm{nr})P(\theta'_\mathrm{nr},\omega'_\mathrm{nr} ,t|  \theta_\mathrm{r},\omega_\mathrm{r}) \sin(\theta'_\mathrm{nr}-\theta_\mathrm{r})\bigg], \label{eq: midway 1}
\end{align}
where in obtaining the second equality, we have substituted the form of $h_r$ from Eq.~\eqref{eq: hx K1 K2 bare}.  Let us compute the terms in the last equality above one by one. We focus on computing the first term in Eq.~\eqref{eq: midway 1}. Since the probability $P (\theta_\mathrm{r},\theta_\mathrm{nr},\omega_\mathrm{r},\omega_\mathrm{nr},t)$ is periodic in $\theta_\mathrm{r},\theta_\mathrm{nr}$, we may expand it in a Fourier series similar to Eq.~\eqref{eq: fourier} with the additional fact that the Fourier coefficient $\mathcal{P}_{l,m}$ is now a function of time, i.e.,  we have $\mathcal{P}_{l,m}(\omega_\mathrm{r},\omega_\mathrm{nr}, t)$. In its term, we may express the order parameter as $z_\mathrm{r}=\int d \omega_\mathrm{r}d\omega_\mathrm{nr} g(\omega_\mathrm{r})g(\omega_\mathrm{nr})\mathcal{P}_{-1,0}(\omega_\mathrm{r},\omega_\mathrm{nr}, t)$. To perform the frequency integrals, we analytically continue the functions in the integrand in the complex-$\omega_\mathrm{r}$ plane, and then convert the integral into a contour integral in that plane. Following a similar argument as presented in Ref.~\cite{Ott_2008}, here also we obtain that $|\mathcal{P}_{-1,0}(\omega_\mathrm{r},\omega_\mathrm{nr}, t)|\to 0$ as $\omega_\mathrm{r}\to +i\infty$. Hence, we close the contour in the upper-half complex-$\omega_\mathrm{r}$ plane. The frequency distribution $g$ being Lorentzian, it has a simple pole in the upper-half plane at $\omega_\mathrm{r}=i\sigma$. Hence, using the residue theorem, we obtain $z_\mathrm{r}=\int d\omega_\mathrm{nr} g(\omega_\mathrm{nr})\mathcal{P}_{-1,0}(i\sigma,\omega_\mathrm{nr}, t)$. Using this, we may reduce the first term in Eq.~\eqref{eq: midway 1} as follows:
\begin{align}
   &i \int d\theta_\mathrm{r}d \omega_\mathrm{r}d\theta_\mathrm{nr}d\omega_\mathrm{nr} e^{i \theta_\mathrm{r}} g(\omega_\mathrm{r})g(\omega_\mathrm{nr})P\omega_\mathrm{r} \nonumber\\
   &=i \int d \omega_\mathrm{r}d\omega_\mathrm{nr} g(\omega_\mathrm{r})g(\omega_\mathrm{nr})\mathcal{P}_{-1,0}(\omega_\mathrm{r},\omega_\mathrm{nr}, t)\omega_\mathrm{r} \nonumber\\
    &=i \int d\omega_\mathrm{nr} g(\omega_\mathrm{nr})\Big[\mathcal{P}_{-1,0}(i\sigma,\omega_\mathrm{nr}, t)\big(i\sigma\big)\Big]\nonumber\\
    &= -\sigma z_\mathrm{r}.
\end{align}

The other term is
\begin{align}
    &i\int d\theta_\mathrm{r}d \omega_\mathrm{r}d\theta_\mathrm{nr}d\omega_\mathrm{nr} e^{i \theta_\mathrm{r}} g(\omega_\mathrm{r})g(\omega_\mathrm{nr})P (\theta_\mathrm{r},\theta_\mathrm{nr},\omega_\mathrm{r},\omega_\mathrm{nr},t)\nonumber\\
   & \times\bigg[K_1f \int d\theta'_\mathrm{r}d\omega'_\mathrm{r} g(\omega'_\mathrm{r}) P(\theta'_\mathrm{r},\omega'_\mathrm{r} ,t| \theta_\mathrm{r} ,\omega_\mathrm{r})\sin(\theta'_\mathrm{r}-\theta_\mathrm{r})\bigg] \nonumber\\
   &=iK_1f\int d\theta_\mathrm{r}d \omega_\mathrm{r}d\theta_\mathrm{nr}d\omega_\mathrm{nr}d\theta'_\mathrm{r}d\omega'_\mathrm{r}  e^{i \theta_\mathrm{r}}g(\omega_\mathrm{r})g(\omega_\mathrm{nr}) g(\omega'_\mathrm{r}) \nonumber\\
   &\times P (\theta_\mathrm{r},\theta_\mathrm{nr},\omega_\mathrm{r},\omega_\mathrm{nr},t) P(\theta'_\mathrm{r},\omega'_\mathrm{r} ,t| \theta_\mathrm{r} ,\omega_\mathrm{r})\sin(\theta'_\mathrm{r}-\theta_\mathrm{r}) \nonumber\\
   &=\frac{K_1f}{2}\int d\theta_\mathrm{r}d \omega_\mathrm{r}d\theta_\mathrm{nr}d\omega_\mathrm{nr}d\theta'_\mathrm{r}d\omega'_\mathrm{r}  g(\omega_\mathrm{r})g(\omega_\mathrm{nr}) g(\omega'_\mathrm{r}) \nonumber\\
   &\times P (\theta_\mathrm{r},\theta_\mathrm{nr},\omega_\mathrm{r},\omega_\mathrm{nr},t) P(\theta'_\mathrm{r},\omega'_\mathrm{r} ,t| \theta_\mathrm{r} ,\omega_\mathrm{r})\bigg[e^{i \theta'_\mathrm{r}}-e^{-i( \theta'_\mathrm{r}-2\theta_\mathrm{r})}\bigg].
\end{align}
To proceed further, we assume that the approximation~\eqref{eq: con approx}, 
assumed to be valid in the stationary state, is valid at all times, which immediately gives
\begin{align}
    P(\theta'_\mathrm{r},\omega'_\mathrm{r} ,t| \theta_\mathrm{r} ,\omega_\mathrm{r}) \approx P(\theta'_\mathrm{r},\omega'_\mathrm{r} ,t).
\end{align}
Under the above approximation, we obtain
\begin{align}
    &\frac{K_1f}{2}\int d\theta_\mathrm{r}d \omega_\mathrm{r}d\theta_\mathrm{nr}d\omega_\mathrm{nr}d\theta'_\mathrm{r}d\omega'_\mathrm{r}  g(\omega_\mathrm{r})g(\omega_\mathrm{nr}) g(\omega'_\mathrm{r}) \nonumber\\
   &\times P (\theta_\mathrm{r},\theta_\mathrm{nr},\omega_\mathrm{r},\omega_\mathrm{nr},t) P(\theta'_\mathrm{r},\omega'_\mathrm{r} ,t| \theta_\mathrm{r} ,\omega_\mathrm{r})\bigg[e^{i \theta'_\mathrm{r}}-e^{-i( \theta'_\mathrm{r}-2\theta_\mathrm{r})}\bigg] \nonumber\\
   &= \frac{K_1f}{2}\bigg[ \int d\theta'_\mathrm{r}d\omega'_\mathrm{r}  g(\omega'_\mathrm{r}) P(\theta'_\mathrm{r},\omega'_\mathrm{r} ,t)e^{i \theta'_\mathrm{r}} \nonumber\\
   &-\int d\theta'_\mathrm{r}d\omega'_\mathrm{r}  g(\omega'_\mathrm{r}) P(\theta'_\mathrm{r},\omega'_\mathrm{r} ,t)e^{-i \theta'_\mathrm{r}}\int d\theta_\mathrm{r}d \omega_\mathrm{r}d\theta_\mathrm{nr}d\omega_\mathrm{nr} g(\omega_\mathrm{r})\nonumber\\
   &\times g(\omega_\mathrm{nr})P (\theta_\mathrm{r},\theta_\mathrm{nr},\omega_\mathrm{r},\omega_\mathrm{nr},t)e^{i2 \theta_\mathrm{r}}\bigg] \nonumber\\
   &= \frac{K_1f}{2}\bigg[z_\mathrm{r}-z^*_\mathrm{r} \Big\langle e^{i2 \theta_\mathrm{r}} \Big\rangle \bigg].
\end{align}

Doing a similar calculation, we obtain
\begin{align}
    &i\int d\theta_\mathrm{r}d \omega_\mathrm{r}d\theta_\mathrm{nr}d\omega_\mathrm{nr} e^{i \theta_\mathrm{r}} g(\omega_\mathrm{r})g(\omega_\mathrm{nr})P (\theta_\mathrm{r},\theta_\mathrm{nr},\omega_\mathrm{r},\omega_\mathrm{nr},t)\bigg[\nonumber\\
   & K_1\bar{f} \int d\theta'_\mathrm{nr}d\omega'_\mathrm{nr} g(\omega'_\mathrm{nr}) P(\theta'_\mathrm{nr},\omega'_\mathrm{nr} ,t| \theta_\mathrm{r} ,\omega_\mathrm{r})\sin(\theta'_\mathrm{nr}-\theta_\mathrm{r})\bigg] \nonumber\\
   &= -\frac{K_1\bar{f}}{2}\bigg[z_\mathrm{nr}-z^*_\mathrm{nr} \Big\langle e^{i2 \theta_\mathrm{r}} \Big\rangle \bigg].
\end{align}
Thus, the third term in Eq.~\eqref{eq: FP Kuramoto Finite Resetting} finally gives
\begin{align}
    &\int d\theta_\mathrm{r}d \omega_\mathrm{r}d\theta_\mathrm{nr}d\omega_\mathrm{nr} e^{i \theta_\mathrm{r}} g(\omega_\mathrm{r})g(\omega_\mathrm{nr})\left[-\frac{\partial (Ph_\mathrm{r})}{\partial \theta_\mathrm{r}}\right] \nonumber\\
    &= -\sigma z_\mathrm{r}+\frac{K_1}{2}\Big[f z_\mathrm{r} + \bar{f}z_\mathrm{nr} \Big]-\frac{K_1}{2}\Big[f z^{*}_\mathrm{r} + \bar{f}z^{*}_\mathrm{nr} \Big]\Big\langle e^{i2 \theta_\mathrm{r}} \Big\rangle.
\end{align}

We now focus on the contribution from the fourth term in Eq.~\eqref{eq: FP Kuramoto Finite Resetting}, which gives
\begin{equation}
    \int d\theta_\mathrm{r}d \omega_\mathrm{r}d\theta_\mathrm{nr}d\omega_\mathrm{nr} e^{i \theta_\mathrm{r}} g(\omega_\mathrm{r})g(\omega_\mathrm{nr})\left[-\frac{\partial (Ph_\mathrm{nr})}{\partial \theta_\mathrm{nr}}\right] =0,
\end{equation}
due to the $2\pi$-periodicity of $P$. The fifth term in Eq.~\eqref{eq: FP Kuramoto Finite Resetting} simply gives $-\lambda z_\mathrm{r}$. The sixth term in Eq.~\eqref{eq: FP Kuramoto Finite Resetting} gives
\begin{align}
    &\int d\theta_\mathrm{r}d \omega_\mathrm{r}d\theta_\mathrm{nr}d\omega_\mathrm{nr} e^{i \theta_\mathrm{r}} g(\omega_\mathrm{r})g(\omega_\mathrm{nr})\bigg[\lambda \Big\{ \alpha \delta(\theta_\mathrm{r}) \nonumber\\
    & + (1-\alpha) \delta(\theta_\mathrm{r}-\pi)\Big\}\int d \theta'_\mathrm{r}d \omega'_\mathrm{r} g(\omega'_\mathrm{r}) P(\theta'_\mathrm{r},\theta_\mathrm{nr},\omega'_\mathrm{r},\omega_\mathrm{nr},t)\bigg]\nonumber\\
    &=\lambda\int d\theta_\mathrm{r} e^{i \theta_\mathrm{r}}\Big[ \alpha \delta(\theta_\mathrm{r}) + (1-\alpha) \delta(\theta_\mathrm{r}-\pi)\Big] \nonumber\\
    &= \lambda(2\alpha-1)=\lambda r_0.
\end{align}

Hence, combining everything, we obtain
\begin{align}
     \frac{d z_\mathrm{r}}{dt} &=-(D+\sigma)z_\mathrm{r}+\frac{K_1}{2}\Big[f z_\mathrm{r} + \bar{f}z_\mathrm{nr} \Big]\nonumber\\
     &-\frac{K_1}{2}\Big[f z^{*}_\mathrm{r}
     + \bar{f}z^{*}_\mathrm{nr} \Big]\Big\langle e^{i2 \theta_\mathrm{r}} \Big\rangle+\lambda\big(r_0-z_\mathrm{r}\big).
\end{align}
Doing a similar computation starting from
\begin{eqnarray}
    \frac{d z_\mathrm{nr}}{dt}  &=& \int d\theta_\mathrm{r}d \omega_\mathrm{r}d\theta_\mathrm{nr}d\omega_\mathrm{nr} e^{i \theta_\mathrm{nr}} g(\omega_\mathrm{r})g(\omega_\mathrm{nr})\frac{\partial P}{\partial t}, \label{eq: d znr dt}
\end{eqnarray}
we further obtain
\begin{align}
     \frac{d z_\mathrm{nr}}{dt} &=-(D+\sigma)z_\mathrm{nr}+\frac{K_1}{2}\Big[f z_\mathrm{r} + \bar{f}z_\mathrm{nr} \Big]\nonumber\\
     &-\frac{K_1}{2}\Big[f z^{*}_\mathrm{r}
     + \bar{f}z^{*}_\mathrm{nr} \Big]\Big\langle e^{i2 \theta_\mathrm{nr}} \Big\rangle.
\end{align}

If we further approximate 
\begin{align}
    \Big\langle e^{i2 \theta_\mathrm{r}} \Big\rangle &\approx \Big\langle e^{i \theta_\mathrm{r}} \Big\rangle^2=z^2_\mathrm{r},\\
    \Big\langle e^{i2 \theta_\mathrm{nr}} \Big\rangle&\approx \Big\langle e^{i \theta_\mathrm{nr}} \Big\rangle^2=z^2_\mathrm{nr},
\end{align}
then we get the closed-form evolution equation
\begin{align}
     \frac{d z_\mathrm{r}}{dt} &=-(D+\sigma)z_\mathrm{r}+\frac{K_1}{2}\Big[f z_\mathrm{r} + \bar{f}z_\mathrm{nr} \Big]\nonumber\\
     &-\frac{K_1}{2}\Big[f z^{*}_\mathrm{r}
     + \bar{f}z^{*}_\mathrm{nr} \Big]z^2_\mathrm{r}+\lambda\big(r_0-z_\mathrm{r}\big),\\
     \frac{d z_\mathrm{nr}}{dt} &=-(D+\sigma)z_\mathrm{nr}+\frac{K_1}{2}\Big[f z_\mathrm{r} + \bar{f}z_\mathrm{nr} \Big]-\frac{K_1}{2}\Big[f z^{*}_\mathrm{r}
     \nonumber \\
     &+ \bar{f}z^{*}_\mathrm{nr} \Big]z^2_\mathrm{nr}.
\end{align}

In the absence of noise ($D=0$), and in the stationary state, the order parameters are obtained from the simultaneous solution of the following equations:
\begin{align}
    &-\sigma z_\mathrm{r}+\frac{K_1}{2}\Big[f z_\mathrm{r} + \bar{f}z_\mathrm{nr} \Big]-\frac{K_1}{2}\Big[f z^{*}_\mathrm{r}
     + \bar{f}z^{*}_\mathrm{nr} \Big]z^2_\mathrm{r}\nonumber\\
     &+\lambda\big(r_0-z_\mathrm{r}\big)=0,\\
     &-\sigma z_\mathrm{nr}+\frac{K_1}{2}\Big[f z_\mathrm{r} + \bar{f}z_\mathrm{nr} \Big]-\frac{K_1}{2}\Big[f z^{*}_\mathrm{r}
     + \bar{f}z^{*}_\mathrm{nr} \Big]z^2_\mathrm{nr}=0.
\end{align}
These equations were obtained independently in Ref.~\cite{PhysRevE.109.064137} using the Ott-Antonsen ansatz, and we show here that they can also be derived without recourse to the ansatz.
\section{Correction in $\gamma_2$ for the case of $K_1\neq0$ and $K_2\neq0$ \label{app: B}}

Here we will show that close to $\gamma_1=0$, the correction in $\gamma_2$ appears as $\mathcal{O}( |\gamma_1|^2)$. In Eq.~\eqref{eq:gamma2_background}, we have derived $\gamma_2^{(0)}$ at $\gamma_1=0$. To evaluate the contribution $\delta \gamma_2=\gamma_2-\gamma_2^{(0)}$, induced by $ \gamma_1$ close to $\gamma_1=0$, we rewrite Eq.~\eqref{eq: gamma general} as
\begin{align}
    \gamma_2 \equiv \frac{K_2}{2}\left(f z^\mathrm{st}_{2,\mathrm{r}}+\bar{f} z^\mathrm{st}_{2,\mathrm{nr}}\right),\label{eq:gamma_2_with_z2r_z2nr} 
\end{align}
where $z^\mathrm{st}_{2,\mathrm{r}}$ and $z^\mathrm{st}_{2,\mathrm{nr}}$ are given by Eqs.~\eqref{eq; z st non-reset subsystem L2 2} and \eqref{eq; z st non-reset subsystem M2 2}. Note that $z^\mathrm{st}_{2,\mathrm{r}}$ and $z^\mathrm{st}_{2,\mathrm{nr}}$ have the form $z^\mathrm{st}_{2,\mathrm{r}}=\int_{-\infty}^{+\infty} d \omega_\mathrm{r} g(\omega_\mathrm{r})I_\mathrm{r}(\omega_\mathrm{r})$ and $z^\mathrm{st}_{2,\mathrm{nr}}=\int_{-\infty}^{+\infty} d \omega_\mathrm{nr} g(\omega_\mathrm{nr})I_\mathrm{nr}(\omega_\mathrm{nr})$, where the integrands $I_\mathrm{r}$ and $I_\mathrm{nr}$ are give by
\begin{align}
I_\mathrm{r}&\equiv\Gamma^{*}_{2,2}+4\pi^2 \Delta^{*}_2\nonumber \\
& +\Gamma^{*}_{1,2} \left[\frac{\Gamma^{*}_{1,1}+\Gamma_{1,1}\Gamma^{*}_{2,1} + 4\pi^2\left(\Delta^{*}_1+\Delta_1\Gamma^{*}_{2,1}\right)}{\left(1-\Gamma^{*}_{2,1}\Gamma_{2,1}\right)}\right],\nonumber\\
&\label{eq:z2r_vs_z2nr}\\
I_\mathrm{nr}&\equiv\Lambda^{*}_{1,2} \left[\frac{\Lambda^{*}_{1,1}+\Lambda_{1,1}\Lambda^{*}_{2,1} }{1-\Lambda^{*}_{2,1}\Lambda_{2,1}}\right] +\Lambda^{*}_{2,2}.\nonumber
\end{align}
Now, as explained below Eqs.~\eqref{eq:Gamma_1l_0}, at $\gamma_1=0$, the continued fractions $\Gamma_{1,l}$ and $\Lambda_{1,l}$ vanish; similarly $\Delta_1$ vanish since $b_1=0$
for $\alpha=1/2$. Therefore, both the square-bracketed terms in Eq.~\eqref{eq:z2r_vs_z2nr} vanish at $\gamma_1=0$.

Let us now focus on the case where $\gamma_1$ is small but nonzero. From the expression of $\Gamma_{1,l}$ in Eq.~\eqref{eq: gamma 1 l}, a Taylor expansion in the small parameter $\gamma_1$ shows that, to leading order, $\Gamma_{1,l} \approx \mathcal{O}(|\gamma_1|)$. For odd $l$ with $\alpha = 1/2$, we have $b_l=0$. Using this in Eq.~\eqref{eq: Delta l} and performing Taylor expansion in the small parameter $\gamma_1$ shows that, to leading order we have $\Delta_l \approx \mathcal{O}(|\gamma_1|)$ for odd $l$. Using these results, we conclude that the third term in the definition of $I_\mathrm{r}$ of the order $|\gamma_1|^2$ in the leading order. Using similar argument in Eq.~\eqref{eq: gamma 2 l} we obtain that
\begin{align}
    \Gamma_{2,2}&= \Gamma_{2,2}^{(0)}+\mathcal{O}(\delta\gamma_2)+\mathcal{O}(|\gamma_1|^2),\\
    \Delta_2 &= \Delta_2^{(0)}+\mathcal{O}(\delta\gamma_2)+\mathcal{O}(|\gamma_1|^2).
\end{align}
Using these results into the definition of $I_\mathrm{r}$, we obtain that
\begin{align}
I_\mathrm{r} =\big[\Gamma_{2,2}^{(0)} \big]^{*}+4\pi^2 \big[\Delta_2^{(0)}\big]^{*}+\mathcal{O}(\delta\gamma_2)+\mathcal{O}(|\gamma_1|^2).\label{eq:z2r_expanded}
\end{align}
Using similar small $\gamma_1$ expansion in Eqs.~\eqref{eq: lambda 1 m},~\eqref{eq: lambda 2 m}, and~\eqref{eq: Pi m} we obtain that
\begin{equation}
    I_\mathrm{nr} = \big[\Lambda_{2,2}^{(0)}\big]^{*} + \mathcal{O}(\delta\gamma_2)+\mathcal{O}(|\gamma_1|^2). \label{eq:z2nr_expanded}
\end{equation}
Now, using Eqs.~\eqref{eq:z2r_expanded},~\eqref{eq:z2nr_expanded}
in Eq.~\eqref{eq:gamma_2_with_z2r_z2nr} and writing
$\gamma_2 = \gamma_2^{(0)}+\delta\gamma_2$ in the lhs (left hand side) of the equation yield
\begin{widetext}
\begin{align}
    \gamma_2^{(0)} + \delta\gamma_2
    &= \frac{K_2 f}{2}
       \!\biggl(\big[\Gamma_{2,2}^{(0)}\big]^{*}
               +4\pi^2\big[\Delta_2^{(0)}\big]^{*}
               +\mathcal{O}(\delta\gamma_2)+\mathcal{O}(|\gamma_1|^2)\biggr)
    +\frac{K_2\bar{f}}{2}
     \!\biggl(\big[\Lambda_{2,2}^{(0)}\big]^{*}
             +\mathcal{O}(\delta\gamma_2)
             +\mathcal{O}(|\gamma_1|^2)\biggr).
\end{align}
\end{widetext}
Note that the zeroth-order terms cancel exactly using
Eq.~\eqref{eq:gamma2_background}, which leaves
\begin{align}
    \delta\gamma_2
    = \mathcal{O}(\delta\gamma_2) + \mathcal{O}(|\gamma_1|^2),
    \label{eq:delta_gamma2_eq}
\end{align}
Equation.~\eqref{eq:delta_gamma2_eq} readily gives $\delta\gamma_2 \sim \mathcal{O}(|\gamma_1|^2)$. Consequently, to first order in $\delta\gamma_1$, the quantity $\gamma_2$ may be held at $\gamma_2^{(0)}$ with corrections only appearing in the $\gamma_1$ self-consistency equation at order
$\mathcal{O}(|\gamma_1|^2)$.

\bibliography{manuscript}
\bibliographystyle{unsrt}

\end{document}